\documentclass[aps,prb,reprint,nofootinbib,superscriptaddress,twocolumn]{revtex4-2}
\usepackage[T1]{fontenc}
\usepackage[utf8]{inputenc}
\usepackage{amsmath}
\usepackage{amsfonts}
\usepackage{amssymb}
\usepackage{newtxtext,newtxmath}
\usepackage{graphicx}
\usepackage{xcolor}
\usepackage{hyperref}
\hypersetup{colorlinks=true,linkcolor=blue,urlcolor=blue,citecolor=blue}
\usepackage{sublabel}
\usepackage{lipsum}

\usepackage{soul}

\begin{document}

\title{Parameter space geometry of the quartic oscillator and the double well potential: Classical and quantum description}

\author{Diego Gonzalez}
\email{dgonzalezv@ipn.mx}
\affiliation{Escuela Superior de Ingeniería Mecánica y Eléctrica - Instituto Politécnico Nacional (ESIME - IPN), 07738, CDMX, Mexico}
\affiliation{Departamento de F\'{i}sica, Cinvestav, Avenida Instituto Polit\'{e}cnico Nacional 2508, San Pedro Zacatenco,\\
		 07360 Gustavo A. Madero, Ciudad de M\'exico, M\'exico}
   
\author{Jorge Ch\'avez-Carlos}
\email{jorge.chavez_carlos@uconn.edu}
\affiliation{Department of Physics, University of Connecticut, Storrs, Connecticut 06269, USA.}

\author{Jorge G. Hirsch}
\email{hirsch@nucleares.unam.mx}
\affiliation{Instituto de Ciencias Nucleares, Universidad Nacional Aut\'onoma de M\'exico, Apartado Postal 70-543, Ciudad de M\'exico, 04510, M\'exico}

\author{J. David Vergara}
\email{vergara@nucleares.unam.mx}
\affiliation{Instituto de Ciencias Nucleares, Universidad Nacional Aut\'onoma de M\'exico, Apartado Postal 70-543, Ciudad de M\'exico, 04510, M\'exico}

\date{\today}

\begin{abstract}
	
We compute both analytically and numerically the geometry of the parameter space of the anharmonic oscillator employing the quantum metric tensor and its scalar curvature. A novel semiclassical treatment based on a Fourier decomposition allows to construct classical analogues of the quantum metric tensor and of the expectation values of the transition matrix elements. A detailed comparison is presented between exact quantum numerical results, a perturbative quantum description and the semiclassical analysis. They are shown to coincide for both positive and negative quadratic potentials, where the potential displays a double well. Although the perturbative method is unable to describe the region where the quartic potential vanishes, it is remarkable that both the perturbative and semiclassical formalisms recognize the negative oscillator parameter at which the ground state starts to be delocalized in two wells.

\end{abstract} 

\maketitle

\section{Introduction}

Exploring the quantum parameter space's geometry has led to captivating insights into various physical systems \cite{Provost, Zanardi2007Information,  SarkarGeodesics2012, GuReview, Carollo2020}. Within these investigations, a prominent element is the quantum geometric tensor (QGT), encompassing the quantum metric tensor (QMT) in its real part and the Berry curvature in its imaginary part. This tensor holds significance in quantum information processing applications, such as adiabatic and holonomic quantum computing \cite{Zanardi1999}. In these applications, minimizing errors corresponds to traversing geodesics on the control parameter manifold \cite{Rezakhani2010}. The QGT plays a vital role in describing quantum phase transitions (QPT) in the thermodynamic limit and provides valuable information about the precursors to these transitions when dealing with systems comprising a finite number of particles. Furthermore, following the ideas of Ozawa and Goldman \cite{Ozawa1, Ozawa2}, recently has been possible to measure the QMT experimentally \cite{Tan2019, Yu2020}, showing that the QGT could be an essential tool to describe quantum phenomena by example, the observations  of the quantum geometry enabled an evaluation of the quantum Cramér-Rao bound (QCRB) \cite{Cappellaro2022}.

In this study, we extensively examine the geometry of the quantum parameter space for the anharmonic oscillator. To our knowledge, a systematic investigation of the geometric properties of the quantum parameter space of the anharmonic oscillator has not been undertaken. Therefore, the primary objective of this article is to fill this gap by conducting a thorough analysis of the quantum geometric tensor of this system.

The anharmonic oscillator is a system that has been analyzed from many points of view. In the original articles by Bender and Wu \cite{Bender1, Bender2}, it was shown that, using semiclassical methods of WKB, the large $n-$ behavior in the series expansion of the energy levels of the perturbation calculus is divergent. Later this study was followed by the perturbative analysis of Lipatov \cite{Lipatov} and Br\'ezin, et al \cite{Brezin}, showing that  the presence of the instanton connecting the two minima of the potential is responsible for the non-Borel summability of the perturbation expansion of the ground state energy. The anharmonic oscillator has served as a test system, useful to extend the results to quantum field theory, and its large order behavior has been analyzed using instanton contributions to the path integral representation of Green's functions \cite{Zinn1, Zinn2, Zinn3, Zinn4}. These studies were extended considering the contributions of multi-instantons to the ground state in \cite{Zinn5} and to excited states in \cite{Zinn6}. 

In Ref. \cite{Caswell}, it was shown that a summable perturbation series exists for the double-well potential employing an effective coupling. Subsequently, in \cite{Turbiner1} a fast convergence in perturbation theory was introduced using a simple uniform approximation of the logarithmic derivative of the ground state eigenfunction. This technique transforms the one-dimensional Schr\"odinger equation into a Riccati form, using the logarithmic derivative of the wave function \cite{Zinn7}. Using this approach, in \cite{Turbiner2} approximate eigenfunctions were obtained for the quartic anharmonic oscillator, and also for the double-well potential \cite{Turbiner3,Nader2023}. 

In the present article, we extend these studies of the anharmonic oscillator by employing a different approach:
a geometric analysis of the parameter space. To do this, we introduce a notion of distance in this space using the quantum metric tensor developed by Provost et al \cite{Provost} and Zanardi, et al \cite{Zanardi2007Information}. With the help of this metric, we can reconstruct all the geometric information of the parameter space, including the  scalar curvature, which shows quite interesting behaviors for both positive and negative oscillator parameters.

We employ three techniques to construct the quantum metric tensor for the ground state of the anharmonic oscillator, which has a double-well potential in the case of negative oscillator parameters. We obtain an exact numerical description performing a diagonalization in a truncated harmonic oscillator basis \cite{Okun}, including a careful analysis of the convergence of all the relevant observables, along with the perturbative method outlined by Zanardi et al \cite{Zanardi2007Information}. These numerical results are compared with a perturbative procedure of nonlinearization to express the wave function as a power series in $\lambda$~\cite{Turbiner1984}, and with a semiclassical procedure employing Fourier series \cite{Diego5}. This particular approach involves constructing a classical analog of the quantum metric tensor \cite{Alvarez2017}, offering the advantage of enabling the derivation of the classical equivalent of the matrix element of an operator between the ground state and any excited state.

\section{Geometry of the parameter space}

In this section, we introduce some basic aspects involved in parameter space associated with quantum and classical systems, as well as fix some notations.   

\subsection{Quantum metric tensor}

We start by considering a one-dimensional quantum system with a Hamiltonian $\hat{H}(\hat{q},\hat{p};x)$ that depends on a set of $\mathfrak{m}$ real adiabatic parameters denoted by $x=\{x^{i}\}$ $(i=1,...,\mathfrak{m})$. It is assumed that the Hamiltonian has at least one 
eigenvector $|\Psi_n(x)\rangle$ with nondegenerate eigenvalue $E_n(x)$. Using this eigenvector, the components of the quantum metric tensor (QMT) defined in the $m$-dimensional parameter space $\mathcal{M}$ of the system are given by~\cite{Provost}
 \begin{equation}
g^{(n)}_{ij} := {\rm Re} \left( \langle\partial_{i}\Psi_n|\partial_{j}\Psi_n\rangle-\langle\partial_{i}\Psi_n|\Psi_n\rangle\langle \Psi_n|\partial_{j}\Psi_n\rangle \right),\label{QMT}
 \end{equation}
where $\partial_{i}:=\frac{\partial}{\partial x^{i}}$. This metric provides the distance  between the eigenvectors $|\Psi_n(x)\rangle$ and $|\Psi_n(x+\delta x)\rangle$ with infinitesimally different parameters, namely $\delta\ell^{2}=g_{ij}^{(n)}(x)\delta x^{i}\delta x^{j}$. For the purposes of this work, it is convenient to introduce the perturbative form of this metric~\cite{Zanardi2007Information}
\begin{equation}
g^{(n)}_{ij} = {\rm Re} \sum_{m\neq n}\frac{\langle \Psi_n|\hat{O}_{i}|\Psi_m\rangle\langle \Psi_m|\hat{O}_{j}|\Psi_n\rangle}{(E_{m}-E_{n})^{2}},\label{QMTpert}
\end{equation}
where the operators $\hat{O}_{i}\equiv \partial_{i}\hat{H}$. Notice that to evaluate the expression \eqref{QMTpert} all the  matrix elements must be known. The advantage of this expression is that it shows that the components of the QMT are singular at the points $x^{*} \in \mathcal{M}$ of the parameter space such that $E_{m}(x^{*})=E_{n}(x^{*})$. This means that at the critical points of the QPT, the components of the QMT are singular. However, in some cases, a more detailed analysis is required \cite{Zanardi2007Scaling,Carollo2020}. For the purposes of this study, we set $n=0$ and write the QMT \eqref{QMTpert} as 
\begin{equation}
    g^{(0)}_{ij} = {\rm Re} \sum_{m=1}^{\infty}G^{(m)}_{ij}, \label{QMTpert2}
\end{equation}
with
\begin{equation}
G^{(m)}_{ij} := \frac{B^{(m)}_i (B^{(m)}_j)^*}{(E_{m}-E_{0})^{2}},\label{QMTmterm}
\end{equation}
where $*$ denotes the complex conjugate and we have defined the transtion matrix elements $B^{(m)}_i := \langle \Psi_0|\partial_{i}\hat{H}|\Psi_m\rangle$.

To find out whether the resulting singularities are genuine or merely a result of the parameter space’s coordinates, we can resort to the scalar curvature $R$, which does not depend on the coordinates used.

In particular, for a two-dimensional space endowed
with a metric tensor (the QMT, in this case), the scalar curvature has the simple form \cite{Sokol}
\begin{equation}
R=\frac{1}{\sqrt{|g|}}({\cal A}+{\cal B}),\label{scalar}
\end{equation}
where  $g=\det[g_{ij}]$ is the determinant of the metric and the quantities ${\cal A}$ and ${\cal B}$ are defined as
\begin{align}
{\cal A} & :=\partial_{1}\left(\frac{g_{12}}{g_{11}\sqrt{|g|}}\partial_{2}g_{11}-\frac{1}{\sqrt{|g|}}\partial_{1}g_{22}\right),\nonumber \\
{\cal B} & :=\partial_{2}\left(\frac{2}{\sqrt{|g|}}\partial_{1}g_{12}-\frac{1}{\sqrt{|g|}}\partial_{2}g_{11}-\frac{g_{12}}{g_{11}\sqrt{|g|}}\partial_{1}g_{11}\right).
\end{align} 

\subsection{Classical metric tensor}

Let us now consider a one-dimensional  classical integrable system described by a Hamiltonian $H(q,p;x)$ depending the set of real adiabatic parameters $x=\{x^{i}\}$ $(i=1,\dots,\mathfrak{m}$). The natural coordinates for this type of systems are the action-angle variables $\{I,\phi\}$, which allow to write the Hamiltonian as $H(I;x)=H(q(\phi,I;x),p(\phi,I;x);x)$ and can be used to define the torus average of a function $f(I,\phi;x)$ as
\begin{equation}\label{classavg}
	\langle f(\phi,I;x)\rangle=\frac{1}{(2\pi)^{N}}\intop_{0}^{2\pi}\mathrm{d}^{N}\phi\,f(\phi,I;x).
\end{equation} 

In this setting, the classical analog of the quantum metric tensor~\eqref{QMT} is\footnote{See Ref.~\cite{GonzalezPRE} for an alternative expression of this metric.}~\cite{GonzalezAnnalen}
\begin{align}
	g_{ij}(I;x)	=&-\intop_{-\infty}^{0}{\rm d}t_{1}\intop_{0}^{\infty}{\rm d}t_{2}\,\bigg( \langle{\cal O}_{i}(t_{1}){\cal O}_{j}(t_{2})\rangle_{0} \nonumber\\
	&-\langle{\cal O}_{i}(t_{1})\rangle_{0}\langle{\cal O}_{j}(t_{2})\rangle_{0}\bigg),\label{CMT}
\end{align}
where $\langle \cdot \rangle_{0}$ means that the classical average \eqref{classavg} is taken over the initial ($t=0$) angle variable $\phi_{0}$ and ${\cal O}_{i}(t)$ are $\mathfrak{m}$ time-dependent functions defined as
\begin{equation}\label{deformationFunc}
	{\cal O}_{i}(t)\equiv{\cal O}_{i}(q(t),p(t);x)=\left( \partial_i H \right)_{q,p}.
\end{equation}
This classical metric tensor (CMT) provides a measure of the distance, on the parameter space  $\mathcal{M}$, between the points $[q(x),p(x)]$ and $[q(x+\delta x),p(x+\delta x)]$ with infinitesimally different parameters, i.e., $\delta\ell^{2}=g_{ij}(I;x)\delta x^{i}\delta x^{j}$.

Since the functions ${\cal O}_{i}(t)$ are periodic in the angle variable $\phi$, they can be expressed as a Fourier series as
\begin{equation}
	{\cal O}_{i}(t)=\sum_{m'=-\infty}^{\infty}\beta^{(m')}_{i} {\rm e}^{{\rm i} m' \phi(t)} ,\label{FourierO}
\end{equation}
where $\phi(t)=\phi_{0}+\omega t$ with $\omega=\partial H/\partial I$ the angular frequency of the system, ``${\rm i}$'' is the imaginary unit, and $\beta_{i}^{(m')}$ are the time-independent Fourier coefficients
\begin{equation}
	\beta^{(m')}_{i}(I;x)\equiv\langle {\cal O}_{i}(\phi,I;x) {\rm e}^{-{\rm i} m'\phi} \rangle. \label{Fouriercoeff}
\end{equation}
Using the functions $\beta^{(m')}_{i}$, it can be shown that the components of  CMT \eqref{CMT} can be equivalently written as\cite{Diego5} 
\begin{equation}
	g_{ij}(I;x)=\sum_{\substack{m'=-\infty\\
    m'\neq0}}^{\infty}\frac{\beta_{i}^{(m')}\, \beta_{j}^{(-m')}}{(m'\omega)^{2}} .\label{CMTpert}
\end{equation}
This expression can be regarded as the perturbative form of the CMT, and in this sense it is analogous to the expression~\eqref{QMTpert} of the QMT. The appealing feature of \eqref{CMTpert} is that it does not require to solve the initial conditions problem, what makes it suitable for this study. To determine if the resulting singularities of this metric are genuine or not, we can also compute the associate scalar curvature \eqref{scalar} of the two-dimensional case.

Notice that the CMT \eqref{CMTpert} can be written in a similar fashion as the QMT \eqref{QMTpert2}. In fact, using $(\beta_{i}^{(m')})^*=\beta_{i}^{(-m')}$ which follows from \eqref{Fouriercoeff}, the CMT can be expressed as 
\begin{equation}
    g_{ij}(I;x)= {\rm Re} \sum_{m'=1}^{\infty} \mathcal{G}^{(m')}_{ij},\label{CMTpert2}
\end{equation}
with
\begin{equation}
	\mathcal{G}^{(m')}_{ij}:=\frac{\beta_{i}^{\prime(m')}\, (\beta_{j}^{\prime(m')})^*}{(m\omega)^{2}},
\end{equation}
where we have defined $\beta_{i}^{\prime (m')}:=\sqrt{2}\beta_{i}^{(m')}$.

\section{Quartic oscillator and double well potential}

The Hamiltonian of the quantum quartic oscillator is
\begin{equation}\label{Hquartic}
\hat{H}=\frac{1}{2}\hat{p}^{2}+\frac{k}{2}\hat{q}^{2}+\frac{\lambda}{4!} \hat{q}^{4},
\end{equation}
where $k$ and $\lambda$ are the system parameters, which we adopt as our adiabatic parameters. Then, the associated parameter space corresponds to a two-dimensional manifold with coordinates $\{x^{i}\}=(k,\lambda)$, \, $i=1,2$. Throughout this paper we consider $\lambda>0$. In the case $k>0$, the potential has a single minimum, while in the case $k<0$ it has two minima and is known as the  double-well potential.

\subsection{Quantum analysis}

\subsubsection{QMT from a numerical calculation}
The quantum Hamiltonian (\ref{Hquartic}) is easily represented with the creation and annihilation operators considering the following equations,
\begin{equation}
    \hat{q}=\sqrt{\frac{\hbar}{2m\omega}}(\hat{a}+\hat{a}^\dagger), \;\; \hat{p}=-i\sqrt{\frac{m\omega\hbar}{2}}(\hat{a}-\hat{a}^\dagger)
\end{equation}
where $\omega=\sqrt{k/m}$.
Consequently the Hamiltonian becomes into, 

\begin{equation}
\hat{H}=\dfrac{\hbar\omega}{2}\left( 1+2\hat{a}^\dagger\hat{a}\right)+\dfrac{\hbar\lambda}{4\cdot 4! m^2\omega^2}\left(\hat{a}^\dagger+\hat{a} \right)^4
\label{eq:hqaadc3}
\end{equation}
the respective matrix elements of $\hat{H}$ can be represented in the Fock basis and are given by,
\begin{align}
\langle n'|\hat{H}|n\rangle =\left[\left(\frac{\omega\hbar}{2}\right)\left(1 + 2 n\right) + \left( \frac{\lambda \hbar^2}{96m^2 \omega^2}\right) \left(6 n^2 + 6 n + 3\right)\right]\delta_{n',n} \nonumber\\
+\sqrt{(n + 1) (n + 2)} \left[  \left(4 n + 6\right)\delta_{n',n+2}\right. \nonumber\\
+\left.\sqrt{(n + 1) (n + 2) (n + 3) (n + 4)}\delta_{n',n+4}\right]\left( \frac{\lambda\hbar^2}{96m^2 \omega^2}\right)+\text{c.c}.
\end{align}

With these matrix elements, a truncated Hamiltonian matrix is built with $n,n'\in[0,N]$, where $N$ is the finite truncated size on the Fock basis, which allows to obtain $N_\text{conv}$ converged eigenstates. 
We encode in Mathematica \cite{Mathematica} the calculations to find the first  $N_\text{conv}$ eigenstates $|\Psi_m\rangle$ of the system, preserving a numerical accuracy of seventy digits of precision.

Using \eqref{Hquartic}, for $k$ and $\lambda$ we get 
\begin{subequations}
\begin{align}
\hat{O}_{1}&\equiv \frac{\partial \hat{H}}{\partial k}= \frac{\hat{q}^2}{2}, \label{O1qu}\\
\hat{O}_{2}&\equiv \frac{\partial \hat{H}}{\partial \lambda} = \frac{\hat{q}^4}{4!}. \label{O2qu}
\end{align}    
\end{subequations}

Both quantities are substituted into (\ref{QMTpert}). The cutoff $N_\text{conv}$ guarantees that the transition matrix elements of  $\hat{O}_{1}$ and $\hat{O}_{2}$ converged. Having obtained numerically the eigenstates, energies, and the transition operators $\hat{O}_{1}$ and $\hat{O}_{2}$ we calculate
numerically the elements of the QMT $g_{ij}$ point by point in the bidimensional parameter space. Employing
%
the command \verb|Interpolation[..., Method->Hermite]|
defined in Mathematica, $g_{i j}$ is represented as a continuous, differentiable function. 
In this way we can use the analytical expression (\ref{scalar}) and 
obtain the Ricci scalar $R$.

The classical limit of the Hamiltonian 
(\ref{Hquartic}) has a double well potential if $k<0$ and a simple well if $k>0$. To visualize the analogue 
behaviour in the $q$ variable for the quantum system, we employ the ground state $|\Psi_0\rangle$ expanded in the quantum harmonic oscillator basis $|\Phi_l\rangle$, so,
\begin{equation}
\Psi_0(q) \equiv \langle q |\Psi_0 \rangle =\sum_l C_{0,l} \langle q |\Phi_l\rangle
\end{equation}
where the coefficients $C_{0,l}=\langle\Phi_l|\Psi_0 \rangle$ are obtained in the numerical diagonalization  and 
\begin{equation}
    \langle q |\Phi_l\rangle =\dfrac{1}{\sqrt{2^l l!}}\left(\dfrac{m\omega}{\pi \hbar}\right)^{1/4}\exp\left( \dfrac{-m \omega q^2}{2 \hbar}\right) H_l\left(\sqrt{m\omega/\hbar}q\right) 
\end{equation}
with $H_l$ the Hermite polynomial of $l$-degree.

\begin{figure}[h!]
\centering
\begin{tabular}{c c}
\includegraphics[width=0.22\textwidth]{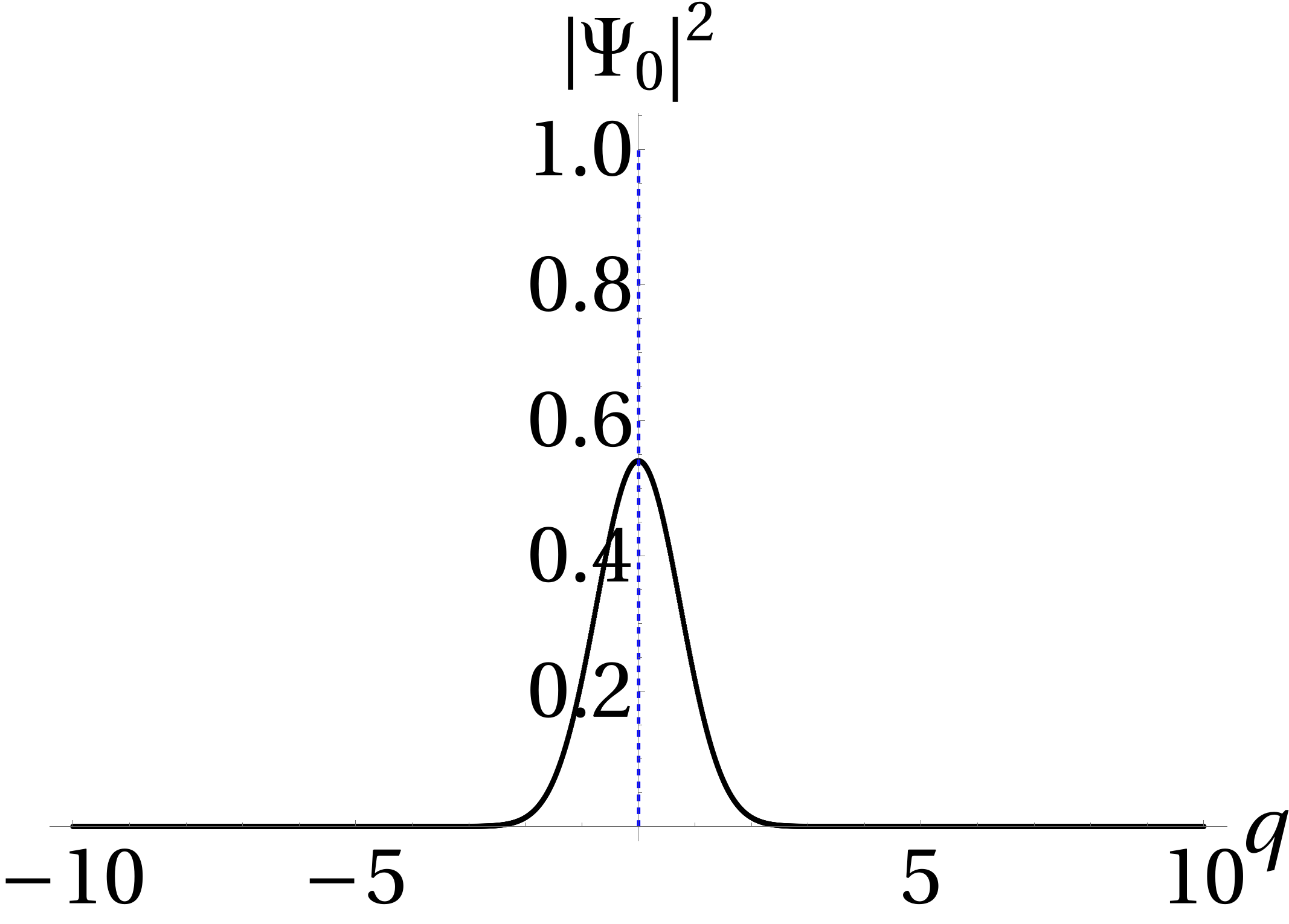} & \includegraphics[width=0.22\textwidth]{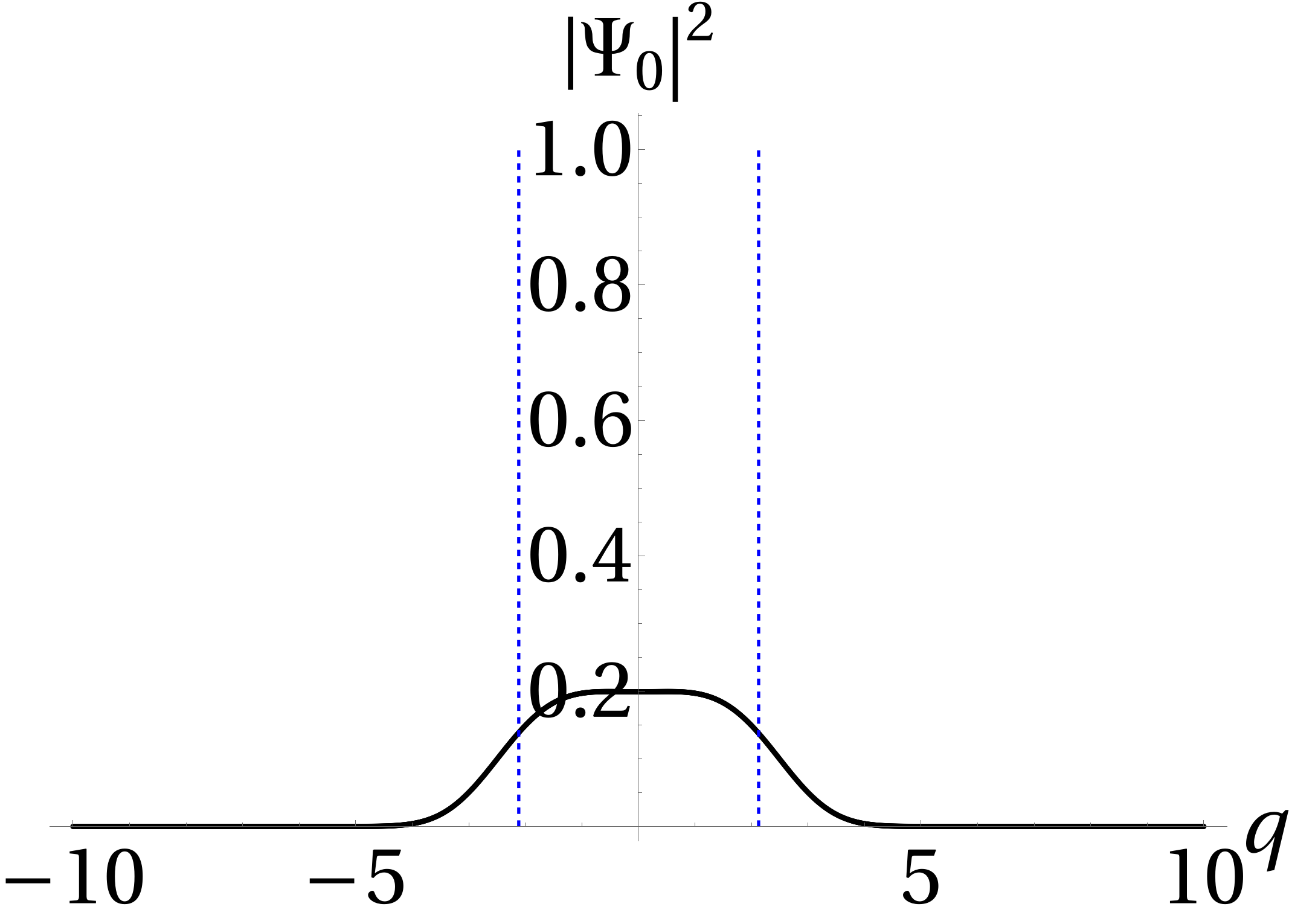} \\
(a) $k=-0.05$ & (b)  $k=-0.15$ \\
\includegraphics[width=0.22\textwidth]{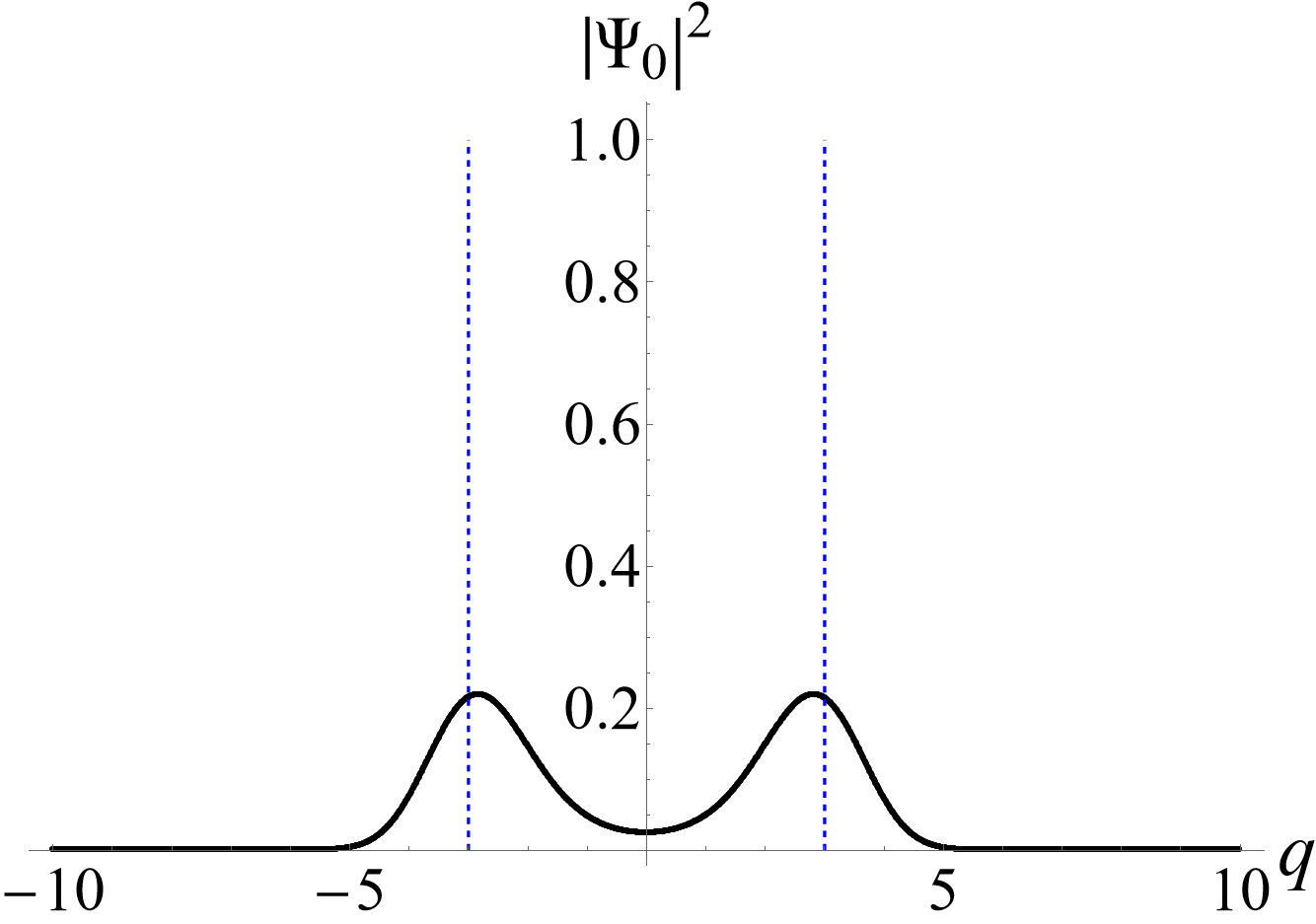} & \includegraphics[width=0.22\textwidth]{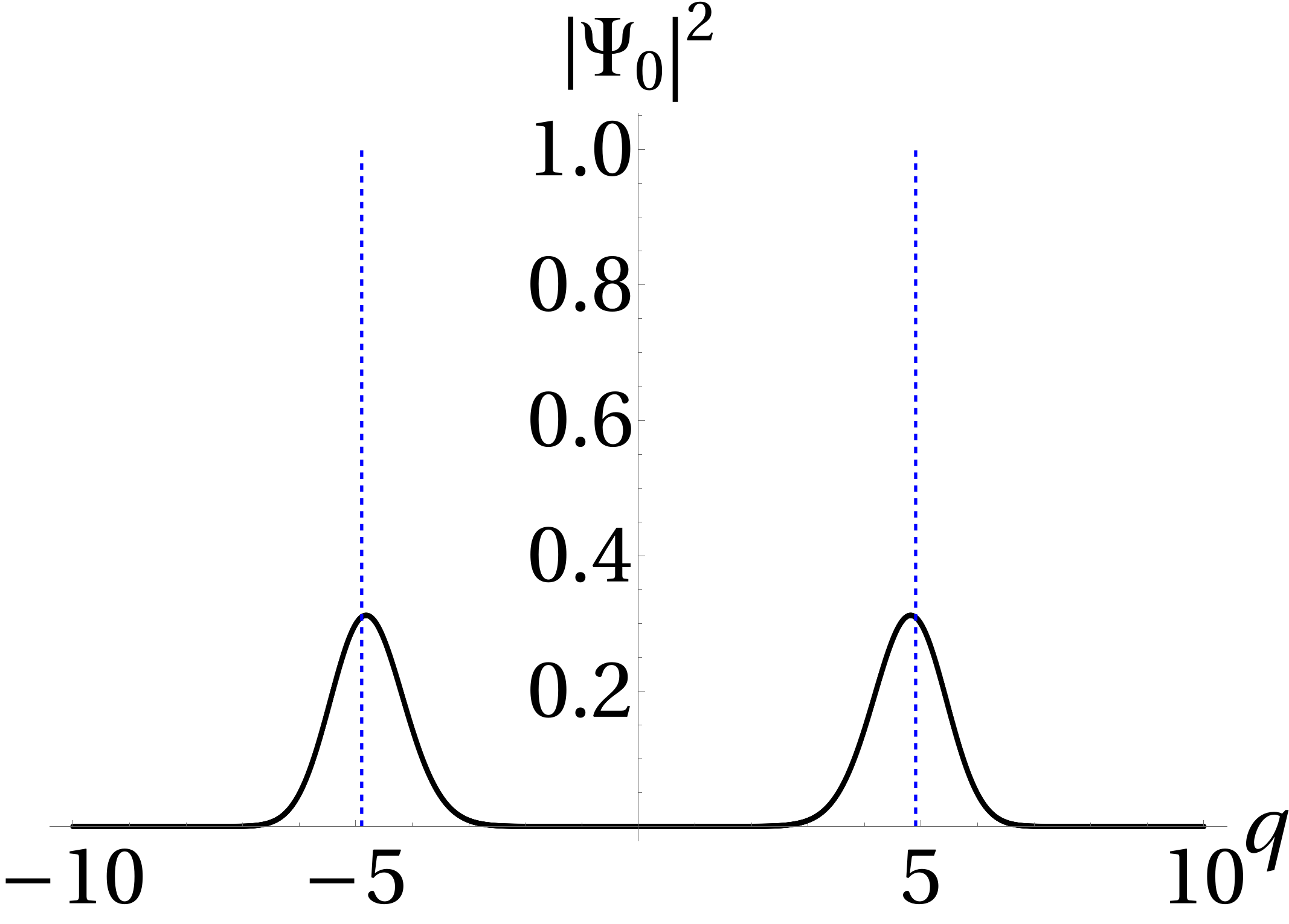}\\
(c)  $k=-0.32$& (d) $k=-0.8$
\end{tabular} 
\caption{$|\Psi_0(q)|^2$ for $\lambda=0.2$. Dashed blue lines are the position for the classical critical point.}
\label{fig:ks}
\end{figure}

\begin{figure}[h!]
\centering
\begin{tabular}{c}
\includegraphics[width=0.45\textwidth]{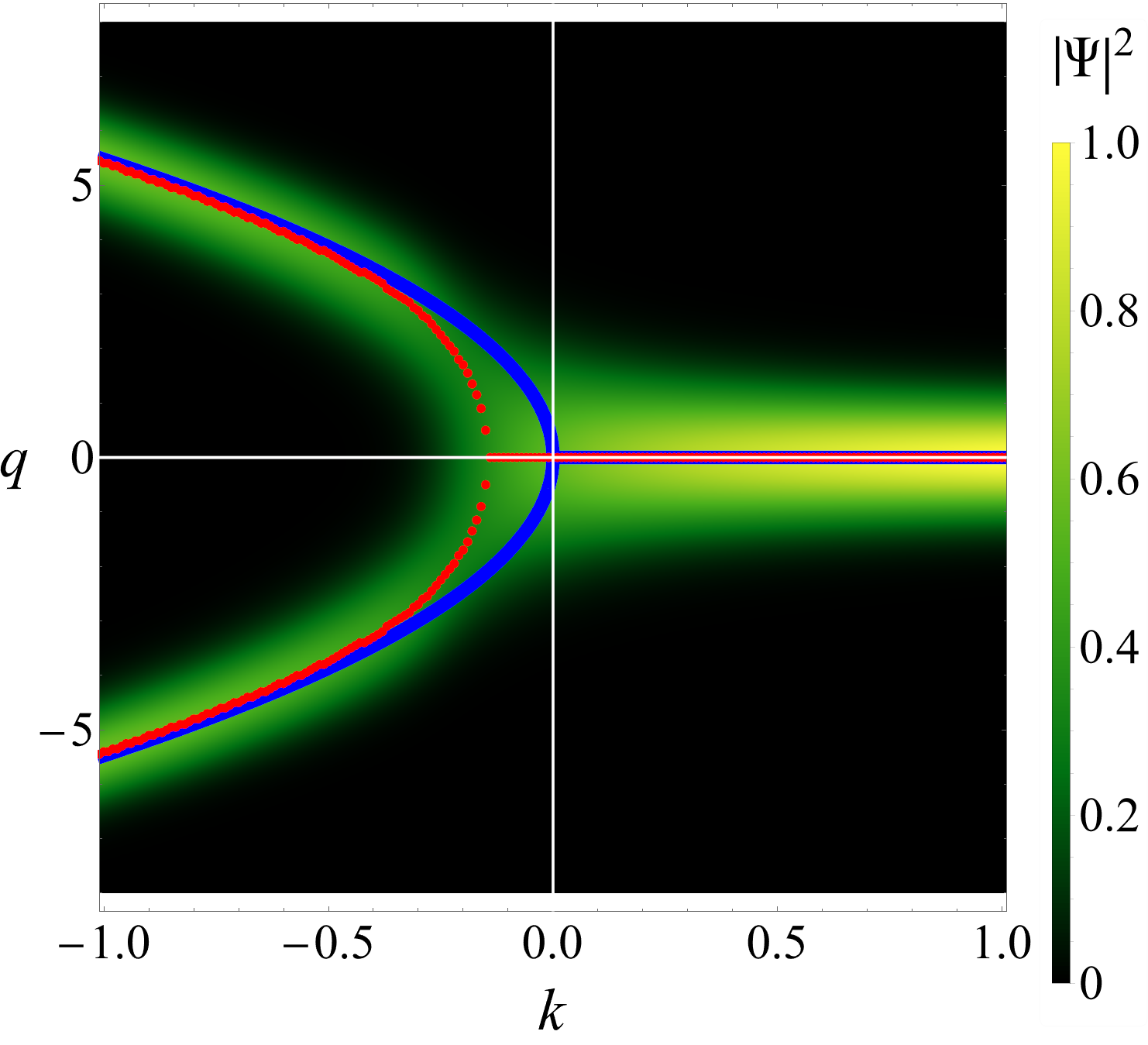}
\end{tabular} 
\caption{$|\Psi_0(q)|^2$ for $\lambda=0.2$ in function of $q$ and $k$, the blue curve is the classical localization of the critical points, and the red curve is the maximal value of $|\Psi(q)|^2$ obtained numerically.}
\label{fig:kl}
\end{figure}

The ground state probability distribution $|\Psi_0(q)|^2$ to find the particle at the position $q$ has a maximum at $q=0$ for $k>0$. For $k<0$, it shows an interesting behavior, displayed in Fig. \ref{fig:ks}. It has one maximum for small values of $|k|$ and, for $\lambda=0.2$, it starts noticing the presence of the two wells at $k\approx -0.15$ and develops two separated probability regions at $k\approx -0.32$.
The delocalization of the probability distribution over the two wells occurs when the ground state energy is lower than the top of the energy barrier between the two wells.
This effect can be best visualized  
in Fig. \ref{fig:kl}, where the blue lines represent the position of the minima of the classical potential, the yellow bands the exact quantum ground state probability distribution, with their maxima depicted with red dots.  




\subsubsection{QMT from a perturbative approach}

Here we consider a perturbative treatment in the parameter $\lambda$ since there is no exact solution to the resulting Schr\"odinger equation. Furthermore, we restrict ourselves to obtain the ground-state wave function and its energy up to the 10th order in $\lambda$. To accomplish this task, we will use the method proposed in Ref.~\cite{Turbiner1984}, suitable for finding corrections to large powers of~$\lambda$. The wave function generated by this approach, shown as an example up to the fourth order in $\lambda$ is 
\begin{align}
   \Psi_0(q)=&e^{-\frac{1}{2} \sqrt{k} q^2}-\frac{\lambda q^2e^{-\frac{1}{2} \sqrt{k} q^2}}{96 k} P_1+\frac{\lambda^2q^2 e^{-\frac{1}{2}
   \sqrt{k} q^2}}{55296 k^{5/2}} P_2\nonumber\\
   &-\frac{\lambda ^3 q^2 e^{-\frac{1}{2} \sqrt{k} q^2} }{5308416 k^4}P_3+\frac{\lambda ^4 q^2
   e^{-\frac{1}{2} \sqrt{k} q^2} }{6115295232 k^{11/2}} P_4+\dots,
\end{align}
where
\begin{subequations}
    \begin{align}
        P_1=&\sqrt{k} q^2+3,\\
        P_2=&3 k^{3/2} q^6+26 k q^4+93 \sqrt{k} q^2+252,\\
        P_3=&k^{5/2} q^{10}+141 k^{3/2} q^6+17 k^2 q^8+813 k q^4\nonumber\\
         &+2916 \sqrt{k}q^2+7992, \\
        P_4=&3 k^{7/2} q^{14}+1198 k^{5/2} q^{10}+82755 k^{3/2} q^6+84 k^3 q^{12}\nonumber\\
         &+11748 k^2 q^8+443064 k q^4+1599552 \sqrt{k} q^2+4447440.
    \end{align}
\end{subequations}
Using the wave function up to the tenth order in $\lambda$ together with the Provost and Valle formula \eqref{QMT}, we get the quantum metric tensor components  for $k>0$
\begin{subequations}\label{QMTkpos}
\begin{align}
    g_{11}&=\frac{1}{k^2} \sum_{\alpha=0}^{10} (-1)^{\alpha} a_\alpha^{(11)} \left(\frac{\hbar\lambda}{k^{3/2}}\right)^\alpha, \\
    g_{12}&=\frac{\hbar}{k^{5/2}} \sum_{\alpha=0}^{10} (-1)^\alpha a_\alpha^{(12)} \left(\frac{\hbar\lambda}{k^{3/2}}\right)^\alpha, \\
    g_{22}&=\frac{\hbar^2}{k^3} \sum_{\alpha=0}^{10} (-1)^\alpha a_\alpha^{(22)} \left(\frac{\hbar\lambda}{k^{3/2}}\right)^\alpha. 
\end{align}
\end{subequations}
where the coefficients $a_\alpha^{(ij)}$ are given in the Table \eqref{Ta111}.
It is evident from the above equations that all the components of the QMT will diverge when $k \rightarrow 0$.
It is also relevant to mention that the quantum perturbative analysis presented in this subsection can only be performed for $k>0$.

\begin{table}[h!] 
\caption{\label{coeffQMpos} Coefficients of the quantum metric tensor \eqref{QMTkpos}.}
 \[
 \begin{array}{|c|c|c|c|}\hline
 \alpha & a_\alpha^{(11)} [\times 10^{-2}] & a_\alpha^{(12)} [\times 10^{-2}] & a_\alpha^{(22)} [\times 10^{-2}] \\ \hline
0 & 3.125 & 0.78125 & 0.21159 \\
 1 & 2.1484 & 0.72428 & 0.25228 \\
 2 & 1.5971 & 0.65121 & 0.26951 \\
 3 & 1.3201 & 0.62212 & 0.29484 \\
 4 & 1.2078 & 0.64218 & 0.34133 \\
 5 & 1.2156 & 0.71847 & 0.42264 \\
 6 & 1.3384 & 0.87078 & 0.56167 \\
 7 & 1.6053 & 1.1413 & 0.80155 \\
 8 & 2.0893 & 1.6137 & 1.2271 \\
 9 & 2.9401 & 2.4552 & 2.0117 \\
 10 & 4.4588 & 4.0081 & 3.5236 \\ \hline
\end{array}
\]\label{Ta111}\end{table}

\subsection{Classical analysis}

In this section, we consider the classical counterpart of quantum Hamiltonian \eqref{Hquartic}, which is given by
\begin{equation}\label{Hclass}
H=\frac12p^{2}+ \frac{k}{2} q^{2}+\frac{\lambda}{4!} q^{4},
\end{equation}
and also take $\{x^{i}\}=(k,\lambda)$ with $i=1,2$ as the set of adiabatic parameters. The classical metric tensor for this system in the case $k>0$ was obtained in \cite{GonzalezPRE}, by using a formulation based on generating functions and resorting to the canonical perturbation theory. However, in \cite{GonzalezPRE} neither the scalar curvature associated with the geometry of the  parameter space nor  the more challenging case of $k<0$ were studied. The aim of the section is to provide a complete analysis of the parameter space of these systems in both cases, $k>0$ and $k<0$, in the framework of the Fourier-base formulation \eqref{CMTpert}.  

To compute the classical metric~\eqref{CMTpert}, we begin by setting the functions~\eqref{deformationFunc}. Using \eqref{Hclass}, for $k$ and $\lambda$ we get 
\begin{subequations}
\begin{align}
	{\cal O}_{1}&\equiv\left( \frac{\partial H}{\partial k}  \right)_{q,p} = \frac{q^2}{2}, \label{O1class}\\
    {\cal O}_{2}&\equiv\left( \frac{\partial H}{\partial \lambda}  \right)_{q,p} = \frac{q^4}{4!}. \label{O2class}
\end{align}    
\end{subequations}
The next step is to obtain the Fourier coefficients \eqref{Fouriercoeff}, which requires first expressing these deformation functions in terms of the system's action-angle variables $\{I,\phi\}$. Since finding the action-angle variables for the Hamiltonian \eqref{Hclass} is not easy, we need to resort to the canonical perturbation theory\cite{dittrich2020}. We carry this out by separately considering the cases $k>0$ and $k<0$. It is essential to point out that functions \eqref{O1class} and \eqref{O2class} hold for both cases.

\subsubsection{Case $k>0$}

In the canonical perturbation theory, a problem is solved by decomposing the Hamiltonian into a well understood system plus a perturbation term. With this in mind, let us decompose the Hamiltonian \eqref{Hclass} as $H=H_0+\lambda H_1$ with
\begin{align}
    H_0&=\frac12p^{2}+ \frac{k}{2} q^{2}, \\
    H_1&=\frac{q^{4}}{4!}.
\end{align}
Assuming $\lambda \ll 1$, $H_0$ can be regarded as the Hamiltonian of the unperturbed problem and has well-known
 action-angle variables $\{I_0,\phi_0\}$, which are related to the variables $\{q,p\}$ as
\begin{subequations}
\begin{align}
q(\phi_{0},I_{0};x)&=\left(\frac{2I_{0}}{\omega_{0}}\right)^{1/2} \sin\phi_{0}, \label{quartic:q}\\
p(\phi_{0},I_{0};x)&=\left(2\omega_{0}I_{0} \right)^{1/2}\cos\phi_{0},\label{quartic:p}
\end{align}
\end{subequations}
where $\omega_{0}=\sqrt{k}$ is the frequency of the unperturbed system. Furthermore, in this setting $H_{1}$ is understood as the perturbative potential. 

\begin{figure}[h!]
\centering
\begin{tabular}{c}
\includegraphics[width=0.45\textwidth]{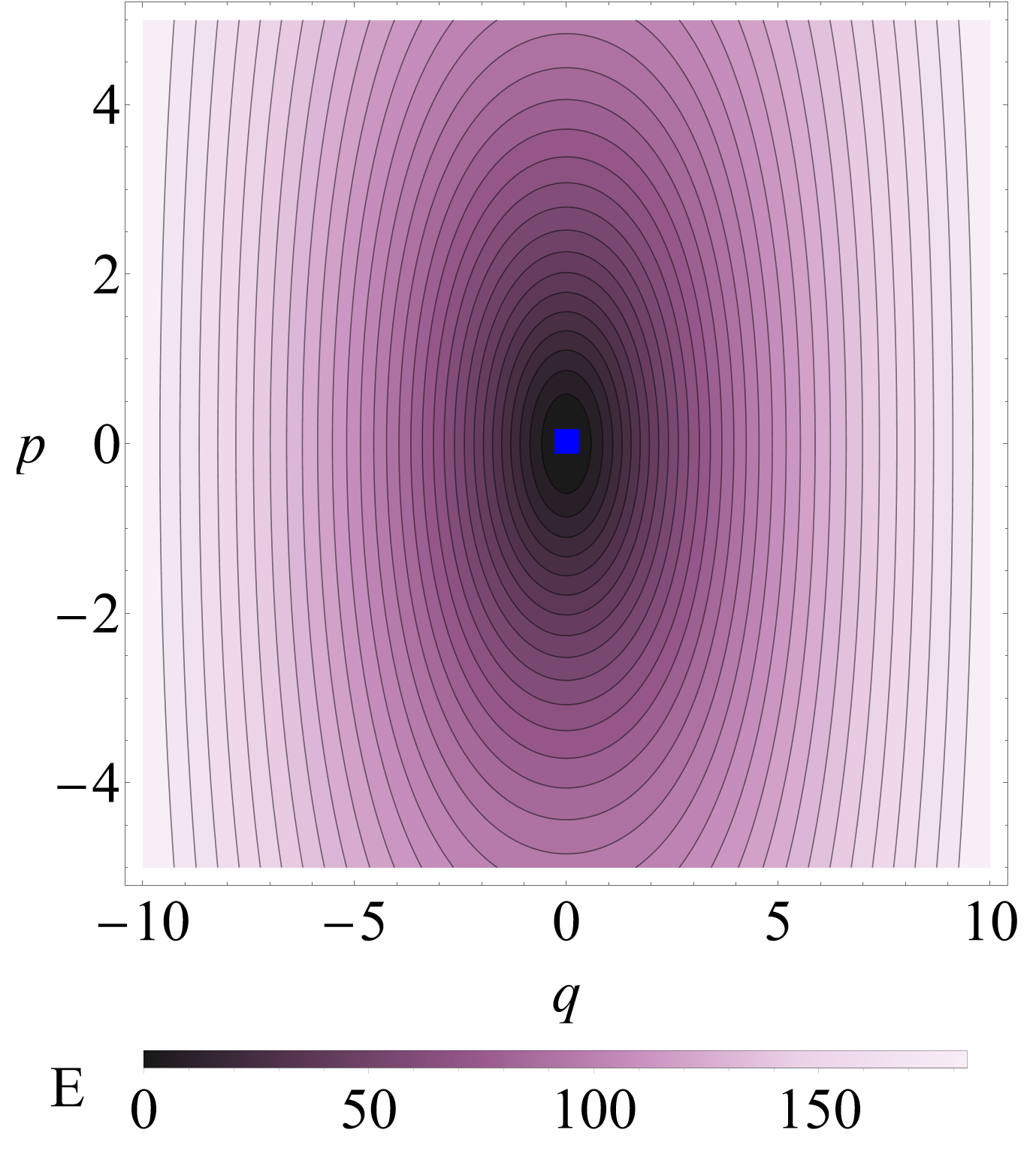}
\end{tabular} 
\caption{Energy surfaces for $k=1$ and $\lambda=0.2$.}
\label{fig:phase-space-k1}
\end{figure}

In the perturbation theory, the type 2 generating function  $W(\phi_{0},I;x)$ of the canonical transformation from $(\phi_{0},I_{0})$ to the action-angle variables $(\phi,I)$ of the total Hamiltonian $H(I;x)$ can be written as
\begin{equation}\label{eqW}
W(\phi_{0},I;x)=\phi_{0}I+\lambda W_{1}(\phi_{0},I;x)+\lambda^{2}W_{2}(\phi_{0},I;x)\dots,
\end{equation}
where $W_{1},W_{2},\dots$ are functions obtained by solving the differential equations~\cite{goldstein2000,dittrich2020}
\begin{equation}\label{quartic:diffW}
\omega_{0}\frac{\partial W_{\mu}(\phi_{0},I;x)}{\partial\phi_{0}}=\left<\Phi_{\mu}(\phi_{0},I;x)\right>_0-\Phi_{\mu}(\phi_{0},I;x).
\end{equation}
Here, $\left< \cdot \right>_{0}$ is the classical average with respect to unperturbed angle $\phi_{0}$ and, in particular, the first three functions $\Phi_{\mu}$ are obtained by $\Phi_{1}=H_{1}$, $\Phi_{2}=\frac{\partial W_{1}}{\partial\phi_{0}} \frac{\partial H_{1}}{\partial I}$ and $\Phi_{3}=\frac{1}{2}\left(\frac{\partial W_{1}}{\partial\phi_{0}}\right)^2\frac{\partial^2 H_{1}}{\partial I^2}+\frac{\partial W_{2}}{\partial\phi_{0}}\frac{\partial H_{1}}{\partial I}$. Using this, we compute the functions $W_{\mu}$ for $\mu=1,\dots,10$, which are provided in Appendix~\eqref{AppGenerating} (see Eq.~\eqref{Wnfunctkpos}).

With the generating function $W$ at hand and bearing in mind the equations of the canonical transformation
\begin{subequations}\label{ctranf}
\begin{align}
 \phi(\phi_{0},I;x)= \frac{\partial W(\phi_{0},I;x)}{\partial I}, \\ 
 I_{0}(\phi_{0},I;x)= \frac{\partial W(\phi_{0},I;x)}{\partial \phi_{0}},  
\end{align}    
\end{subequations}
the variables $\phi$ and $I_{0}$ can be calculated in terms of $\phi_{0}$, $I$, and the parameters $x$. Then, using \eqref{quartic:q} together with the resulting action variable $I_{0}(\phi_{0},I;x)$, the classical deformation functions \eqref{O1class} and \eqref{O2class} can also be expressed in terms of the $\phi_{0}$, $I$, and $x$, i.e. ${\cal O}_{1}={\cal O}_{1}(\phi_{0},I;x)$ and ${\cal O}_{2}={\cal O}_{2}(\phi_{0},I;x)$ which are given in \eqref{O1kpos} and \eqref{O2kpos} up to the fourth order in $\lambda$. Because of this, it is convenient to perform the change of variable $\phi \to \phi_0$ in \eqref{Fouriercoeff}, which allows us to write the expression for the Fourier coefficients as
\begin{equation}
	\beta^{(n')}_{j}(I;x)=\left\langle \frac{\partial \phi(\phi_{0},I;x)}{\partial \phi_{0}} {\cal O}_{j}(\phi_0,I;x) {\rm e}^{-{\rm i} n'\phi(\phi_0,I;x)}  \right\rangle_0. \label{Fouriercoeff2}
\end{equation}
From this expression we obtain the coefficients $\beta^{(n')}_{1}(I;x)$ and $\beta^{(n')}_{2}(I;x)$ for $n'=0,\pm1,\dots,\pm10$, which are given in \eqref{beta1kpos} and \eqref{beta2kpos}, respectively.

Substituting \eqref{beta1kpos} and \eqref{beta2kpos} into \eqref{CMTpert}, the components of the classical metric tensor for $k>0$ are
\begin{subequations}\label{CMTkpos}
\begin{align}
    g^{\rm cl}_{11}&=\frac{I^2}{k^2} \sum_{\alpha=0}^{10} (-1)^{\alpha} b_{\alpha}^{(11)} \left(\frac{I\lambda}{k^{3/2}}\right)^{\alpha}, \\
    g^{\rm cl}_{12}&=\frac{I^3}{k^{5/2}} \sum_{\alpha=0}^{10} (-1)^{\alpha} b_{\alpha}^{(12)} \left(\frac{I\lambda}{k^{3/2}}\right)^{\alpha}, \\
    g^{\rm cl}_{22}&=\frac{I^4}{k^3} \sum_{\alpha=0}^{10} (-1)^{\alpha} b_{\alpha}^{(22)} \left(\frac{I\lambda}{k^{3/2}}\right)^{\alpha}, 
\end{align}
\end{subequations}
where the numerical coefficients $b_{\alpha}^{(11)}$, $b_{\alpha}^{(12)}$, and $b_{\alpha}^{(22)}$ are given in Table~\ref{coeffCMpos}.  
As in the quantum case, all the components of the CMT will also diverge when $k \rightarrow 0$.

\begin{table}[h!]
\caption{\label{coeffCMpos} Coefficients of the classical metric tensor \eqref{CMTkpos}.}
 \[
 \begin{array}{|c|c|c|c|}\hline
 \alpha & b_{\alpha}^{(11)} [\times 10^{-4}] & b_{\alpha}^{(12)} [\times 10^{-4}] & b_{\alpha}^{(22)} [\times 10^{-5}] \\ \hline
0 & 312.5 & 52.083 & 88.162 \\
 1 & 143.23 & 29.772 & 60.357 \\
 2 & 65.07 & 15.191 & 34.176 \\
 3 & 30.272 & 7.6057 & 18.337 \\
 4 & 14.402 & 3.8085 & 9.6513 \\
 5 & 6.9781 & 1.9169 & 5.0451 \\
 6 & 3.4315 & 0.97081 & 2.6326 \\
 7 & 1.7079 & 0.49472 & 1.3745 \\
 8 & 0.85856 & 0.25355 & 0.71878 \\
 9 & 0.4352 & 0.13062 & 0.37665 \\
 10 & 0.22216 & 0.067608 & 0.19781 \\ \hline
\end{array}
\]
\end{table}

\subsubsection{Case $k<0$}

\begin{figure}[h!]
\centering
\begin{tabular}{c}
\includegraphics[width=0.45\textwidth]{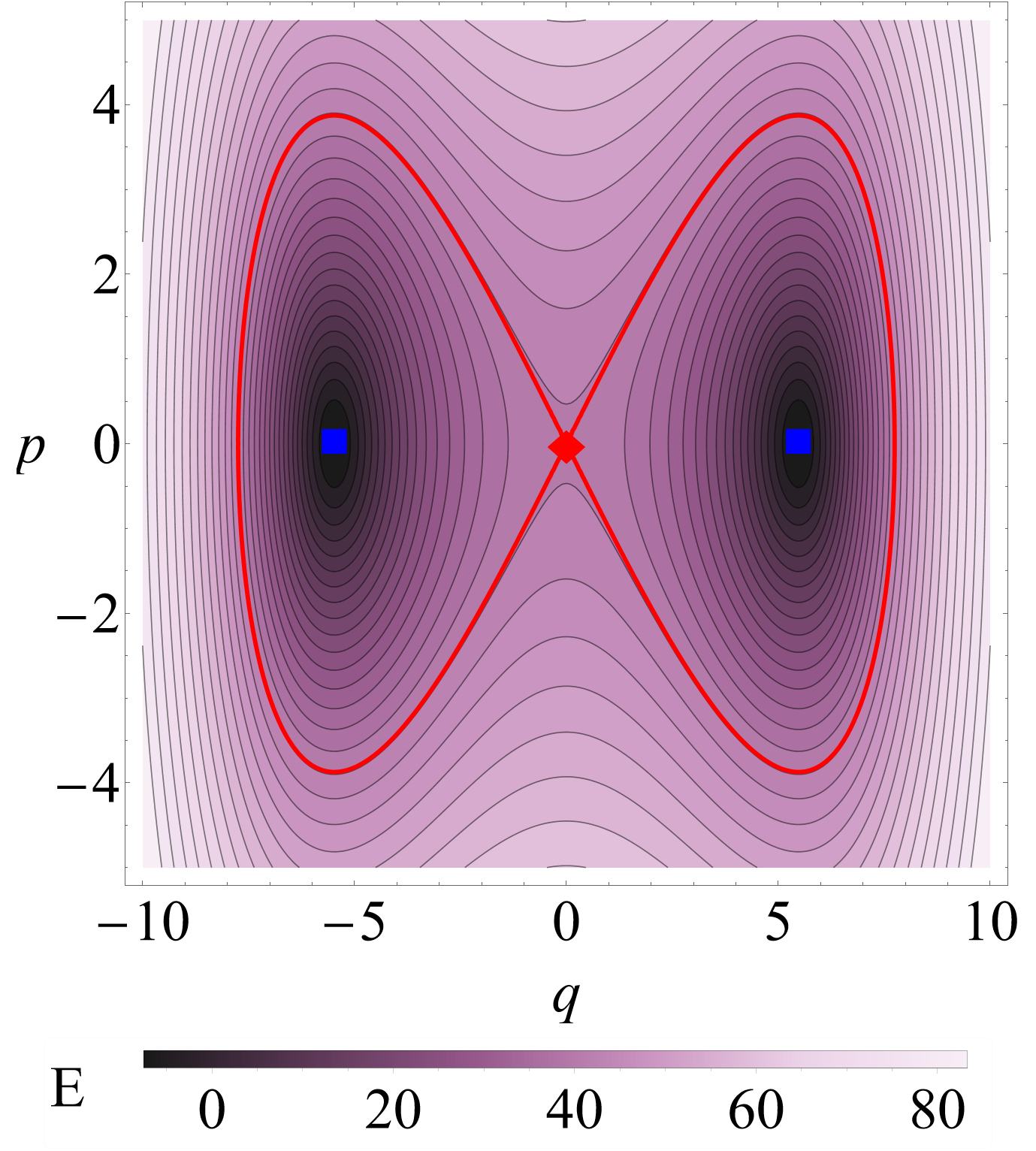}
\end{tabular} 
\caption{Energy surfaces for $k=-1$ and $\lambda=0.2$. Blue points are the stable center
points $\chi_1$ and $\chi_3$, whereas the red one is the unstable center point $\chi_2$.}
\label{fig:phase-space-k1neg}
\end{figure}

In this case, the system presents three fixed points corresponding to vanishing phase space
velocities $(\Dot{q},\Dot{p})$. The points are
\begin{subequations}
    \begin{align}
        \chi_1=&(q,p)=\left( -\sqrt{\frac{-6k}{\lambda}} ,0\right), \\
        \chi_2=&(q,p)=\left(0,0\right), \\
        \chi_3=&(q,p)=\left(\sqrt{\frac{-6k}{\lambda}} ,0\right). 
    \end{align}
\end{subequations}
In Fig. \ref{fig:phase-space-k1neg}, we can see these points. The blue points correspond to $\chi_1$ and $\chi_3$, which are center points as long as $k<0$. Furthermore, the red point corresponds to $\chi_2$ and it is a hyperbolic point. 

Taking into account this, it is convenient to carry out a transformation to a coordinate system centered on $\chi_1$ or $\chi_3$. Then, let us take the point $\chi_1$ and consider the change of coordinates $Q=q+\sqrt{\frac{-6k}{\lambda}}$ and $P=p$. In terms of the new variables the Hamiltonian \eqref{Hclass} reads 
\begin{equation}\label{Hclassneg}
H=\frac12P^{2}-k Q^{2}-\sqrt{\frac{-k\lambda}{6}} Q^{3}+\frac{\lambda}{4!} Q^{4}-\frac{3 k^2}{2 \lambda}.
\end{equation}
Since the constant term $\frac{3 k^2}{2 \lambda}$ does not affect the dynamics of the system, we can get rid of it. However, note that by removing this term we are removing the divergence in energy at $\lambda=0$. Redefining the parameter $\lambda'=\sqrt{\lambda}$, this Hamiltonian can be decomposed as $H=H_0+\lambda' H_1+\lambda'^2 H_2$ with
\begin{align}
    H_0&=\frac12P^{2}-k Q^{2}, \\
    H_1&=-\sqrt{\frac{-k}{6}} Q^{3}, \\
    H_2&=\frac{Q^{4}}{4!}.
\end{align}
Analogously to the previous case, we assume that $\lambda' \ll 1$. In this setting, $H_0$ is  just a harmonic oscillator since $k<0$ and then plays the role of the Hamiltonian of the unperturbed problem with action-angle variables $\{I_0,\phi_0\}$ given by
\begin{subequations}
\begin{align}
Q(\phi_{0},I_{0};x)&=\left(\frac{2I_{0}}{\omega_{0}}\right)^{1/2} \sin\phi_{0}, \label{quartic:Qneg}\\
P(\phi_{0},I_{0};x)&=\left(\frac{-4 k I_{0}}{\omega_{0}} \right)^{1/2}\cos\phi_{0}.\label{quartic:Qneg}
\end{align}
\end{subequations}
Here, $\omega_{0}=\sqrt{-2k}$ is the frequency of the unperturbed system. In addition, the terms $H_{1}$ and $H_{2}$ are regarded as first-order and second-order potentials, respectively. 

Following the same procedure as in the previous case, we first need to obtain the generating function $W$. The functions $W_{1},W_{2},\dots$ involved in \eqref{eqW} are again obtained from \eqref{quartic:diffW}, but with functions $\Phi_{\mu}$ modified by the presence of $H_2$. In particular, the first three functions $\Phi_{\mu}$ are $\Phi_{1}=H_{1}$, $\Phi_{2}=H_{2}+\frac{\partial W_{1}}{\partial\phi_{0}} \frac{\partial H_{1}}{\partial I}$ and $\Phi_{3}=\frac{\partial W_{1}}{\partial\phi_{0}} \frac{\partial H_{2}}{\partial I}+\frac{1}{2}\left(\frac{\partial W_{1}}{\partial\phi_{0}}\right)^2\frac{\partial^2 H_{1}}{\partial I^2}+\frac{\partial W_{2}}{\partial\phi_{0}}\frac{\partial H_{1}}{\partial I}$. The resulting functions $W_{\mu}$ for $\mu=1,\dots,10$  are given in~\eqref{Wnfunctkneg}. Substituting these functions into \eqref{ctranf} we get  $I_{0}$, which together with \eqref{quartic:q} allows us to obtain the classical deformation functions ${\cal O}_{1}={\cal O}_{1}(\phi_{0},I;x)$ and ${\cal O}_{2}={\cal O}_{2}(\phi_{0},I;x)$  through \eqref{O1class} and \eqref{O2class}, respectively. In \eqref{O1kneg} and \eqref{O2kneg} we give ${\cal O}_{1}$ and ${\cal O}_{2}$ up to the first order in $\lambda$. Then, plugging  ${\cal O}_{1}$ and ${\cal O}_{2}$ into \eqref{Fouriercoeff2} we arrive at the corresponding Fourier coefficients $\beta^{(n')}_{i}(I;x)$, explicitly shown in \eqref{beta1kneg} and \eqref{beta2kneg} for $n'=0,1,\dots,10$. The functions $\beta^{(n')}_{i}(I;x)$ for negative $n'$ can be obtained from \eqref{beta1kneg} and \eqref{beta2kneg} by recalling that $\beta_{i}^{(-n')}=(\beta_{i}^{(n')})^*$.

Substituting \eqref{beta1kneg} and \eqref{beta2kneg} into \eqref{CMTpert}, the components of the classical metric tensor for $k>0$ are
\begin{subequations}\label{CMTkneg}
\begin{align}
    g^{\rm cl}_{11}&=\frac{I}{(-k)^{1/2} \lambda} \sum_{\alpha=0}^{6} c_{\alpha}^{(11)} \left(\frac{I\lambda}{(-k)^{3/2}}\right)^{\alpha}, \\
    g^{\rm cl}_{12}&=\frac{(-k)^{1/2}I}{\lambda^2} \sum_{\alpha=0}^{7} c_{\alpha}^{(12)} \left(\frac{I\lambda}{(-k)^{3/2}}\right)^{\alpha}, \\
    g^{\rm cl}_{22}&=\frac{(-k)^{3/2} I}{\lambda^3} \sum_{\alpha=0}^{8} c_{\alpha}^{(22)} \left(\frac{I\lambda}{(-k)^{3/2}}\right)^{\alpha}, 
\end{align}
\end{subequations}
where the coefficients $c_{\alpha}^{(11)}$, $c_{\alpha}^{(12)}$, and $c_{\alpha}^{(22)}$ are given in Table~\ref{coeffCMneg}.
\begin{table}[h!]
\caption{\label{coeffCMneg} Coefficients of the classical metric tensor \eqref{CMTkneg}, up to  order $\lambda^5$.}
 \[
 \begin{array}{|c|c|c|c|}\hline
 {\alpha} & c_{\alpha}^{(11)} [\times 10^{-2}] & c_{\alpha}^{(12)} [\times 10^{-2}] & c_{\alpha}^{(22)} [\times 10^{-2}] \\ \hline
0 & 212.13 & 212.13 & 212.13 \\
 1 & 40.625 & 18.75 & 0 \\
 2 & 17.678 & 9.8823 & 4.4399 \\
 3 & 9.4394 & 5.481 & 2.7389 \\
 4 & 5.5548 & 3.2736 & 1.7019 \\
 5 & 3.4622 & 2.0555 & 1.0902 \\
 6 & 1.8352 & 1.0193 & 0.48949 \\
 7 & \text{} & 0.50029 & 0.18924 \\
 8 & \text{} & \text{} & 0.12832 \\  \hline
\end{array}
\]
\end{table}

\section{Classical and quantum metrics}

\begin{figure}[h!]
\includegraphics[width= 0.89 \columnwidth]{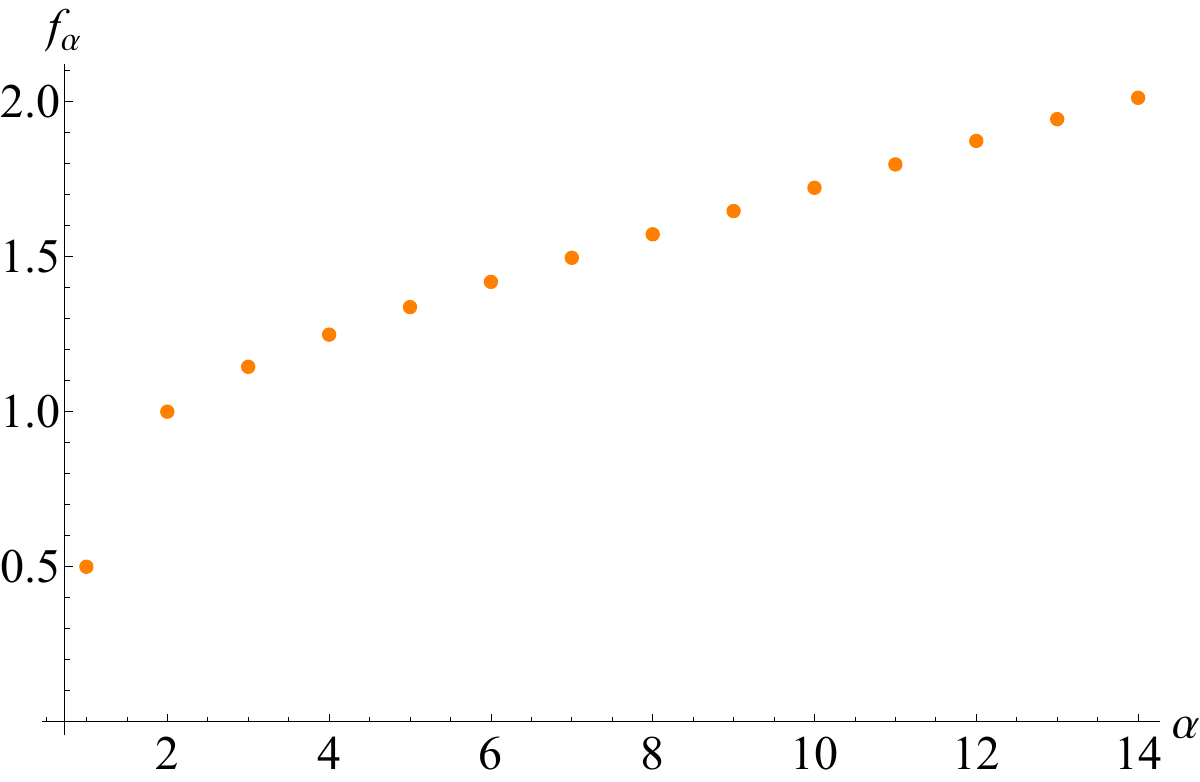}
\caption{Coefficients $f_\alpha$ of the identifications as a function $\alpha$.} \label{identifications}
\end{figure}

Our goal now is to compare the classical and quantum metrics and their scalar curvatures for both cases $k>0$ and $k<0$. With this in mind, we first need to establish the value of the action variable $I$ and its different powers. One way to do this is by using the semiclassical relation between the quantum metric tensor and the classical metric~\cite{GonzalezAnnalen}
\begin{equation}\label{semiclassical}
    \hbar^2 g_{ij}=g^{\rm cl}_{ij}.
\end{equation}
In particular, for the metrics components \eqref{QMTkpos} and \eqref{CMTkpos} of the case $k>0$ we have the relations $\hbar^2 g_{11}=g^{\rm cl}_{11}$, $\hbar^2 g_{12}=g^{\rm cl}_{12}$, and $\hbar^2 g_{22}=g^{\rm cl}_{22}$. For each of these relations, we get an  identification of the $\alpha$-th power of the action variable $I$ as $I^{\alpha}=(f_{\alpha} \hbar)^{\alpha}$ ($\alpha=1,2,\dots,14$) where $f_\alpha$ are the numerical coefficients. However, using \eqref{QMTkpos} and \eqref{CMTkpos} the values obtained for $f_\alpha$ from each relation are slightly different. Then, we take as the value of $f_\alpha$ the average value of the resulting values, obtaining $f_1=0.5$, $f_2=1$, $f_3=1.1447$, $f_4=1.2484$, $f_5=1.3372$, $f_6=1.4186$, $f_7=1.4962$, $f_8=1.5720$, $f_9=1.6470$, $f_{10}=1.7219$, $f_{11}=1.7972$, $f_{12}=1.8730$, $f_{13}=1.9433$, and $f_{14}=2.0120$. In Fig.~\ref{identifications}, we show $f_\alpha$ as a function of $\alpha$. In what follows we will use these identifications of the action variables for both cases $k>0$ and $k<0$, and we will set $\hbar=1$ to perform the comparisons between classical and quantum objects. 

Before proceeding, it is convenient to compute the scalar curvature of each metric. Using \eqref{scalar}, the scalar curvature of the quantum metric tensor \eqref{QMTkpos} is
\begin{equation}\label{pertQR}
    R=\sum_{\alpha=0}^{8} (-1)^{\alpha+1} d_\alpha \left(\frac{\hbar\lambda}{k^{3/2}}\right)^\alpha,
\end{equation}
where the coefficients $d_\alpha$ are given in Table \ref{coeffQMpos}. Analogously, the scalar curvature of the classical metric tensors \eqref{CMTkpos} for $k>0$ and \eqref{CMTkneg} for $k<0$ with the identifications $I^\alpha=(f_\alpha \hbar)^\alpha$ are
\begin{equation}\label{pertCR}
R^{\rm cl} = 
\begin{cases}
\sum_{\alpha=0}^{8} (-1)^{\alpha+1} h_n \left(\frac{\hbar\lambda}{k^{3/2}}\right)^\alpha &\text{if $k>0$}\\
\sum_{\alpha=0}^{3} l_\alpha \left(\frac{\hbar\lambda}{(-k)^{3/2}}\right)^\alpha &\text{if $k<0$},
\end{cases}
\end{equation}
where the coefficients $h_\alpha$ and $l_\alpha$ are provided in Table \ref{coeffQMpos}.

\begin{table}[h!]
\caption{\label{coeffQMpos} Coefficients of the scalar curvatures.}
 \[
\begin{array}{|c|c|c|c|}\hline
\alpha & d_\alpha  & h_\alpha  & l_\alpha \\ \hline
0& 28 & 21.1866 & -4 \\
1& 30.5556 & 0.833929 & 0 \\
2 & 54.6499 & 60.5879 & 1.02388 \\
3& 106.587 & 95.9538 & 1.84957 \\
4& 220.2 & 205.334 & \text{} \\
5& 476.399 & 444.255 & \text{} \\
6& 1073.07 & 1003.07 & \text{} \\
7& 2507.83 & 2348.16 & \text{} \\
8& 6067.94 & 1574.36 & \text{} \\\hline
\end{array}
\]
\end{table}

\begin{figure}[h!]
	\begin{tabular}{c c}
		\includegraphics[width= 0.49 \columnwidth]{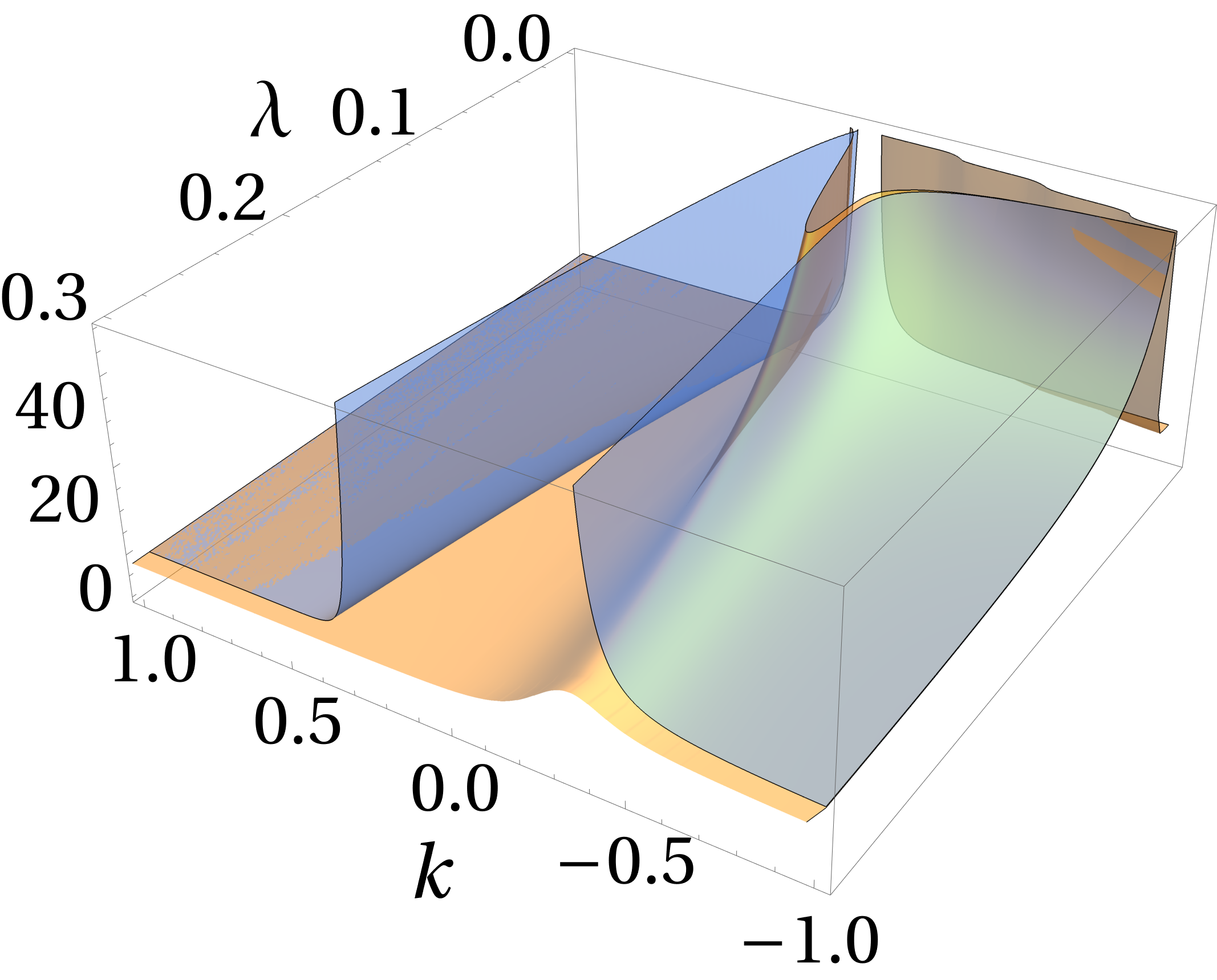} &
		\includegraphics[width= 0.49 \columnwidth]{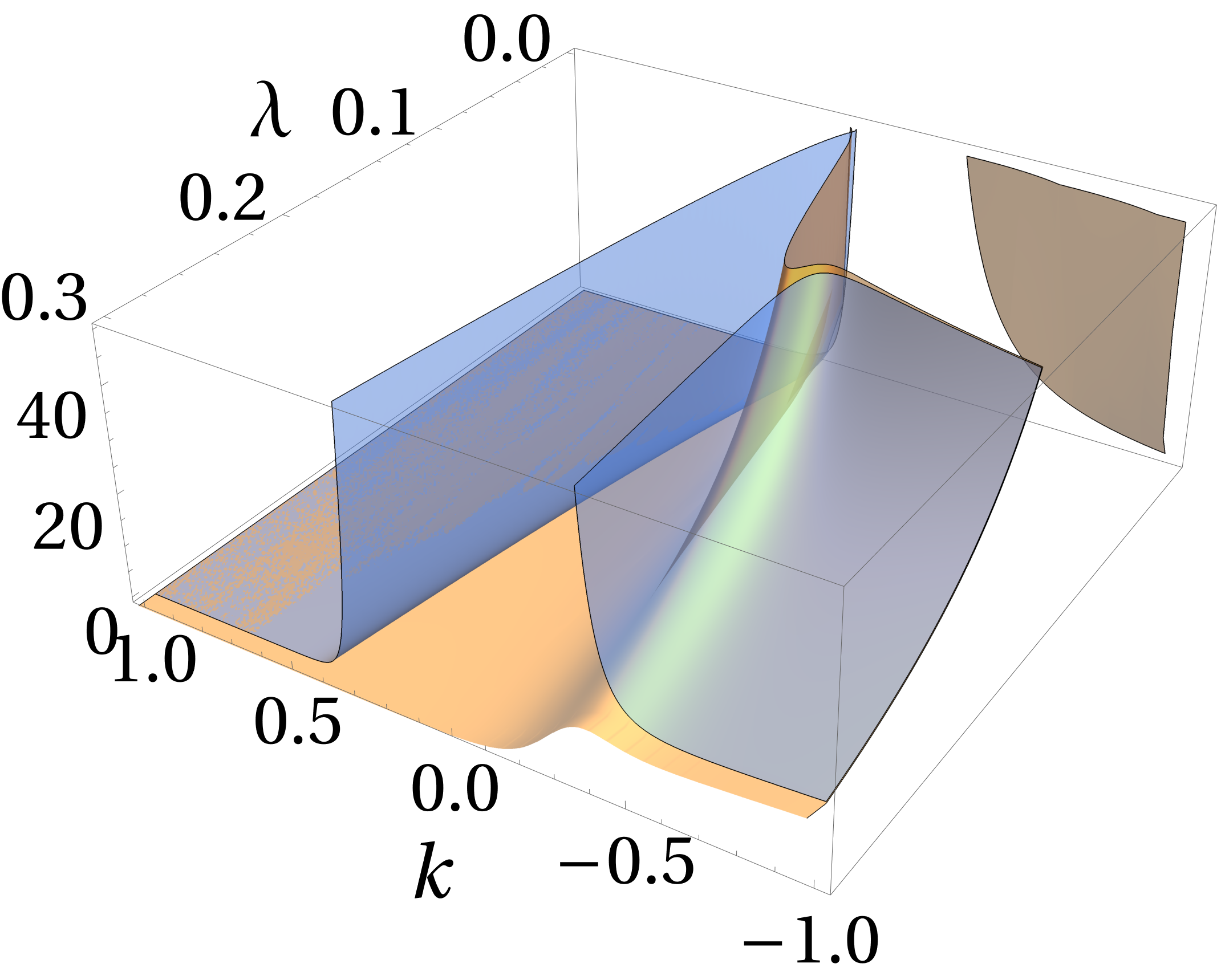} \\
		(a) $g_{11}$ & (b) $g_{12}$ \\
		\includegraphics[width= 0.49 \columnwidth]{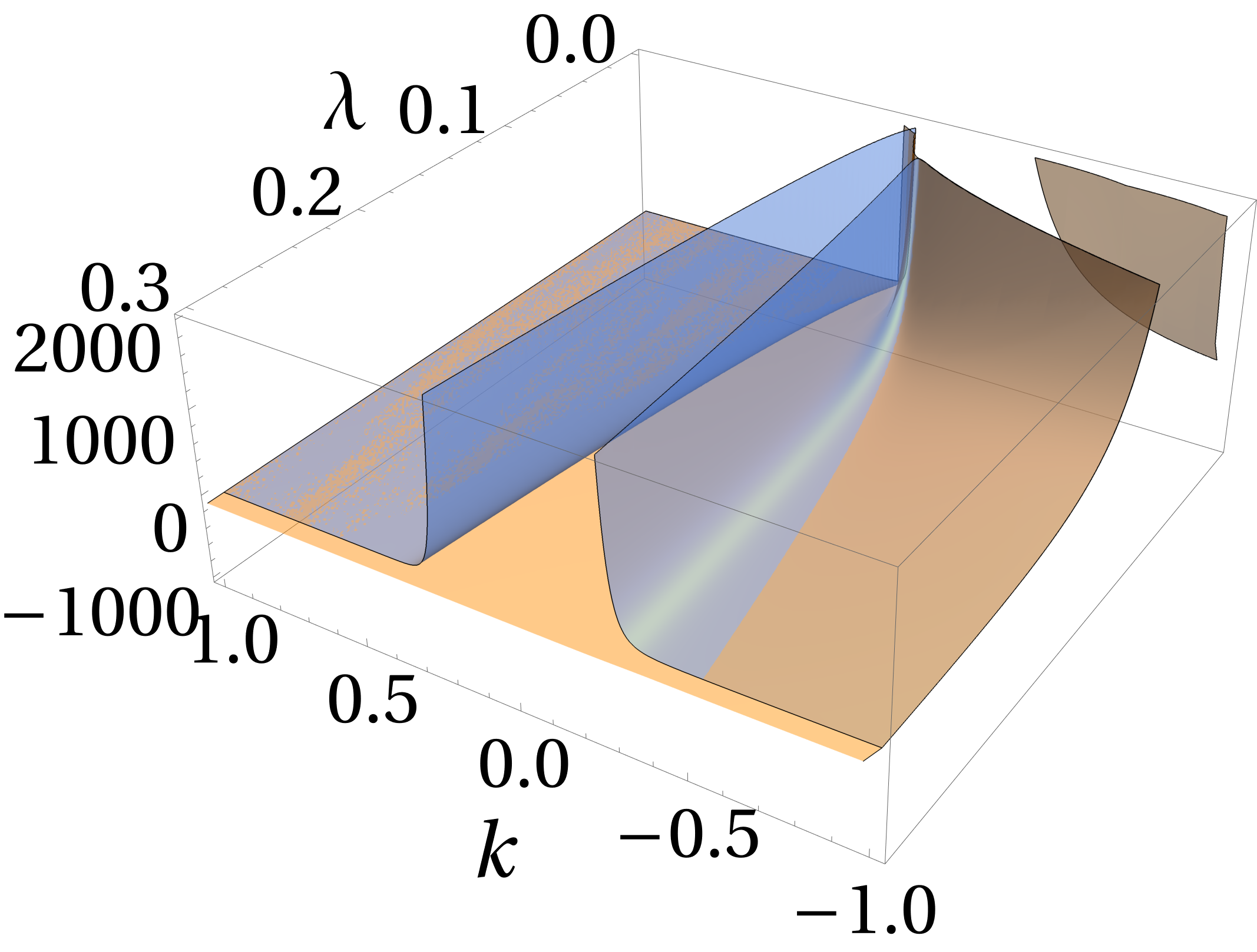} &  
		\includegraphics[width= 0.49 \columnwidth]{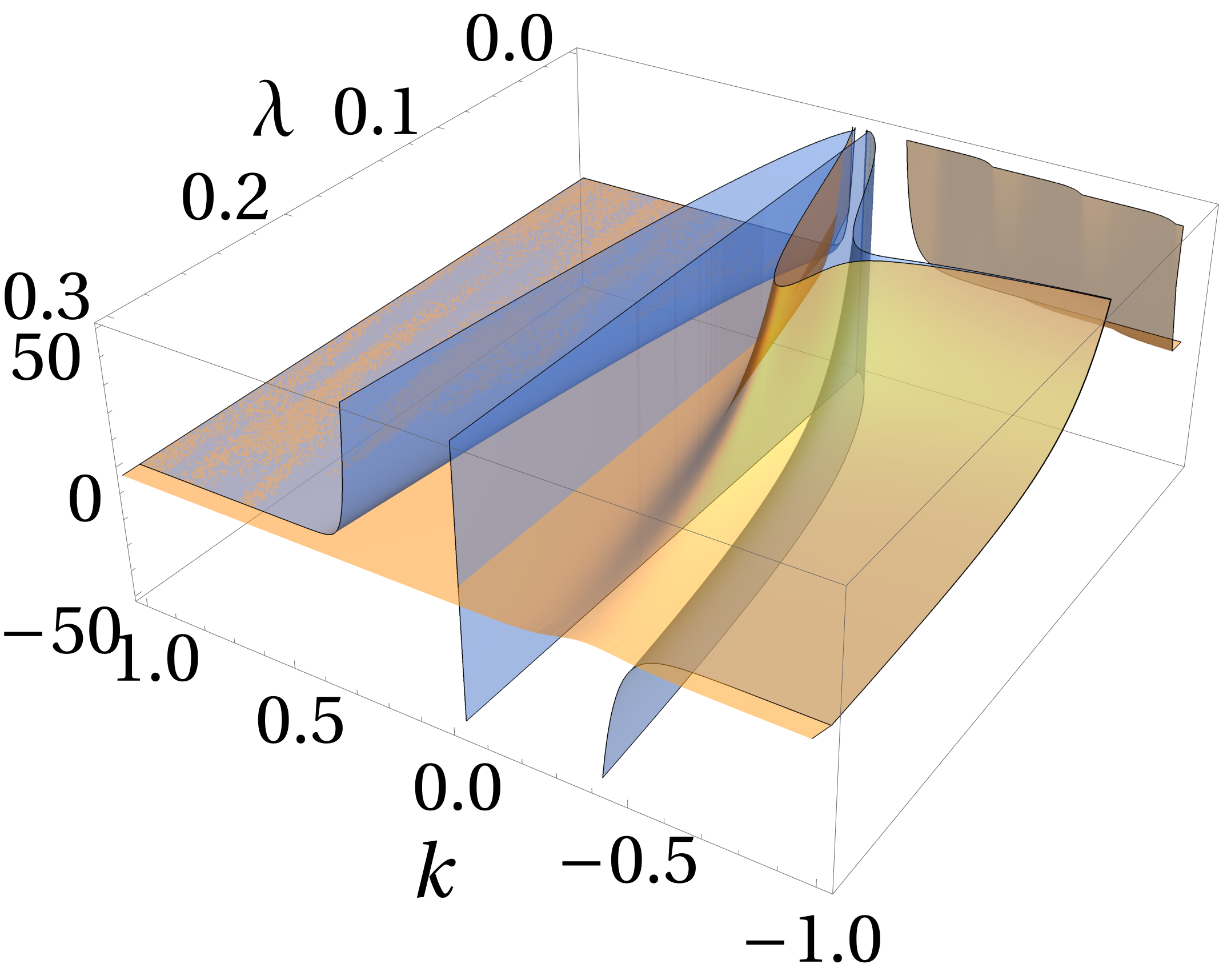} \\
		(c) $g_{22}$ & (d) $g$  \\
		\includegraphics[width= 0.49 \columnwidth]{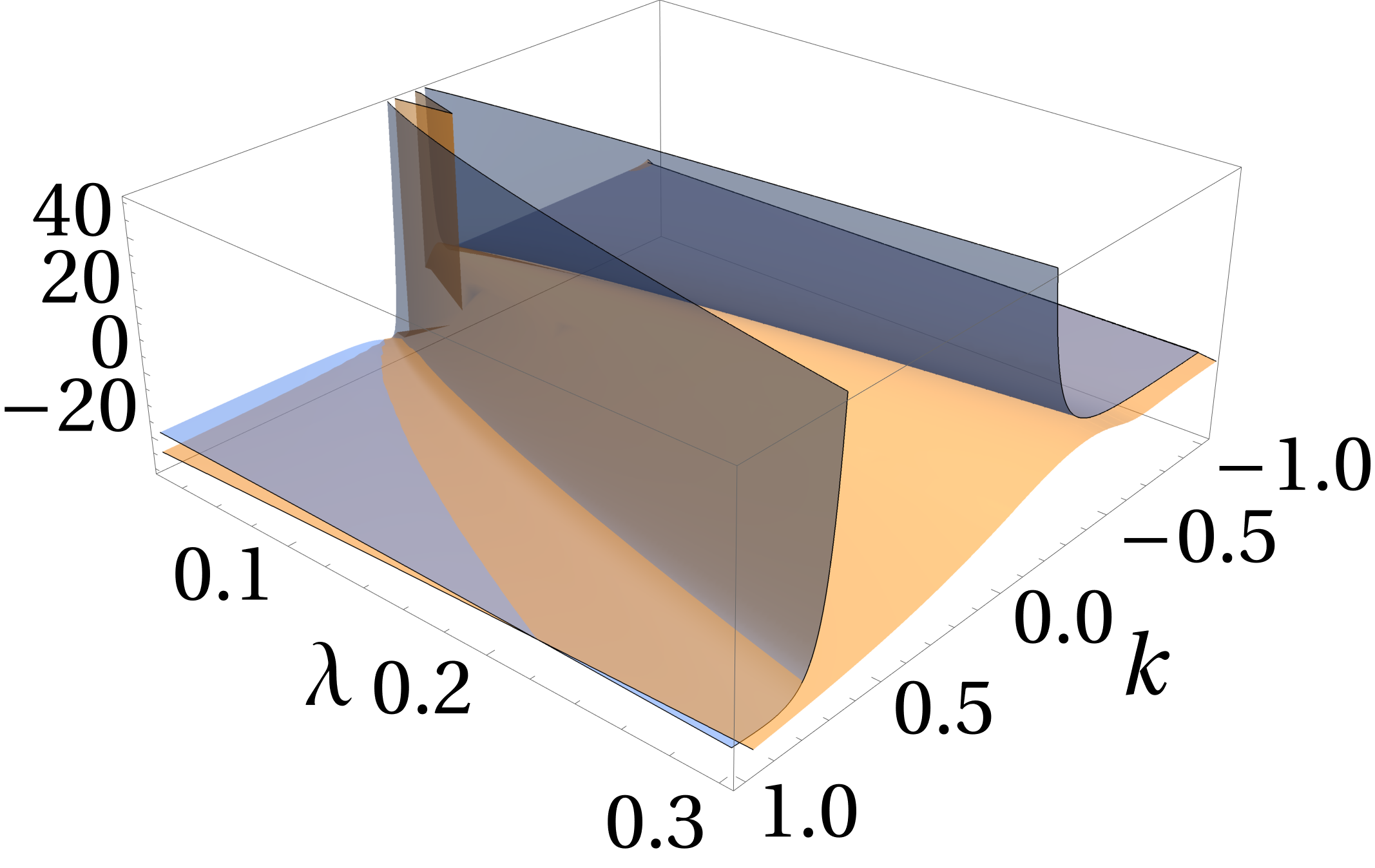} &  
		\includegraphics[width= 0.49 \columnwidth]{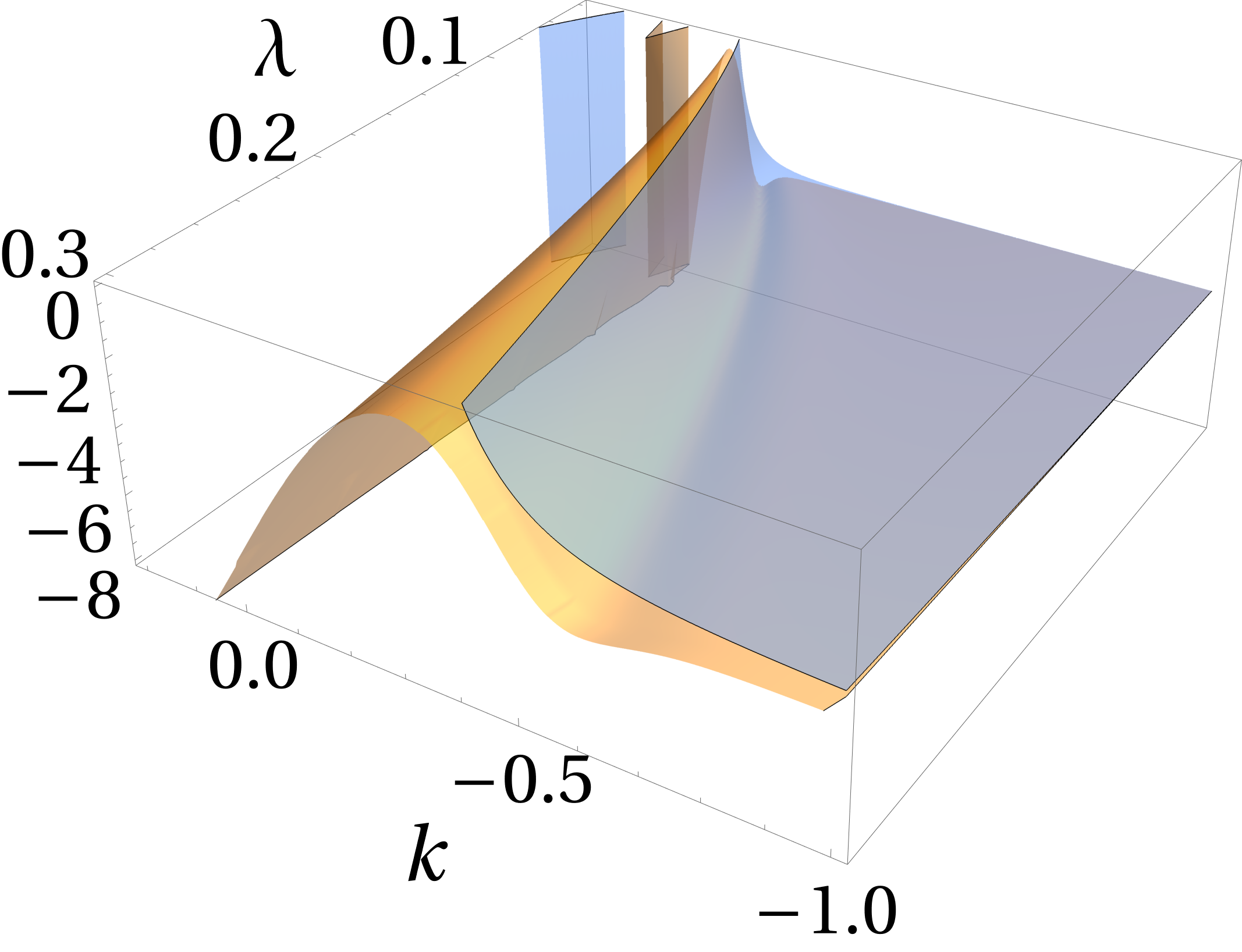} \\
		(e) $R$   & (f) $R$   
	\end{tabular}
	\caption{Comparison of parameter space metric and $R$ obtained from the exact numerical quantum approach (orange) and the classical approach (blue). The agreement is very good except near the region where $k=0$.}
	\label{Fig:QMTandR}
\end{figure}

In Fig. \ref{Fig:QMTandR} we plot the numeric QMT and its scalar curvature obtained from the exact numerical diagonalization and compare the results with the analytic CMT and its scalar curvature obtained from the Fourier approach. We see in Figs. \ref{Fig:QMTandR}(a)-\ref{Fig:QMTandR}(c) that far from the region where $k=0$ the components of the QMT and the CMT have a very close behavior. Note in Figs. \ref{Fig:QMTandR}(d)-\ref{Fig:QMTandR}(e) that this similitude between classical and quantum quantities is also exhibited by the determinant and scalar curvature. This shows that the classical metrics for the regions  $k>0$ and $k<0$ far from $k=0$ agree well with their quantum counterparts. Nevertheless, we can also see from Figs. \ref{Fig:QMTandR}(a)-\ref{Fig:QMTandR}(c) that near the region where $k=0$, the components of the CMT show a divergent behavior, in contrast to their quantum counterparts, which do not diverge but show a peak. In this regard, it is remarkable that the classical metric for $k<0$ behaves in a similar way. From Fig. \ref{fig:kl}, we notice that at $k=0$, the potential abruptly changes from having two minima to having only one if we go in the $-k\to k$ direction or from having one minima to having two minima in the opposite direction. In the classical and perturbative quantum sense, this is interpreted as an exact quantum transition, while in the exact quantum sense, this transition is moderated by tunneling, which gives rise only to a precursor of a quantum phase transition and hence the maximum near $k=0$. However, this transition can be confirmed only in the thermodynamic limit.

\begin{figure}[h!]
\begin{tabular}{c c}
\includegraphics[width= 0.49 \columnwidth]{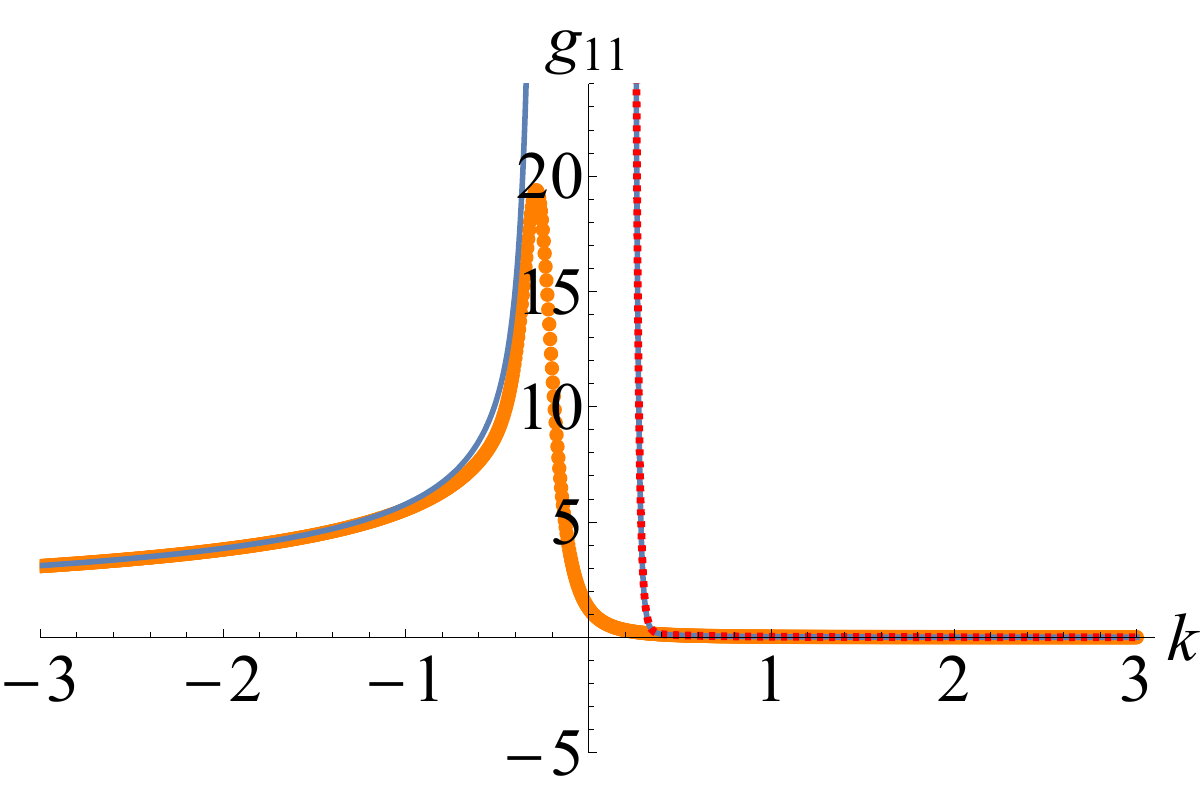} & \includegraphics[width= 0.49 \columnwidth]{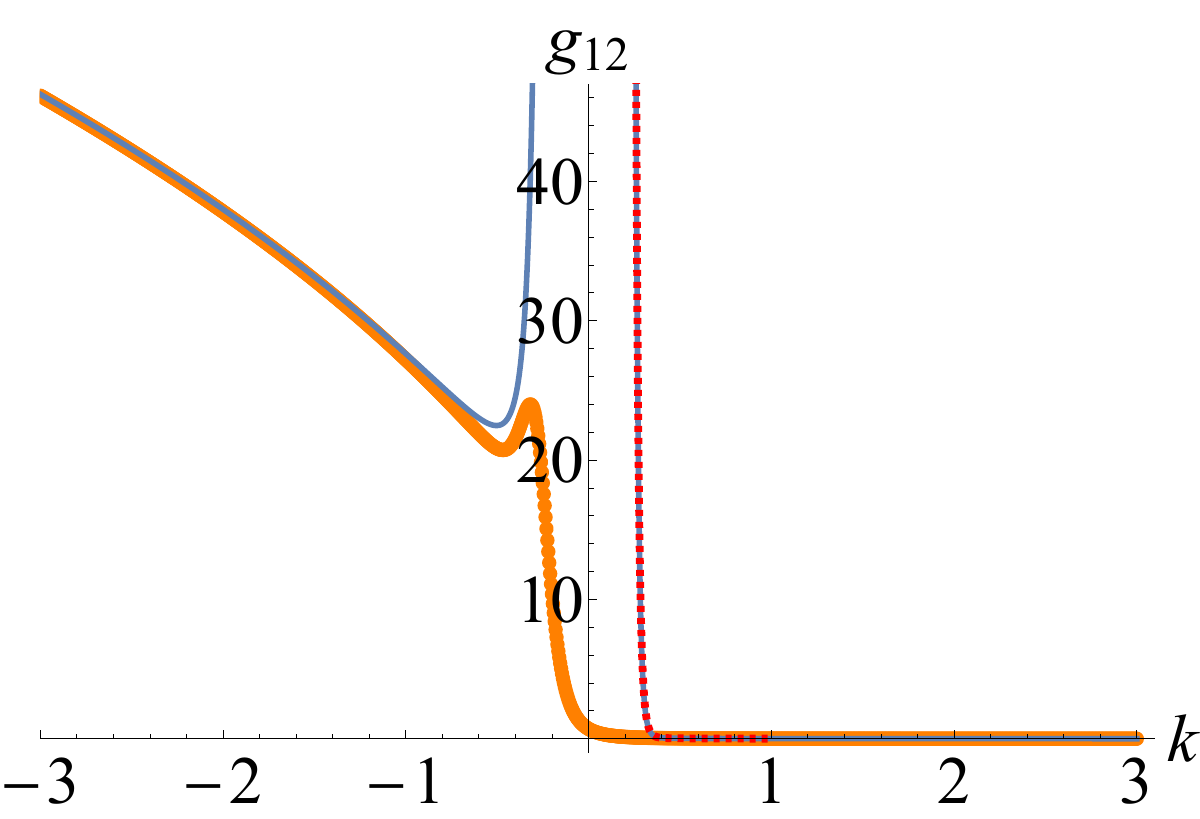} \\
(a) $g_{11}$ & (b) $g_{12}$ \\
\includegraphics[width= 0.49 \columnwidth]{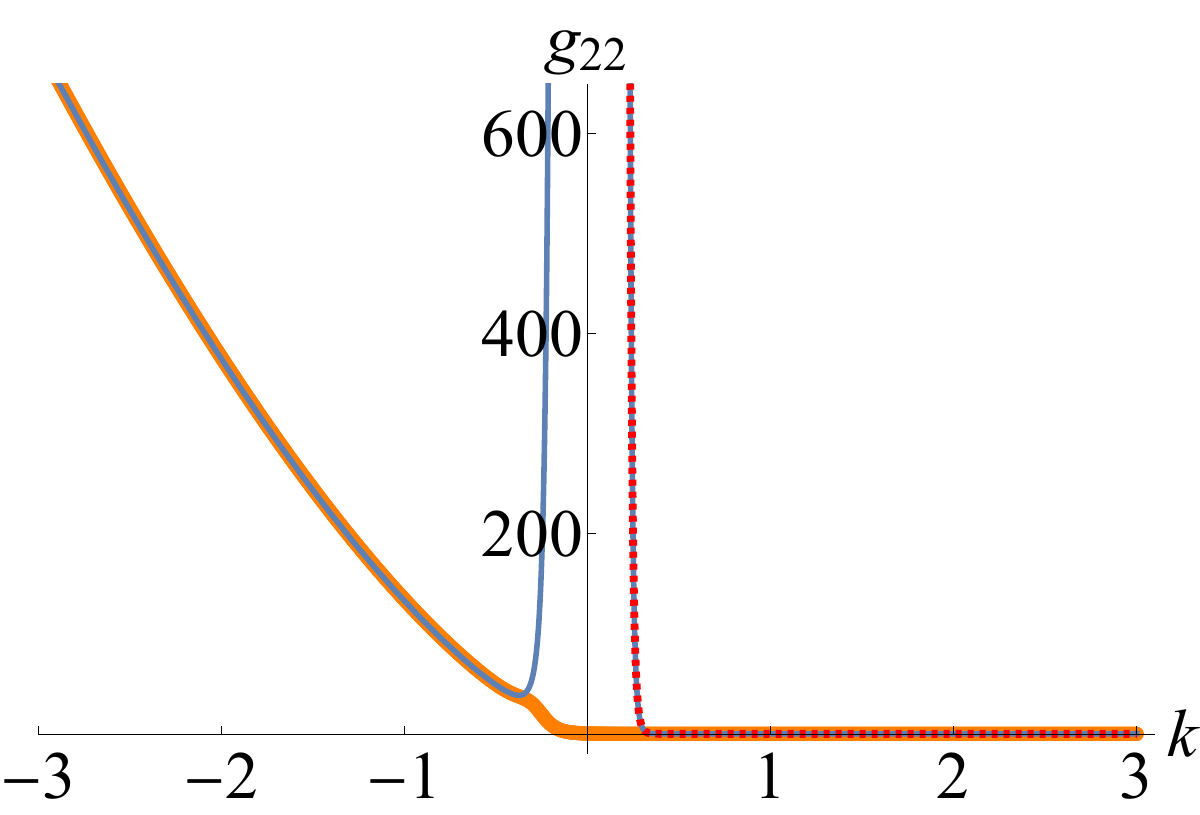} & \includegraphics[width= 0.49 \columnwidth]{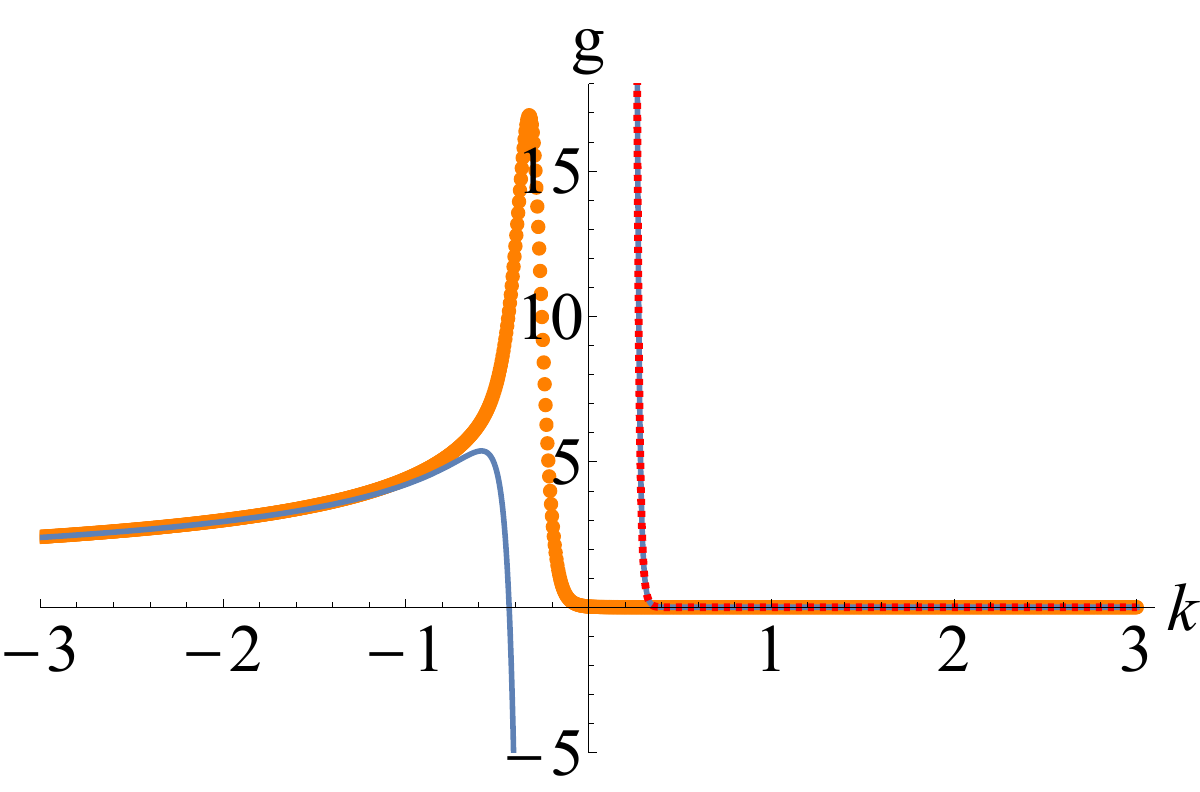} \\
(c) $g_{22}$ & (d) $g$ 
\end{tabular}
\caption{Comparison of the parameter space metrics for $\lambda=0.2$. Orange round markers correspond to the exact numerical QMT, red square markers correspond to the perturbative QTM, and the blue line corresponds to the perturbative CTM.} \label{Fig:QMT02}
\end{figure}

\begin{figure}[h!]
\includegraphics[width= 0.9 \columnwidth]{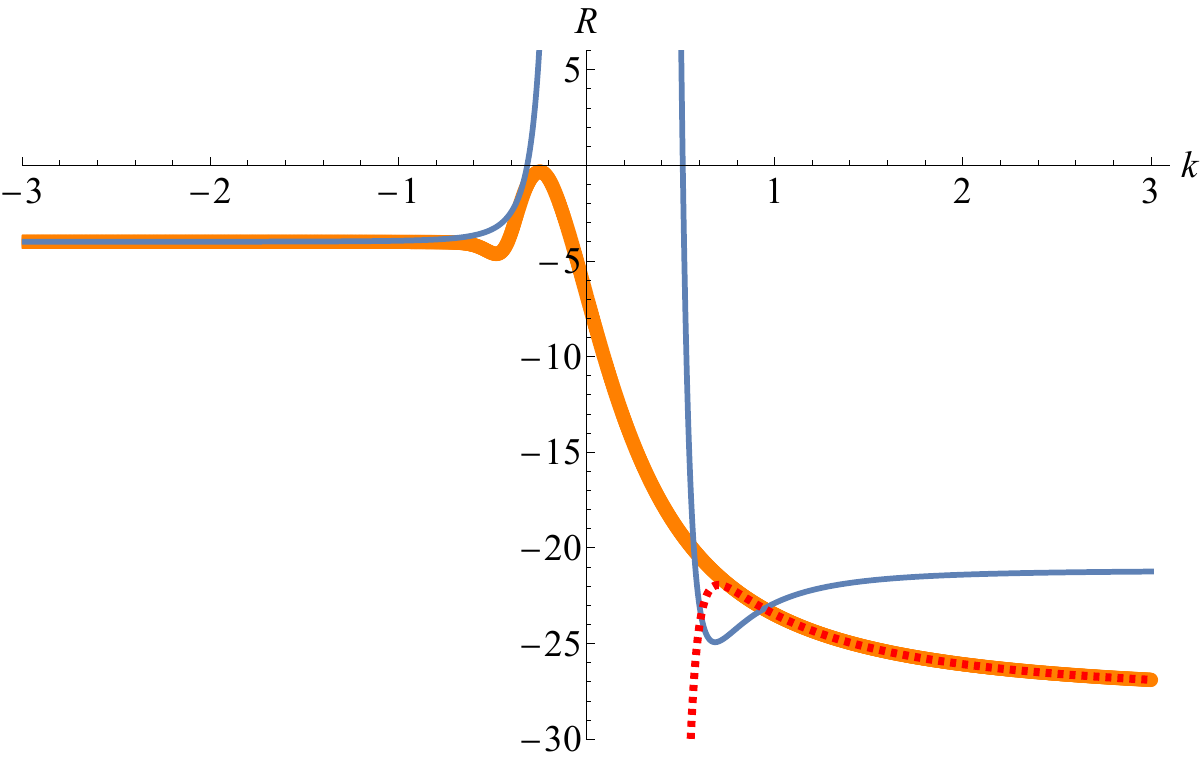} 
\caption{Comparison of the scalar curvatures for $\lambda=0.2$. Orange round markers correspond to the exact quantum numerical $R$, red square markers correspond the perturbative quantum $R$, and the blue line corresponds to the perturbative classical $R$.} \label{Fig:R02}
\end{figure}

In Fig. \ref{Fig:QMT02}, the plots of the numeric QMT, the analytic QMT, and the analytic CMT are shown for $\lambda=0.2$. We see that in fact the (numeric and analytic) quantum and classical metrics present the same behavior for values of $k$ far from $k=0$. Furthermore, it is clear that the analytic classical and quantum metric components diverge at $k=0$, while the numeric QMT remains finite. This behavior is also characteristic of the determinants of the corresponding metrics. From Fig. \ref{Fig:QMT02} we observe that the component $g_{11}$ of the numeric QMT displays peak at $k=-0.285$, whereas the component $g_{12}$ of this metric has a peak at $k=-0.32$ and a local minimum at $k=-0.45$. The analytic CMT counterpart does not possess these peaks; however, its $g_{12}$ component does have a local minimum at $k=-0.504$. Simultaneously, the determinant of the numeric QMT exhibits a peak at $k=-0.325$, while the determinant of the CMT shows a peak at $k=-0.586$. 
In Fig. \ref{Fig:R02}, we show the corresponding scalar curvatures for $\lambda=0.2$. We can see that the scalar curvature of the numeric QMT has a peak at $k=-0.245$ and a local minimum at $k=-0.48$.

The local maxima and minima in both the components of the metric tensor and the scalar curvature, for negative $k$, reflect the appearance of the delocalization of the probability distribution, i.e., the point where the probability density is spread out over the two wells (see Fig.~\ref{fig:kl}). 
On the other hand, the scalar curvature of the CMT presents a divergent behavior that signals the appearance of the aforementioned extreme values (maximum or minimum). Our results then reveal that the CMT can be used to predict the occurrence of peaks in the QMT, which indicate the appearance of the delocalization of the probability distribution. 

Fig \ref{Fig:R02} also shows that both scalar curvatures take the value of $-4$ in the region $k<0$ far form $k=0$, where the ground state wave function corresponds approximately to two well-located Gaussians on each of the wells. The negative constant scalar curvature means that the parameter space associated with the ground state has a hyperbolic geometry in that region. In addition, it is worth noting that, in the region, $k>0$, the scalar curvatures calculated using the analytic QMT and the perturbative CMT show a, analogous divergent behavior for values close to $k=0$. Finally, for $k\gg\lambda$ the system tends to behave like a harmonic oscillator, and for this reason, the components of the QTM and CMT tend to zero; however, in the limit $k\to \infty$ the scalar curvature of the exact numerical QMT tends to $-28$, while the scalar curvature of CMT tends to $21.1866$, as it can be verified from the analytic expressions of these curvatures, \eqref{pertQR} and \eqref{pertCR}.

\begin{figure}[h!]
	\begin{tabular}{c c}
		\includegraphics[width= 0.49 \columnwidth]{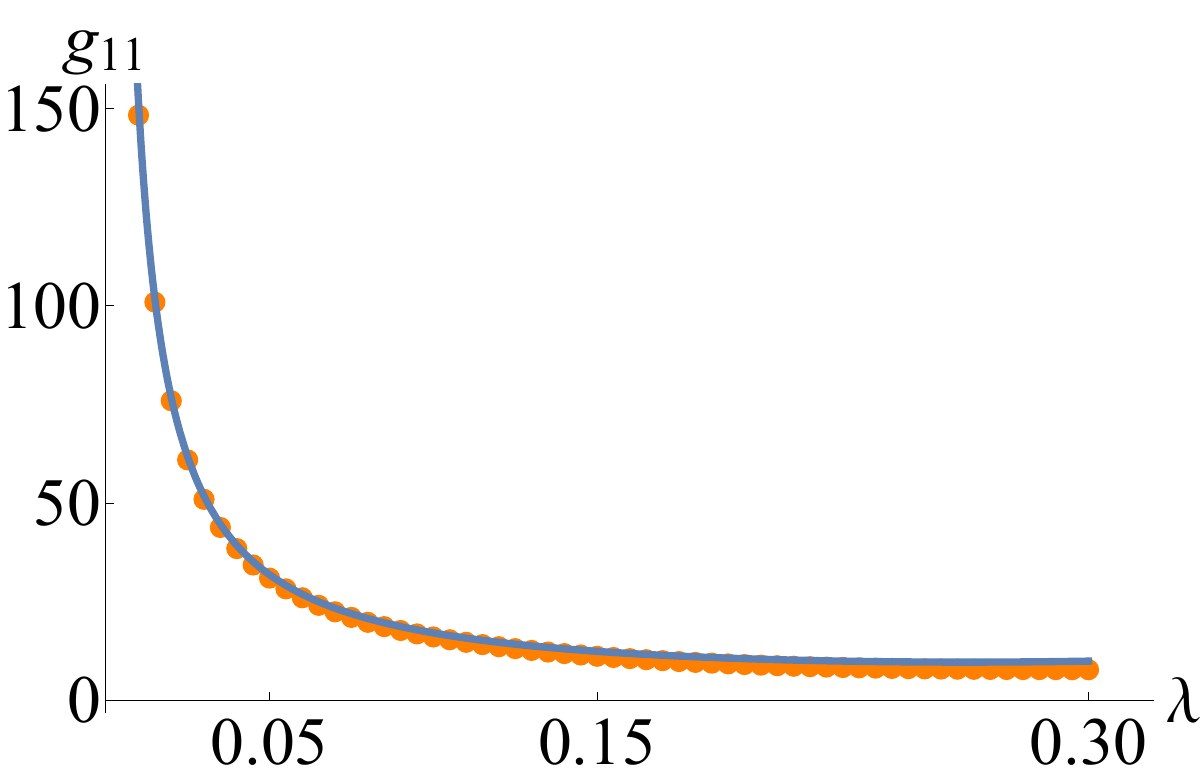} & \includegraphics[width= 0.49 \columnwidth]{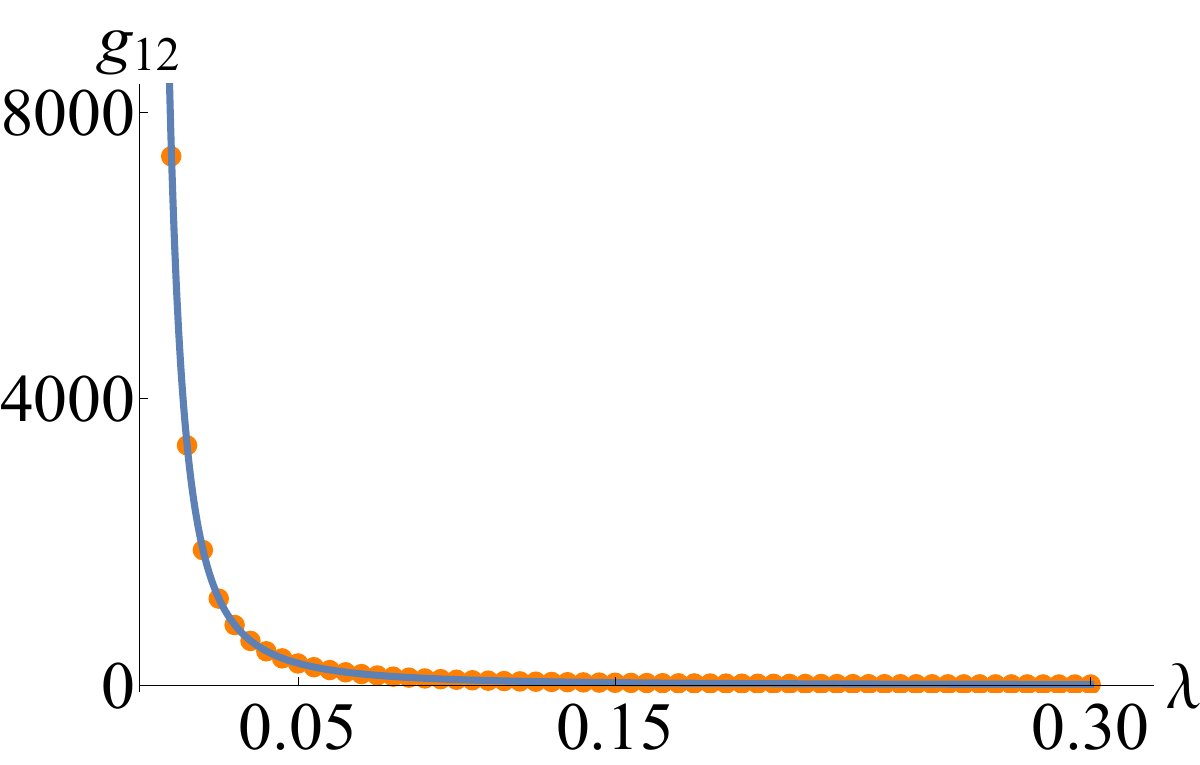} \\
		(a) $g_{11}$ & (b) $g_{12}$ \\
		\includegraphics[width= 0.49 \columnwidth]{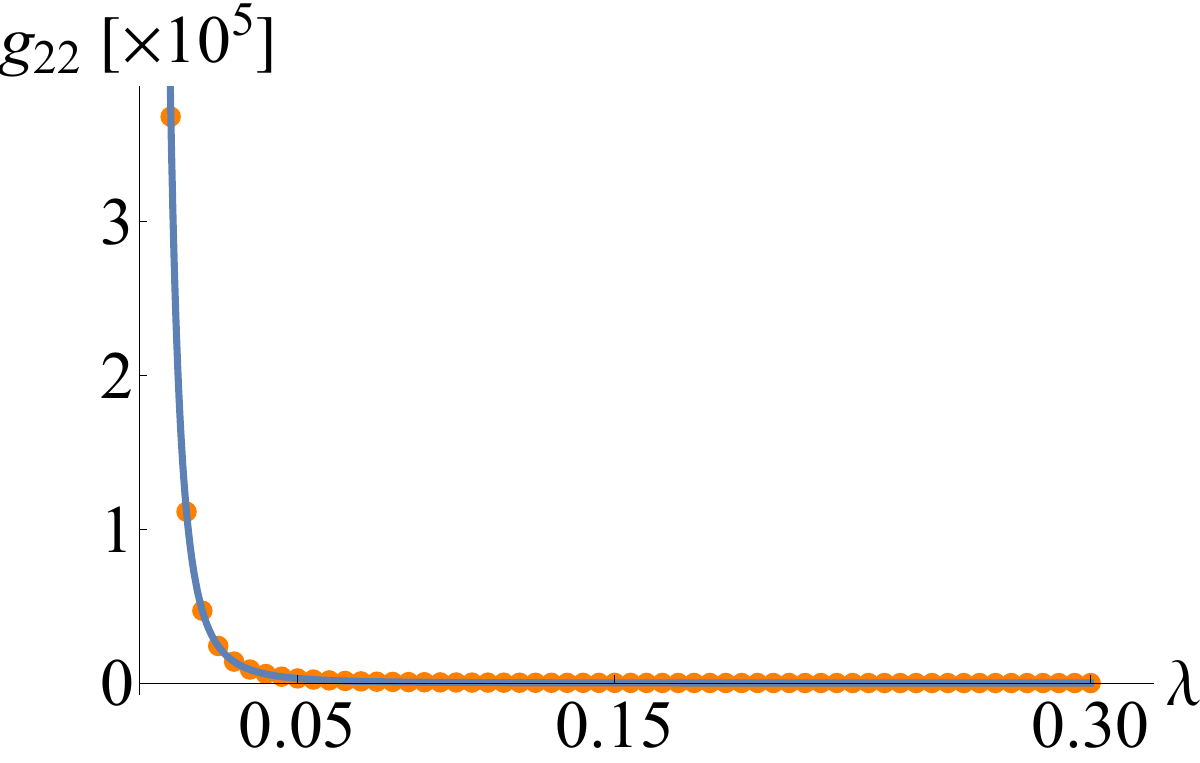} & \includegraphics[width= 0.49 \columnwidth]{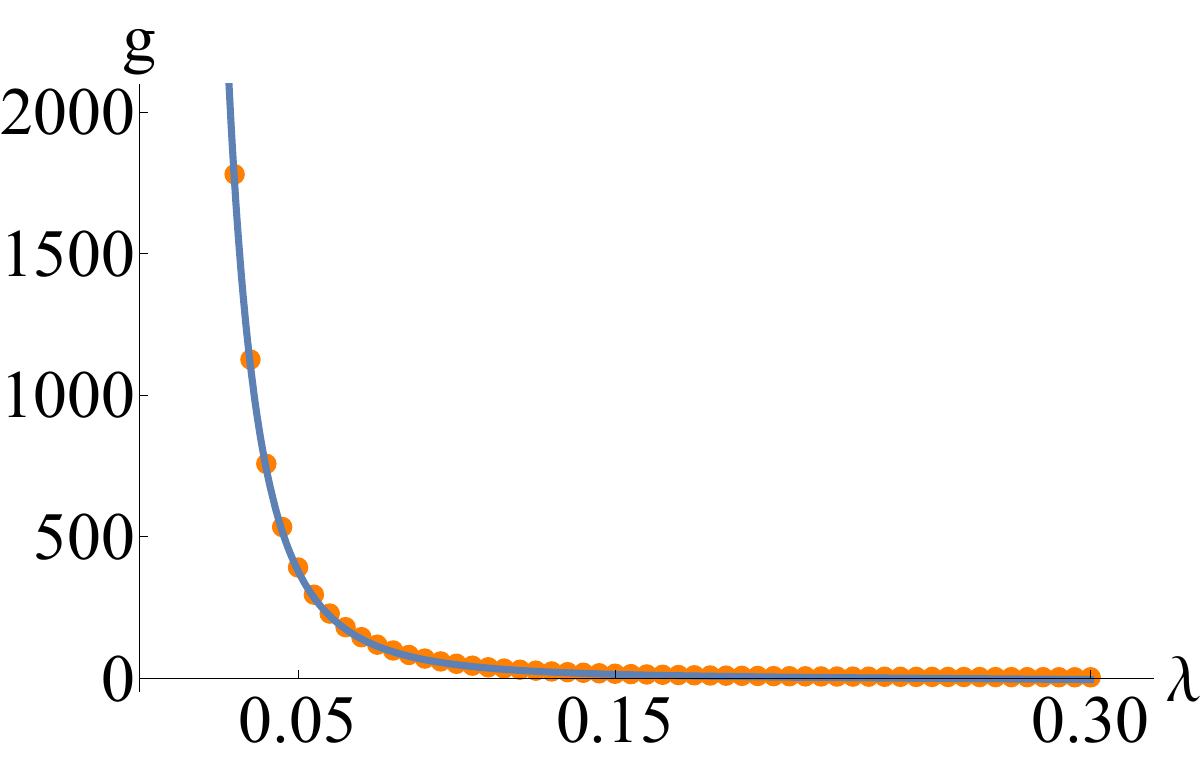} \\
		(c) $g_{22}$ & (d) $g$ 
	\end{tabular}
	\caption{Comparison of the parameter space metrics for $k=-0.5$. Orange round markers correspond to the exact numerical QMT and the blue line corresponds to the perturbative CTM.} \label{Fig:QMT05}
\end{figure}

In Fig. \eqref{Fig:QMT05}, we show the plots of the numerical QMT and analytic CMT for $k=-0.5$. Remarkably, the results show an excellent agreement between the components of QTM and the components of CMT. This confirms the usefulness of the classical framework as a tool to be adopted in order to have a first glance over the information contained in the parameter space of a quantum system.

\begin{figure}[h!]
\includegraphics[width= 0.9 \columnwidth]{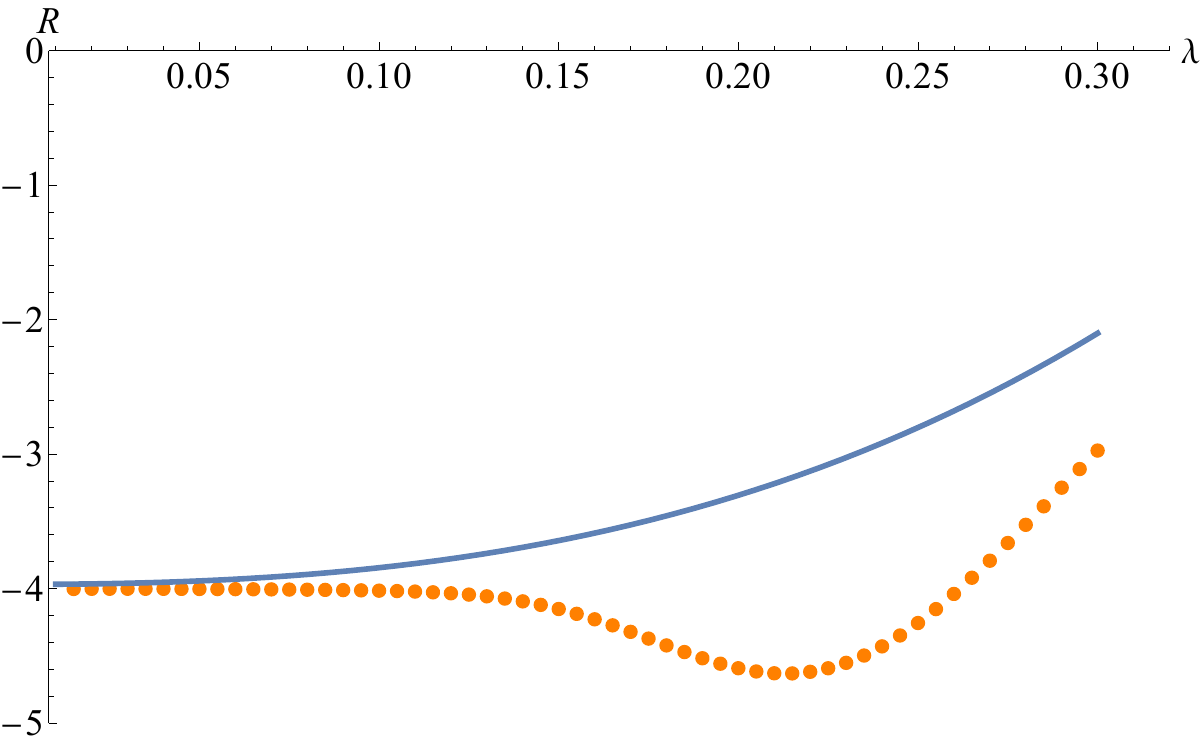} 
\caption{Comparison of the scalar curvatures for $k=-0.5$. Orange round markers correspond to the exact quantum numerical $R$ and the blue line corresponds to the perturbative classical $R$.} \label{Fig:R05}
\end{figure}

In Fig. \ref{Fig:R05}, the plots of the scalar curvatures of the numeric QMT and the analytic CMT are shown for $k=-0.5$. Note that for $\lambda\to0$ both scalar curvatures tend to $-4$, which can also be seen from \eqref{pertCR} in case of $R^{\rm cl}$ with $k<0$. The fact that the scalar curvature has a finite value at $\lambda\to0$ reveals that the divergence present in the QMT and the CMT can be removed by a change of coordinates in the parameter space. It is worth mentioning that in the case $\lambda=0$, the system's Hamiltonian reduces to that of an inverted harmonic oscillator, and then the classical and quantum methods used in this work cannot be applied, at least in the conventional form. Furthermore, from Fig. \ref{Fig:R05} we can see that the scalar curvature of the QMT has a local minimum at $\lambda=0.215$. The origin of this local minimum may be  analogous to the one of Fig. \ref{Fig:R02}, i.e., its appearance corresponds to the separation of the probability distribution into two branches.

\section{Comparison between quantum and classical approaches for the parameter space metric}

\begin{figure}[h!]
\includegraphics[width= 0.9 \columnwidth]{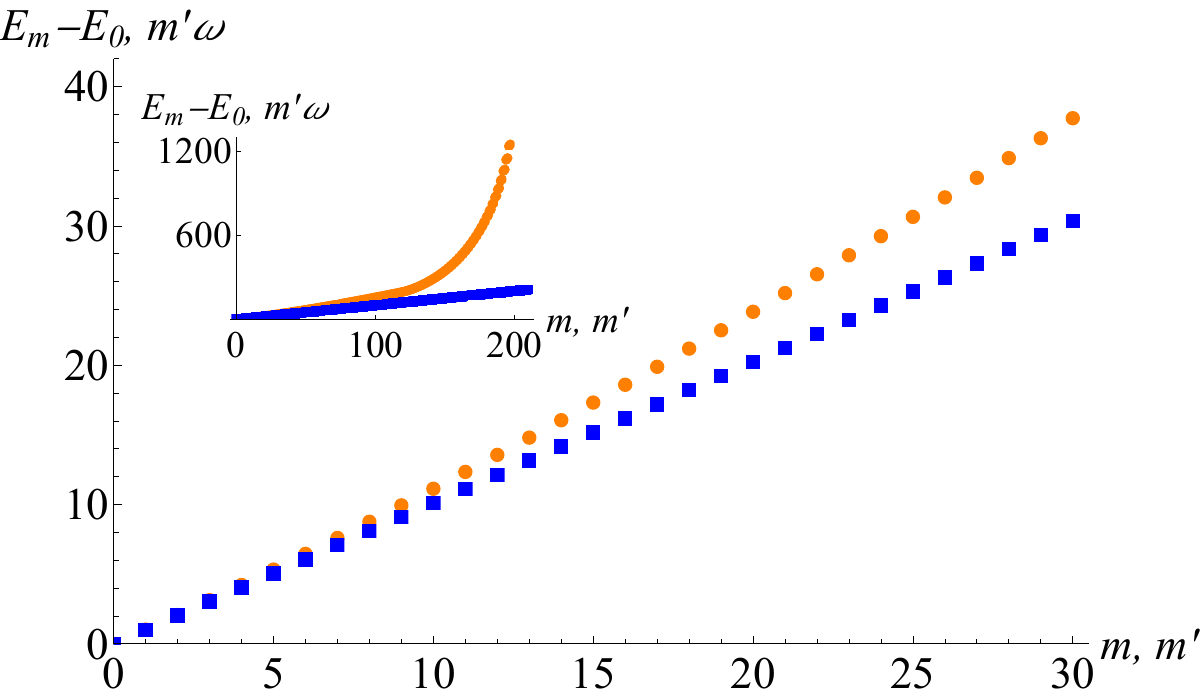} \\
(a) $k=1$ and $\lambda=0.2$\\
\includegraphics[width= 0.9 \columnwidth]{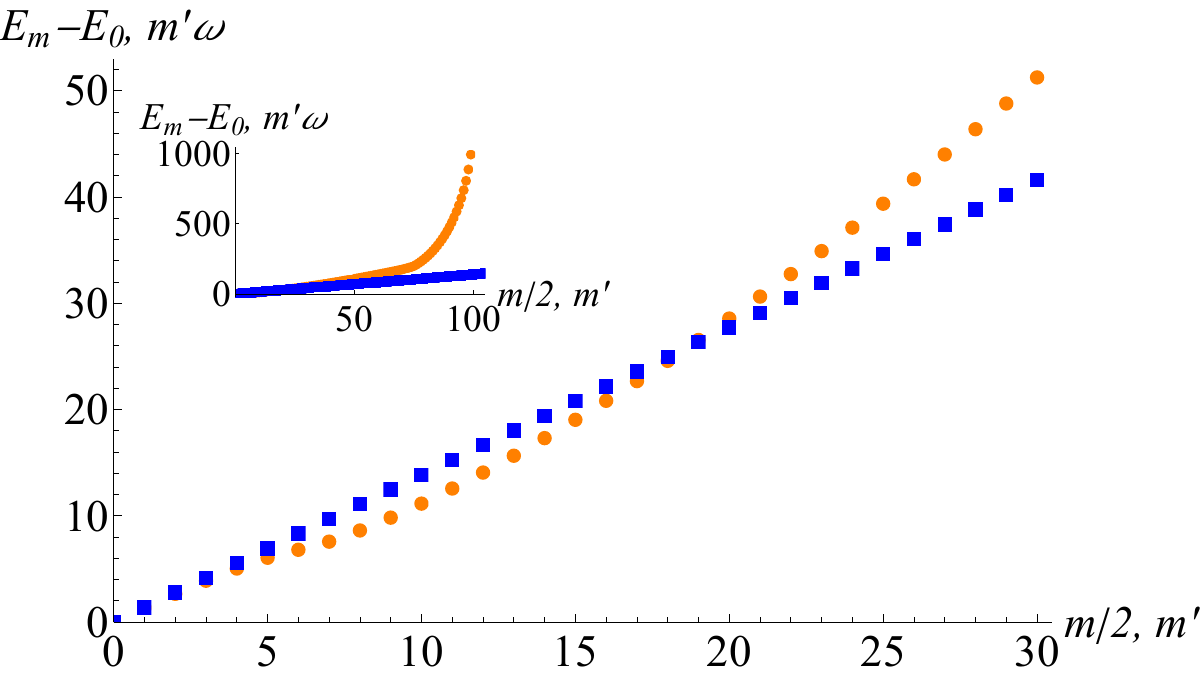} \\
(b) $k=-1$ and $\lambda=0.2$
\caption{Comparison between $E_m-E_0$ and $m'\omega$ . Orange round markers correspond to $E_m-E_0$ and blue square  markers correspond to $m'\omega$.} \label{Fig:E}
\end{figure}

The aim of this section is to carry out a more detailed analysis of the quantum and classical metric tensors, \eqref{QMTpert2} and \eqref{CMTpert2}, and to strengthen the analogy between them. To do this, we use the same identifications for powers of the action variable that were introduced in the previous section. We begin by comparing both metrics \eqref{QMTpert2} and  \eqref{CMTpert2}, which suggests that exact excitation energies $E_m-E_0$ are mimicked in the classical approximation by the harmonic oscillator energies $m'\omega$. In Fig.~\ref{Fig:E}(a) we plot $E_m-E_0$ and $m'\omega$ for $k=1$ and $\lambda=0.2$, finding that for small values of $m$ and $m'$ ($m,m'<10$), the quantities $E_m-E_0$ and $m'\omega$ remain close to each other. However, for large values of $m$ and $m'$ ($m, m'>10$), $E_m-E_0$ deviates from the linear behavior of $m'\omega$, which is a consequence of quantum corrections.

In the case $k=-1$ and $\lambda=0.2$, the first seven values of the energies are $E_0=0$, $E_1=1.04\times 10^{-11}$, $E_2=1.360866$, $E_3=1.360866$, $E_4=2.661983$, $E_5=2.661984$, $E_6=3.893522$, and $E_7=3.893550$. From this it is clear that $E_m \approx E_{m+1}$ for even $m$, which is a quasi-degeneration that emerges as a consequence of the double well potential. In contrast, in the classical case, all ``excitation'' energies $m'\omega$ appear only once. This is because, in the classical perturbation formalism, we have considered only one of the wells, the one associated with the fixed point $\chi_1$. We have verified that employing the well involving the fixed point $\chi_2$, the resulting CMT is exactly the same. In Fig.~\ref{Fig:E}(b) we plot $E_{m}-E_0$ and $m'\omega$ as functions of $m/2$ and $m'$, respectively. We can see that both quantities present a similar behavior for small values of $m$ ($m/2<20$) and $m'$ ($m'<20$).

\begin{figure}[h!]
\includegraphics[width= 0.9 \columnwidth]{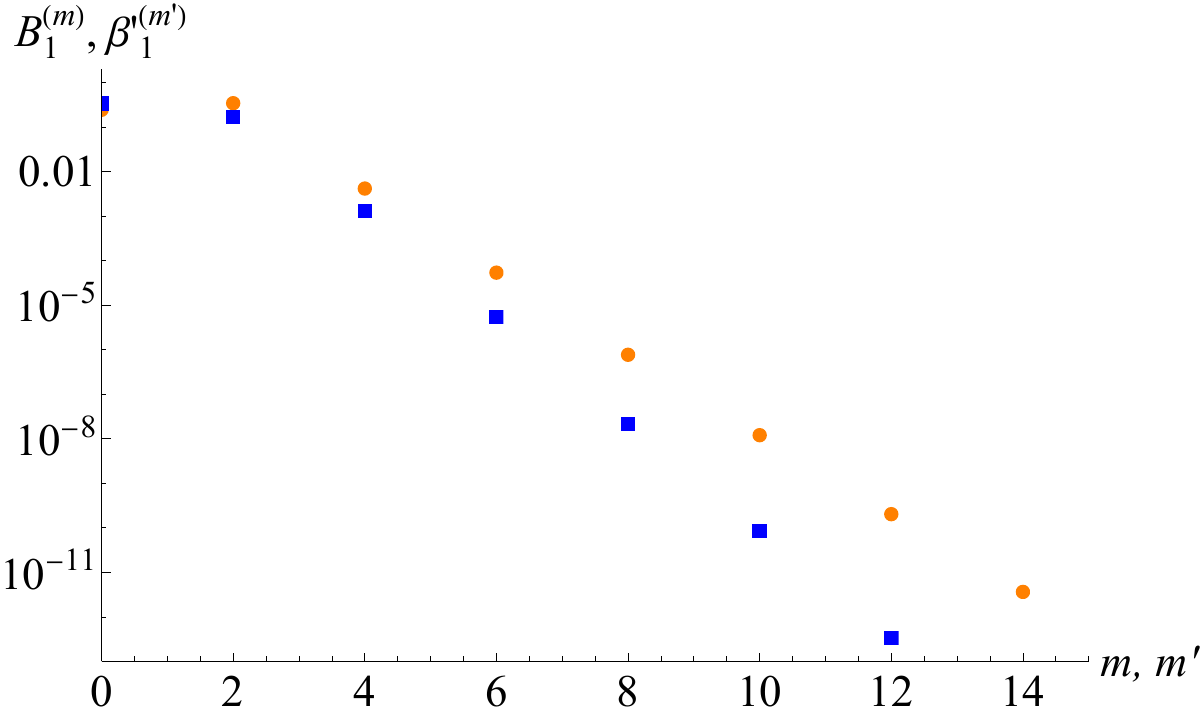} \\
(a) \\
\includegraphics[width= 0.9 \columnwidth]{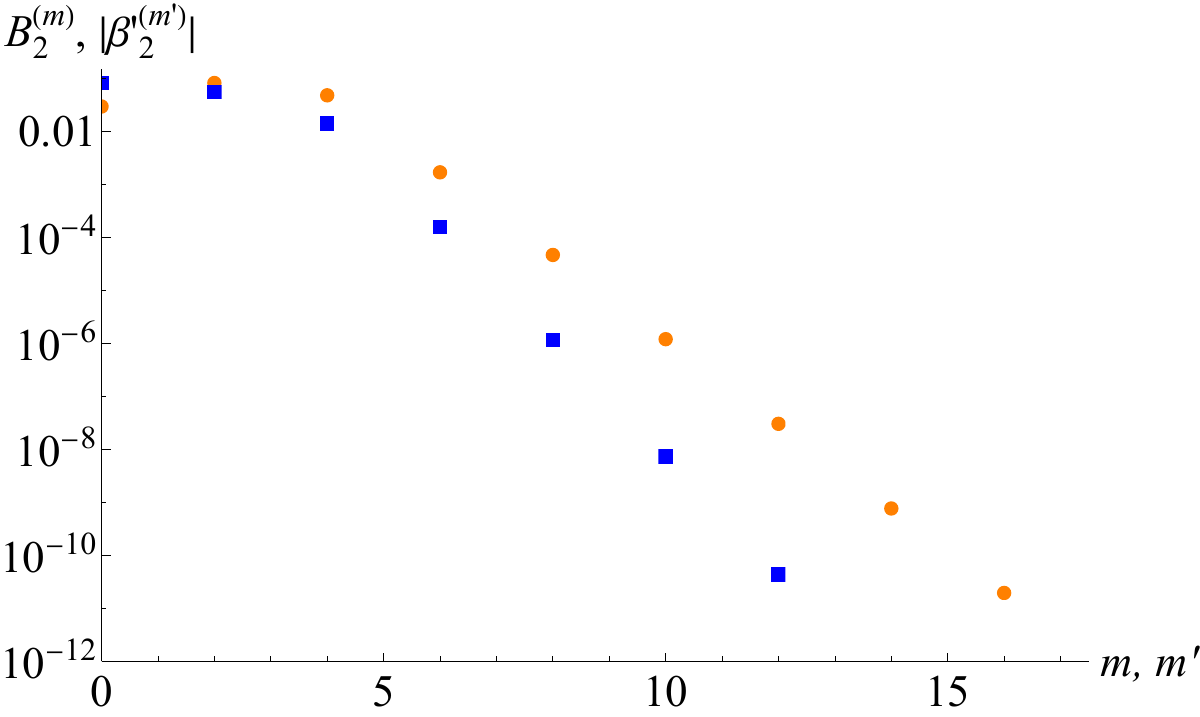} \\
(b)
\caption{Comparison between $B^{(m)}_i$ and $\beta'^{(m')}_i$, with $k=1$ and $\lambda=0.2$. Orange round markers correspond to $B^{(m)}_i$ and blue square markers correspond to $\beta'^{(m')}_i$.} \label{Fig:Bipos}
\end{figure}

The comparison between the metrics \eqref{QMTpert2} and \eqref{CMTpert2} also suggests that  classical functions $\beta'^{(m')}_i$ must play an analogous role to that of the functions $B^{(m)}_i$ of the quantum case. In Fig.~\ref{Fig:Bipos} we plot $B^{(m)}_i$ and $|\beta'^{(m')}_i|$ as functions of $m$ and $m'$, setting $k=1$ and $\lambda=0.2$. From this plot we can see that both functions exhibit a very similar behavior, approaching each other for small values of $m$ and $m'$. Also, we can notice that the classical functions $\beta'^{(m')}_i$ tend to zero faster than their quantum counterparts $B^{(m)}_i$. Remarkably and reinforcing the analogy, for odd $m$ and $m'$ both functions  $B^{(m)}_i$ and $\beta'^{(m')}_i$ are zero. This is the reason why only the values of $B^{(m)}_i$ and $\beta'^{(m')}_i$ for even $m$ and $m'$ appear in the plot.

\begin{figure}[h!]
\includegraphics[width= 0.9 \columnwidth]{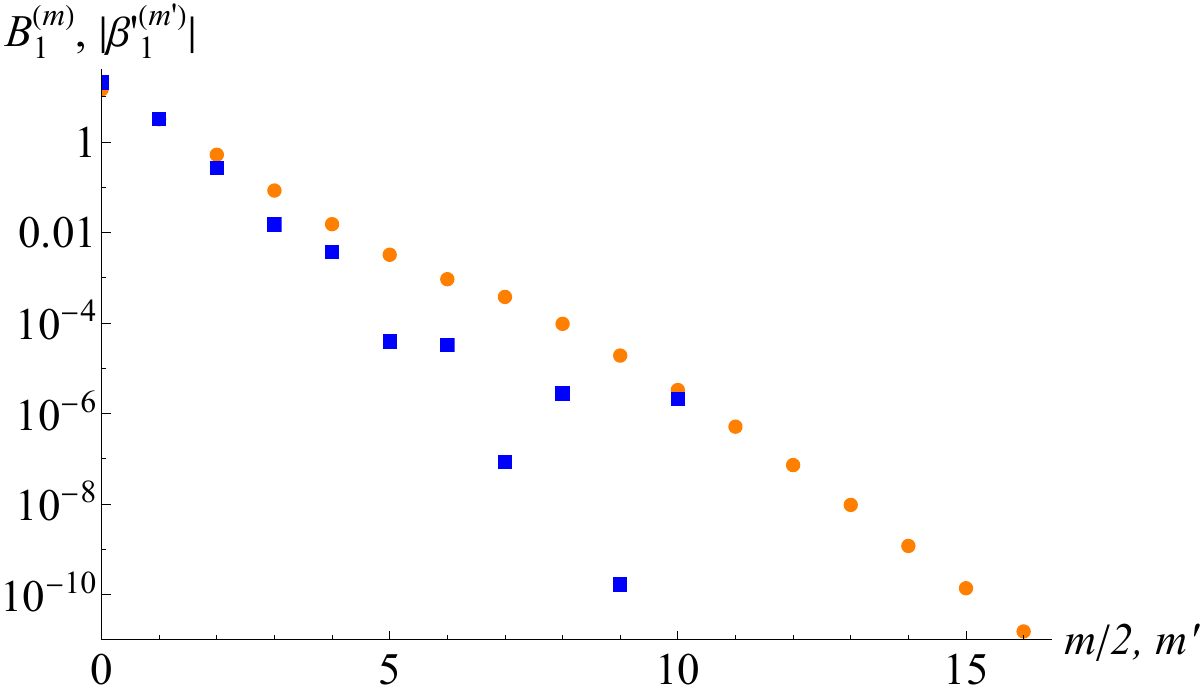} \\
(a) \\
\includegraphics[width= 0.9 \columnwidth]{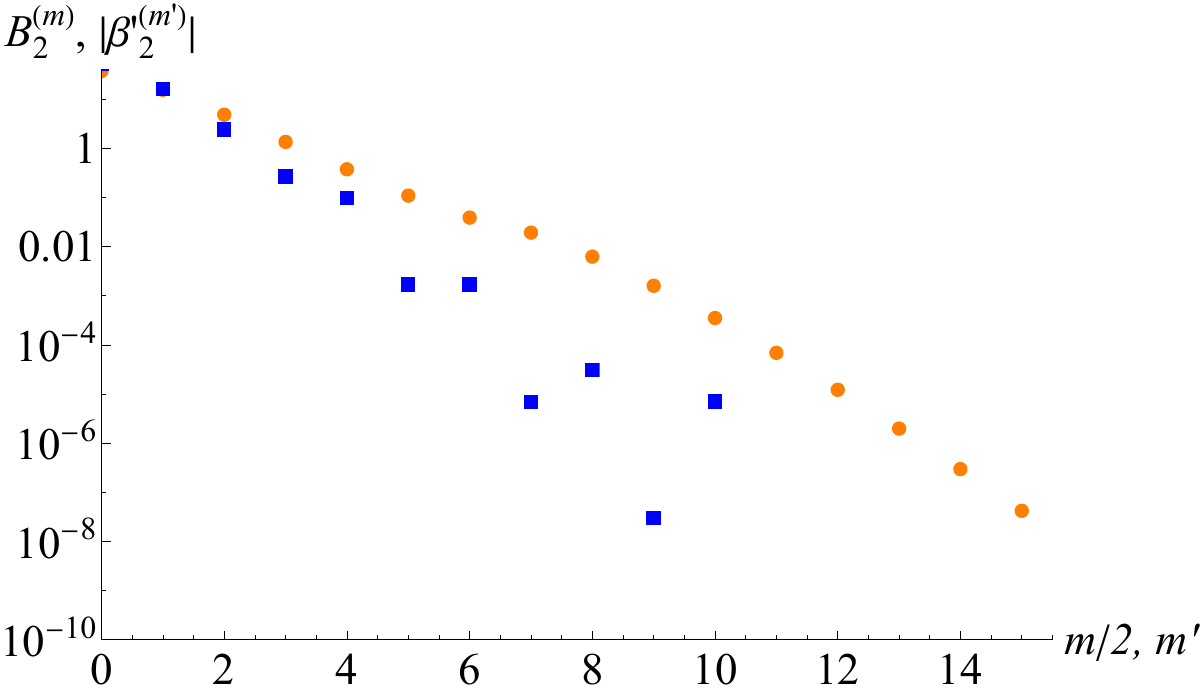} \\
(b)
\caption{Comparison between $B^{(m)}_i$ and $\beta'^{(m')}_i$, with $k=-1$ and $\lambda=0.2$. Orange round markers correspond to $B^{(m)}_i$ and blue square markers correspond to $|\beta'^{(m')}_i|$.} \label{Fig:Bineg}
\end{figure}

In the case  $k=-1$ and $\lambda=0.2$, the functions $\beta'^{(m')}_i$ are real for even $m'$ and pure imaginary for odd $m'$. However, for the same values of $k$ and $\lambda$, the functions $B^{(m)}_i$ are real for even $m$ and vanishes for odd $m$. In Fig. \ref{Fig:Bineg} we plot $B^{(m)}_i$ and $|\beta'^{(m')}_i|$ as functions of $m/2$ and $m'$, respectively. From this plot we can appreciate that both functions, $B^{(m)}_i$ and $|\beta'^{(m')}_i|$, exhibit analogous behavior. We have plotted $B^{(m)}_i$ as a function of $m/2$ because of the quasi-degeneration resulting from the double well potential, which does not have a classical counterpart.

\begin{figure}[h!]
\begin{tabular}{c c}
\includegraphics[width= 0.49 \columnwidth]{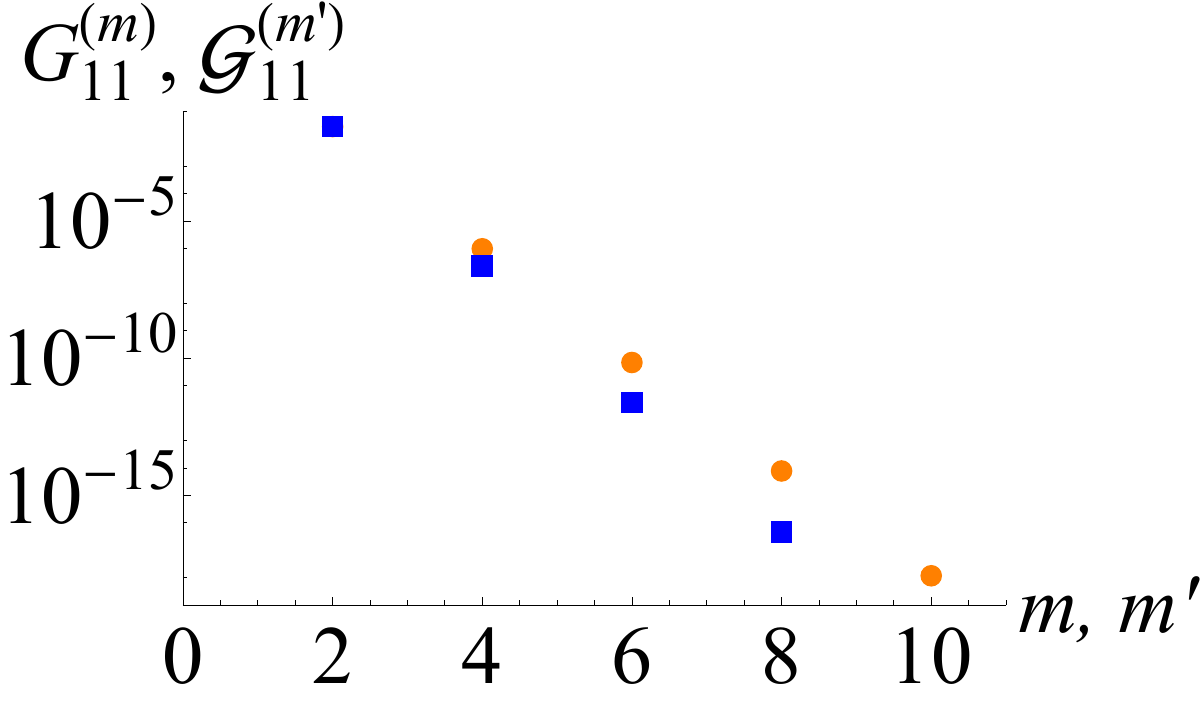} & \includegraphics[width= 0.49 \columnwidth]{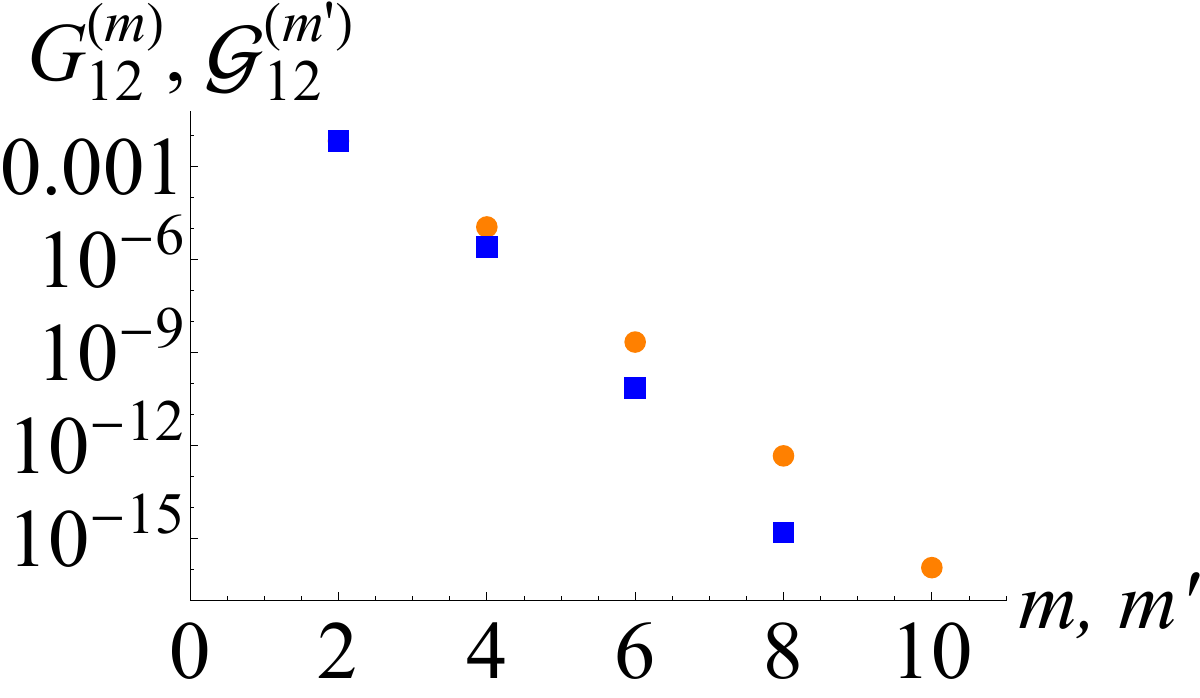} \\
(a) & (b) \\
\includegraphics[width= 0.49 \columnwidth]{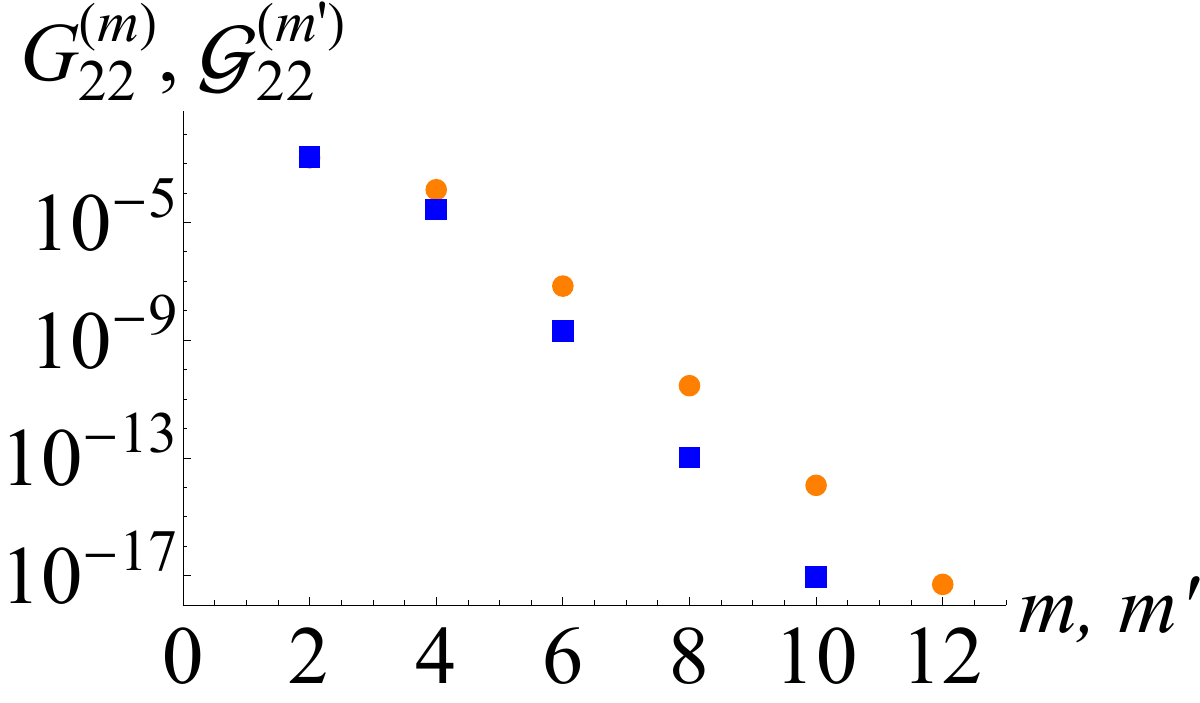} & \\
(c) & 
\end{tabular}
\caption{Comparison between $G^{(m)}_{ij}$ and $\mathcal{G}^{(m')}_{ij}$, with $k=1$ and $\lambda=0.2$. Orange round markers correspond to $G^{(m)}_{ij}$ and blue square markers correspond to $\mathcal{G}^{(m')}_{ij}$.} \label{Fig:Gpos}
\end{figure}

To aid in a better understanding of this, in Figs.~\ref{Fig:Gpos} we show $G^{(m)}_{ij}$ and $\mathcal{G}^{(m')}_{ij}$ for $k=1$ and $\lambda=0.2$. We see that the significant contributions of $G^{(m)}_{ij}$ and $\mathcal{G}^{(m')}_{ij}$ to their metrics occur for $m \leq 4$ and $m' \leq 4$, and that precisely for these values of $m$ and $m'$ there is good agreement between these quantities. For $m>4$ and $m'>4$, the contributions of $G^{(m)}_{ij}$ and $\mathcal{G}^{(m')}_{ij}$ are of the order of $10^{-5}$ or smaller.

\begin{figure}[h!]
\begin{tabular}{c c}
\includegraphics[width= 0.49 \columnwidth]{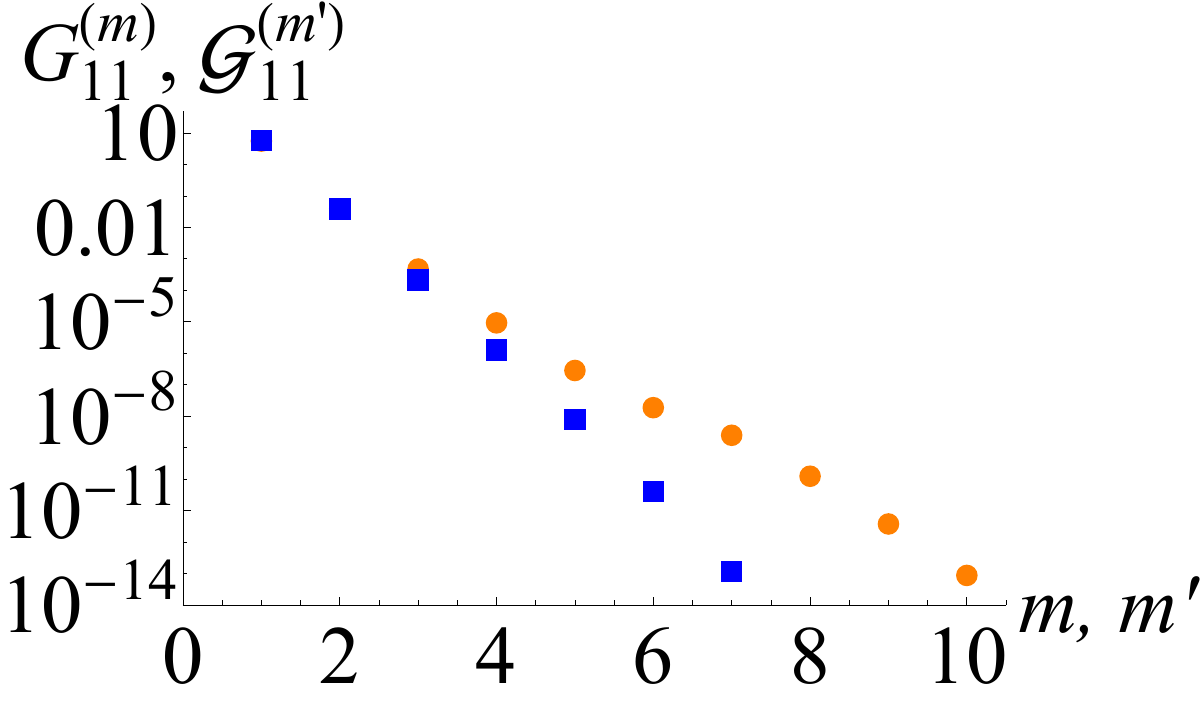} & \includegraphics[width= 0.49 \columnwidth]{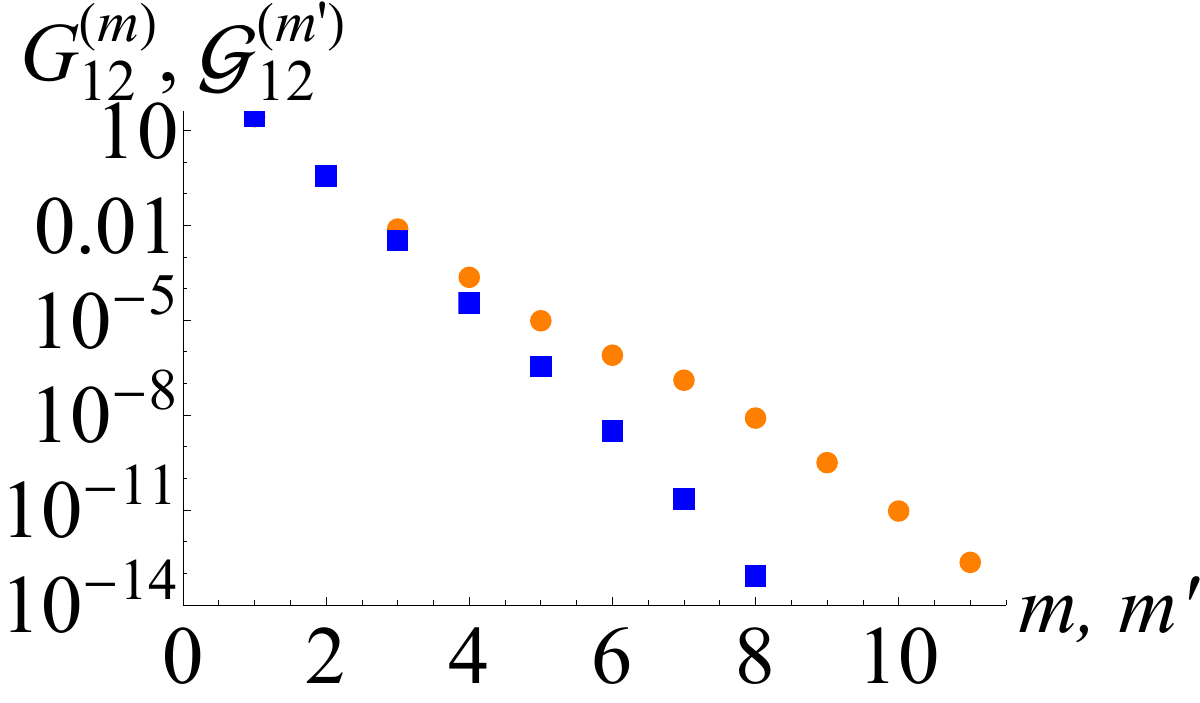} \\
(a) & (b)  \\
\includegraphics[width= 0.49 \columnwidth]{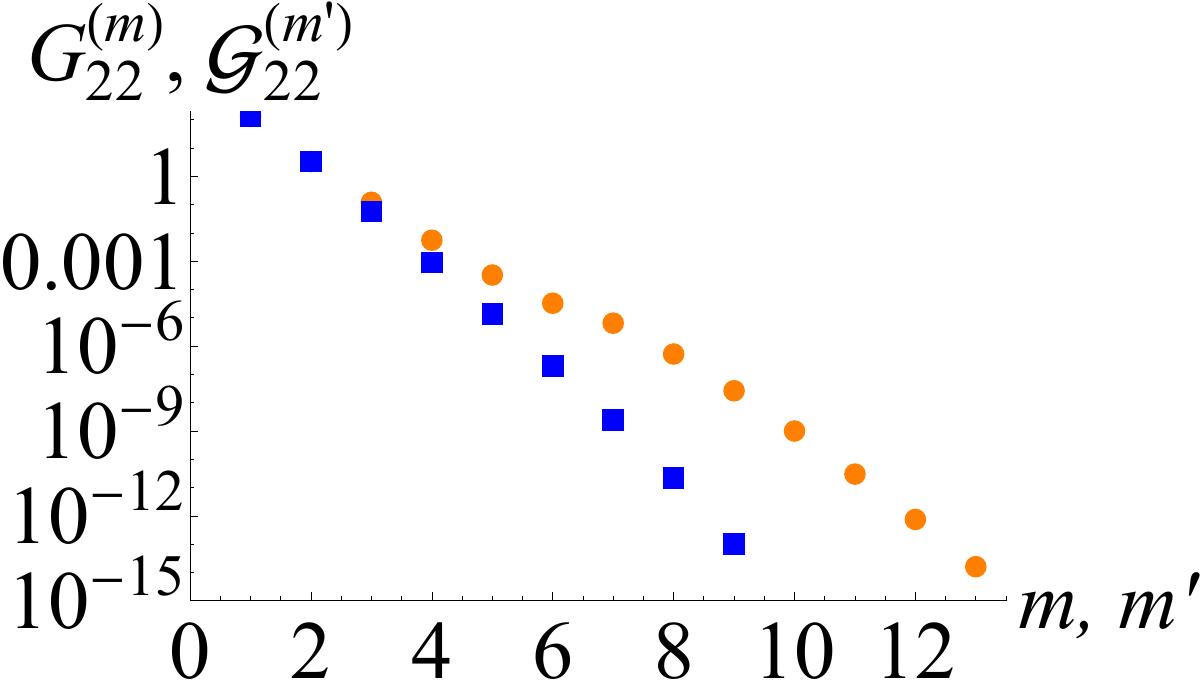} & \\
(c)  & 
\end{tabular}
\caption{Comparison between $G^{(m)}_{ij}$ and $\mathcal{G}^{(m')}_{ij}$, with $k=-1$ and $\lambda=0.2$. Orange round markers correspond to $G^{(m)}_{ij}$ and blue square markers correspond to $\mathcal{G}^{(m')}_{ij}$.} \label{Fig:Gneg}
\end{figure}

In the case $k=-1$ and $\lambda=0.2$, the classical function $\mathcal{G}^{(m')}_{ij}$ is real for all $m'$, since the product ${\beta'}_{i}^{(m')} ({\beta'}_{j}^{(m')})^*$ is real because $\beta^{(m')}_i$ is real for even $m'$ and pure imaginary for odd $m'$. In the quantum case, the function $G^{(m)}_{ij}$ is nonzero for even $m$ and zero for odd $m$. Clearly, this is because $B^{(m)}_i$ is nonzero for even $m$ and zero for odd $m$. In Fig. \ref{Fig:Gneg} we plot $G^{(m)}_{ij}$ and $\mathcal{G}^{(m')}_{ij}$ as functions of $m$ and $m'$ for $k=-1$ and $\lambda=0.2$. Here we can also appreciate that the relevant contributions of $G^{(m)}_{ij}$ and $\mathcal{G}^{(m')}_{ij}$ to the quantum and classical metrics, respectively, occur for $m\leq 5$ and $m'\leq 5$. Certainly, the contributions of  $G^{(m)}_{ij}$ and $\mathcal{G}^{(m')}_{ij}$ are of the order of $10^{-5}$ or smaller for $m>5$ and $m'>5$.

\section{Conclusions}

In this paper we have studied the geometry of the space parameter  of the single-well anharmonic oscillator ($k>0$ ) and the quartic double-well potential ($k<0$) from the classical and quantum points of view. In the quantum setting, we computed the exact QMT and its scalar curvature numerically for both the single-well and double-well potentials, using a basis of oscillator states and performing an exact numerical diagonalization. In the quantum setting, we also obtained analytically the QMT and its scalar curvature for the ground-state of the single-well case ($k>0$), by following a perturbative treatment in the parameter $\lambda$ and employing the method proposed in Ref.~\cite{Turbiner1984} to obtain the ground-state wave function up to 10th order in $\lambda$. In the  classical framework, we computed analytically the CMT and its scalar curvature for both cases, the single-well and double-well potentials, by employing a formulation of the CMT based on Fourier series and introduced in \cite{Diego5}. This approach allowed us to obtain the CMT up to 10th order in $\lambda$ for the single-well system and up to 6th order in $\lambda$ for the double-well system.

We performed a detailed analysis of the exact quantum numerical results and the classical analytical results. This analysis was accomplished by considering identifications for the different powers of the action variable that arise from the semiclassical relation  \eqref{semiclassical} between the analytic classical and quantum metrics of the single-well problem. We found that the QMT and the CMT have a very close behavior except in points near $k=0$, as shown in Fig. \ref{Fig:QMTandR}. Similarly, we found that there is a very good agreement between the corresponding scalar curvatures of these classical and quantum metrics for that region.

For $\lambda=0.2$, it was shown that components of the numeric QMT (except the component $g_{22}$) and its scalar curvature exhibit peaks indicating the  appearance of a delocalization of the probability distribution. The classical metric and its curvature show a divergent behavior at $k=0$ that could be regarded as a sign of the appearance of such peaks. In addition, for $\lambda=0.2$ with $k<0$, far from $k=0$, both classical and quantum scalar curvatures take the value of $-4$, indicating that the associated parameter space has a hyperbolic geometry in that region. In contrast, for $\lambda=0.2$ with $k>0$ far from $k=0$ the scalar curvature of the QMT tends to $-28$, while its classical analog tends to $21.1866$. This discrepancy could be because the ground state wave function is strictly not Gaussian for those values of the parameters.

The case with $k=-0.5$ was also analyzed. We found a remarkable agreement between the
numeric QMT and analytic CMT, which show a  divergent behavior for $\lambda\to0$. However, for this limit, the scalar curvatures of both metrics tend to $-4$, meaning that the singularity at $\lambda\to0$ is apparent and can be removed by performing a change of coordinates (parameters). In this case, it is also worth mentioning that the scalar curvature of the QMT has a local minimum, which is absent in its classical counterpart and could be related to a separation of the probability distribution into two branches.

We also compared in detail the perturbative expression of QMT \eqref{QMTpert2} with the Fourier-based expression of the CMT \eqref{CMTpert2}, and found that the quantum energies $E_m-E_0$ and the classical quantity $m'\omega$ have similar behavior, as well as that the classical functions $\beta'^{(m')}_i$ play an analogous role to that of the functions $B^{(m)}_i$. This allows us to conclude that \eqref{CMTpert2} can be regarded as the classical version of \eqref{QMTpert2}.

\acknowledgments

We acknowledge the support of the Computing Center-ICN, in particular, of E. Palacios, L. D\'\i az, and E. Murrieta.   This work was partially supported by DGAPA-PAPIIT Grants No. IN105422 and IN109523. D.G. acknowledges the financial support under Project SIP-20230323 of IPN. J.C.-C was funded by the NSF CCI grant (award Number 2124511).

\appendix

\


\begin{widetext}

\section{Generating functions and Fourier coefficients} \label{AppGenerating}
\subsection{Case $k>0$}
Functions $W_{\mu}$ for $\mu=1,\dots,10$:
\begin{subequations}\label{Wnfunctkpos}
\begin{align}
    W_1=&-\frac{I^2}{192 k^{3/2}} \left(-8 \sin \left(2 \phi _0\right)+\sin \left(4 \phi _0\right)\right), \\
    W_2=&\frac{I^3}{55296 k^3} \left(-384 \sin \left(2 \phi _0\right)+132 \sin \left(4 \phi _0\right)-32 \sin \left(6 \phi _0\right)+3 \sin \left(8 \phi _0\right)\right),\\
    W_3=&\frac{I^4}{5308416 k^{9/2}} \left(9264 \sin \left(2 \phi _0\right)-4101 \sin \left(4 \phi _0\right)+1624 \sin \left(6 \phi _0\right)-441 \sin \left(8 \phi _0\right)+72 \sin \left(10 \phi _0\right)-5 \sin \left(12 \phi _0\right)\right), \\
    W_4=&\frac{I^5}{21233664 k^6} \left(-11408 \sin \left(2 \phi _0\right)+5644 \sin \left(4 \phi _0\right)-2768 \sin \left(6 \phi _0\right)+\frac{2171}{2} \sin \left(8 \phi _0\right)-320 \sin \left(10 \phi _0\right)+65 \sin \left(12 \phi _0\right) \right. \nonumber\\
    &\left.-8 \sin
   \left(14 \phi _0\right)+\frac{7}{16} \sin \left(16 \phi _0\right)\right), \\
   W_5=&\frac{7 I^6}{122305904640 k^{15/2}} \left(3261840 \sin \left(2 \phi _0\right)-1713090 \sin \left(4 \phi _0\right)+945280 \sin \left(6 \phi _0\right)-450270 \sin \left(8 \phi _0\right) \right. \nonumber\\
   & \left.+176832 \sin \left(10 \phi _0\right)-54575 \sin \left(12 \phi
   _0\right)+12600 \sin \left(14 \phi _0\right)-2025 \sin \left(16 \phi _0\right)+200 \sin \left(18 \phi _0\right)-9 \sin \left(20 \phi _0\right)\right), \\
   W_6=&\frac{I^7}{48922361856 k^9} \left(-3438784 \sin \left(2 \phi _0\right)+\frac{3744541}{2} \sin \left(4 \phi _0\right)-1111556 \sin \left(6 \phi _0\right)+\frac{9550433}{16} \sin \left(8 \phi _0\right)\right. \nonumber\\
   & \left.-278530 \sin \left(10 \phi _0\right)+\frac{436033}{4}
   \sin \left(12 \phi _0\right)-34744 \sin \left(14 \phi _0\right)+\frac{69727}{8} \sin \left(16 \phi _0\right)-\frac{4936}{3} \sin \left(18 \phi _0\right)\right. \nonumber\\
   & \left.+\frac{873}{4} \sin \left(20 \phi _0\right)-18 \sin \left(22 \phi
   _0\right)+\frac{11}{16} \sin \left(24 \phi _0\right)\right),\\
   W_7=&\frac{I^8}{350675489783808 k^{21/2}} \left(9825412576 \sin \left(2 \phi _0\right)-5476716371 \sin \left(4 \phi _0\right)+3414796056 \sin \left(6 \phi _0\right)\right. \nonumber\\
   & \left.-1987448015 \sin \left(8 \phi _0\right)+1040341960 \sin \left(10 \phi _0\right)-475378323 \sin \left(12
   \phi _0\right)+185530096 \sin \left(14 \phi _0\right)\right. \nonumber\\
   & \left.-60516911 \sin \left(16 \phi _0\right)+16101008 \sin \left(18 \phi _0\right)-3387615 \sin \left(20 \phi _0\right)+539784 \sin \left(22 \phi _0\right)-60907 \sin \left(24 \phi
   _0\right)\right. \nonumber\\
   & \left.+4312 \sin \left(26 \phi _0\right)-143 \sin \left(28 \phi _0\right)\right),\\
   W_8=&\frac{I^9}{1803473947459584 k^{12}} \left(-21017661600 \sin \left(2 \phi _0\right)+11907018980 \sin \left(4 \phi _0\right)-\frac{23057409280}{3} \sin \left(6 \phi _0\right)\right. \nonumber\\
   & \left.+4736665695 \sin \left(8 \phi _0\right)-2690310560 \sin \left(10 \phi
   _0\right)+1370942900 \sin \left(12 \phi _0\right)-615537920 \sin \left(14 \phi _0\right)\right. \nonumber\\
   & \left.+\frac{479168405}{2} \sin \left(16 \phi _0\right)-\frac{238607680}{3} \sin \left(18 \phi _0\right)+22109724 \sin \left(20 \phi _0\right)-5031712
   \sin \left(22 \phi _0\right)\right. \nonumber\\
   & \left.+\frac{2729749}{3} \sin \left(24 \phi _0\right)-125312 \sin \left(26 \phi _0\right)+12298 \sin \left(28 \phi _0\right)-\frac{2288}{3} \sin \left(30 \phi _0\right)+\frac{715}{32} \sin \left(32 \phi
   _0\right)\right),\\
   W_9=&\frac{11 I^{10}}{6232805962420322304 k^{27/2}} \left(2838604924368 \sin \left(2 \phi _0\right)-1627087268250 \sin \left(4 \phi _0\right)+1077391402752 \sin \left(6 \phi _0\right)\right. \nonumber\\
   & \left.-692748547848 \sin \left(8 \phi _0\right)+418042653696 \sin \left(10 \phi
   _0\right)-230923291308 \sin \left(12 \phi _0\right)+114908848992 \sin \left(14 \phi _0\right)\right. \nonumber\\
   & \left.-50823988380 \sin \left(16 \phi _0\right)+19735646368 \sin \left(18 \phi _0\right)-6642798660 \sin \left(20 \phi _0\right)+1910480256 \sin
   \left(22 \phi _0\right)\right. \nonumber\\
   & \left.-461485548 \sin \left(24 \phi _0\right)+91604736 \sin \left(26 \phi _0\right)-14510691 \sin \left(28 \phi _0\right)+1758744 \sin \left(30 \phi _0\right)-152685 \sin \left(32 \phi _0\right)\right. \nonumber\\
   & \left.+8424 \sin \left(34
   \phi _0\right)-221 \sin \left(36 \phi _0\right)\right), \\
   W_{10}=&\frac{I^{11}}{4155203974946881536 k^{15}} \bigg(-9186469977056 \sin \left(2 \phi _0\right)+5311969531237 \sin \left(4 \phi _0\right)-3586272994472 \sin \left(6 \phi _0\right)\bigg. \nonumber\\
   & \left.+\frac{9528014994589}{4} \sin \left(8 \phi _0\right)-\frac{7529224160272}{5} \sin \left(10
   \phi _0\right)+884878319954 \sin \left(12 \phi _0\right)-476281378048 \sin \left(14 \phi _0\right)\right. \nonumber\\
   & \left.+\frac{464168079361}{2} \sin \left(16 \phi _0\right)-101347955904 \sin \left(18 \phi _0\right)+\frac{196368227671}{5} \sin \left(20
   \phi _0\right)-13367025784 \sin \left(22 \phi _0\right)\right. \nonumber\\
   & \left.+\frac{63203940789}{16} \sin \left(24 \phi _0\right)-999988518 \sin \left(26 \phi _0\right)+\frac{853040643}{4} \sin \left(28 \phi _0\right)-\frac{187506696}{5} \sin \left(30
   \phi _0\right)\right. \nonumber\\
   & \left.+\frac{42255655}{8} \sin \left(32 \phi _0\right)-571480 \sin \left(34 \phi _0\right)+\frac{177905}{4} \sin \left(36 \phi _0\right)-2210 \sin \left(38 \phi _0\right)+\frac{4199}{80} \sin \left(40 \phi
   _0\right)\right).
\end{align}
\end{subequations}

Classical functions ${\cal O}_{1}(\phi_{0},I;x)$ and ${\cal O}_{2}(\phi_{0},I;x)$ up to the fourth order in $\lambda$ (the complete list can be obtained under request):
\begin{align}\label{O1kpos}
     {\cal O}_{1}=&\frac{I }{k^{1/2}}\sin ^2(\phi_0)-\frac{I^2 \lambda}{192 k^2}(-16 \cos (2 \phi_0)+4 \cos (4 \phi_0)) \sin ^2(\phi_0)+\frac{I^3 \lambda^2 }{55296 k^{7/2}}(-768 \cos (2 \phi_0)+528 \cos (4 \phi_0)-192 \cos (6 \phi_0)\nonumber\\
   &+24 \cos (8
   \phi_0)) \sin ^2(\phi_0)+\frac{I^4 \lambda ^3 }{5308416 k^5}(18528 \cos (2 \phi_0)-16404 \cos (4 \phi_0)+9744 \cos (6 \phi_0)-3528 \cos (8 \phi_0)\nonumber\\
   &+720 \cos (10 \phi_0)-60 \cos (12 \phi_0)) \sin^2(\phi_0)+\frac{I^5 \lambda ^4 }{21233664 k^{13/2}}(-22816 \cos (2 \phi_0)+22576 \cos (4 \phi_0)-16608 \cos (6\phi_0)\nonumber\\
   &+8684 \cos (8 \phi_0)-3200 \cos (10 \phi_0)+780 \cos (12 \phi_0)-112 \cos (14 \phi_0)+7
   \cos (16 \phi_0)) \sin ^2(\phi_0),
\end{align}
\begin{align}\label{O2kpos}
    {\cal O}_{2}=&\frac{I^2}{6k}\sin ^4(\phi_0)-\frac{I^3 \lambda}{576 k^{5/2}}(-16 \cos (2 \phi_0)+4 \cos (4 \phi_0)) \sin ^4(\phi_0)+\frac{I^4 \lambda ^2 }{13824 k^4}(-64 \cos (2 \phi_0)+16 \cos ^2(2 \phi_0)+44 \cos (4 \phi_0)\nonumber\\
  &-8 \cos(2 \phi_0) \cos (4 \phi_0)+\cos ^2(4 \phi_0)-16 \cos (6 \phi_0)+2 \cos (8 \phi_0)) \sin^4(\phi_0)+\frac{I^5 \lambda ^3}{k^{11/2}}\bigg(-\frac{1}{31850496}(-16 \cos (2 \phi_0)\nonumber\\
   &+4 \cos (4 \phi_0)) (-768 \cos (2\phi_0)+528 \cos (4 \phi_0)-192 \cos (6 \phi_0)+24 \cos (8 \phi_0)) \sin ^4(\phi_0)+\frac{1}{15925248}(18528 \cos (2 \phi_0)\nonumber\\
   &-16404 \cos (4 \phi_0)+9744 \cos (6 \phi_0)-3528 \cos (8 \phi_0
   )+720 \cos (10 \phi_0)-60 \cos (12 \phi_0)) \sin ^4(\phi_0)\bigg)\nonumber\\
   &+\frac{I^6 \lambda ^4}{k^7} \bigg(\frac{1}{18345885696}(-768 \cos (2 \phi_0)+528 \cos (4 \phi_0)-192 \cos (6 \phi_0)+24 \cos (8 \phi_0))^2 \sin^4(\phi_0)\nonumber\\
   &-\frac{1}{3057647616}(-16 \cos (2 \phi_0)+4 \cos (4 \phi_0)) (18528 \cos (2 \phi_0)-16404 \cos(4 \phi_0)+9744 \cos (6 \phi_0)\nonumber\\
   &-3528 \cos (8 \phi_0)+720 \cos (10 \phi_0)-60 \cos (12 \phi_0))\sin ^4(\phi_0)+\frac{1}{63700992}(-22816 \cos (2 \phi_0)+22576 \cos (4 \phi_0)\nonumber\\
   &-16608 \cos (6 \phi_0)+8684 \cos (8 \phi_0)-3200 \cos (10 \phi_0)+780 \cos (12 \phi_0)-112 \cos (14 \phi_0)+7 \cos (16 \phi_0)) \sin ^4(\phi_0)\bigg).
\end{align}

Coefficients $\beta^{(n')}_{1}(I;x)$ and $\beta^{(n')}_{2}(I;x)$ for $n'=0,\pm1,\dots,\pm10$:
\begin{subequations}\label{beta1kpos}
    \begin{align}
    \beta_1^{(0)}=& \frac{I}{2 \sqrt{k}}-\frac{I^2 \lambda }{16 k^2}+\frac{85 I^3 \lambda ^2}{4608 k^{7/2}}-\frac{125 I^4 \lambda ^3}{18432 k^5}+\frac{39193 I^5 \lambda ^4}{14155776 k^{13/2}}-\frac{204281 I^6 \lambda ^5}{169869312 k^8}+\frac{17750261 I^7
   \lambda ^6}{32614907904 k^{19/2}}\nonumber\\
   &-\frac{132347765 I^8 \lambda ^7}{521838526464 k^{11}}+\frac{435757124275 I^9 \lambda ^8}{3606947894919168 k^{25/2}}-\frac{281528168779 I^{10} \lambda ^9}{4809263859892224
   k^{14}}+\frac{79655597681647 I^{11} \lambda ^{10}}{2770135983297921024 k^{31/2}},\\
    \beta_1^{(\pm2)}=& -\frac{I}{4 \sqrt{k}}+\frac{5 I^2 \lambda }{192 k^2}-\frac{83 I^3 \lambda ^2}{12288 k^{7/2}}+\frac{3965 I^4 \lambda ^3}{1769472 k^5}-\frac{94969 I^5 \lambda ^4}{113246208 k^{13/2}}+\frac{5502505 I^6 \lambda ^5}{16307453952
   k^8}-\frac{446236787 I^7 \lambda ^6}{3131031158784 k^{19/2}}\nonumber\\
   &+\frac{1040920667 I^8 \lambda ^7}{16698832846848 k^{11}}-\frac{403835109703 I^9 \lambda ^8}{14427791579676672 k^{25/2}}+\frac{2221131730877 I^{10} \lambda
   ^9}{173133498956120064 k^{14}}-\frac{198708866965765 I^{11} \lambda ^{10}}{33241631799575052288 k^{31/2}},\\
    \beta_1^{(\pm4)}=& \frac{I^2 \lambda }{192 k^2}-\frac{11 I^3 \lambda ^2}{4608 k^{7/2}}+\frac{479 I^4 \lambda ^3}{442368 k^5}-\frac{3557 I^5 \lambda ^4}{7077888 k^{13/2}}+\frac{3888053 I^6 \lambda ^5}{16307453952 k^8}-\frac{15026351 I^7 \lambda
   ^6}{130459631616 k^{19/2}}\nonumber\\
   &+\frac{2121579677 I^8 \lambda ^7}{37572373905408 k^{11}}-\frac{50528138287 I^9 \lambda ^8}{1803473947459584 k^{25/2}}+\frac{19446977965327 I^{10} \lambda ^9}{1385067991648960512
   k^{14}}-\frac{235859488654553 I^{11} \lambda ^{10}}{33241631799575052288 k^{31/2}},\\
    \beta_1^{(\pm6)}=& -\frac{I^3 \lambda ^2}{12288 k^{7/2}}+\frac{13 I^4 \lambda ^3}{196608 k^5}-\frac{395 I^5 \lambda ^4}{9437184 k^{13/2}}+\frac{66187 I^6 \lambda ^5}{2717908992 k^8}-\frac{791329 I^7 \lambda ^6}{57982058496 k^{19/2}}+\frac{62666879 I^8
   \lambda ^7}{8349416423424 k^{11}}\nonumber\\
   &-\frac{4366456033 I^9 \lambda ^8}{1068725302198272 k^{25/2}}+\frac{113442334381 I^{10} \lambda ^9}{51298814505517056 k^{14}}-\frac{11751695020477 I^{11} \lambda ^{10}}{9849372385059274752 k^{31/2}},\\
    \beta_1^{(\pm8)}=&\frac{I^4 \lambda ^3}{884736 k^5}-\frac{7 I^5 \lambda ^4}{5308416 k^{13/2}}+\frac{545 I^6 \lambda ^5}{509607936 k^8}-\frac{36581 I^7 \lambda ^6}{48922361856 k^{19/2}}+\frac{28045 I^8 \lambda ^7}{57982058496 k^{11}}-\frac{8418031 I^9
   \lambda ^8}{28179280429056 k^{25/2}}\nonumber\\
   &+\frac{3875661979 I^{10} \lambda ^9}{21641687369515008 k^{14}}-\frac{218564097493 I^{11} \lambda ^{10}}{2077601987473440768 k^{31/2}},\\
    \beta_1^{(\pm10)}=&-\frac{5 I^5 \lambda ^4}{339738624 k^{13/2}}+\frac{365 I^6 \lambda ^5}{16307453952 k^8}-\frac{2165 I^7 \lambda ^6}{97844723712 k^{19/2}}+\frac{1357895 I^8 \lambda ^7}{75144747810816 k^{11}}-\frac{63775105 I^9 \lambda
   ^8}{4809263859892224 k^{25/2}}\nonumber\\
   &+\frac{6300898345 I^{10} \lambda ^9}{692533995824480256 k^{14}}-\frac{99135644645 I^{11} \lambda ^{10}}{16620815899787526144 k^{31/2}},\\
    \beta_1^{(\pm12)}=&\frac{I^6 \lambda ^5}{5435817984 k^8}-\frac{5 I^7 \lambda ^6}{14495514624 k^{19/2}}+\frac{1679 I^8 \lambda ^7}{4174708211712 k^{11}}-\frac{225889 I^9 \lambda ^8}{601157982486528 k^{25/2}}+\frac{11872213 I^{10} \lambda
   ^9}{38474110879137792 k^{14}}\nonumber\\
   &-\frac{107583905 I^{11} \lambda ^{10}}{461689330549653504 k^{31/2}},\\
    \beta_1^{(\pm14)}=&-\frac{7 I^7 \lambda ^6}{3131031158784 k^{19/2}}+\frac{749 I^8 \lambda ^7}{150289495621632 k^{11}}-\frac{48293 I^9 \lambda ^8}{7213895789838336 k^{25/2}} \nonumber\\
   & +\frac{1625743 I^{10} \lambda ^9}{230844665274826752 k^{14}}-\frac{851071067
   I^{11} \lambda ^{10}}{132966527198300209152 k^{31/2}}.
    \end{align}
\end{subequations}

\begin{subequations}\label{beta2kpos}
    \begin{align}
   \beta_2^{(0)}=& \frac{I^2}{16 k}-\frac{17 I^3 \lambda }{1152 k^{5/2}}+\frac{125 I^4 \lambda ^2}{24576 k^4}-\frac{3563 I^5 \lambda ^3}{1769472 k^{11/2}}+\frac{145915 I^6 \lambda ^4}{169869312 k^7}-\frac{1044133 I^7 \lambda ^5}{2717908992
   k^{17/2}}+\frac{185286871 I^8 \lambda ^6}{1043677052928 k^{10}}\nonumber\\
   &-\frac{18945961925 I^9 \lambda ^7}{225434243432448 k^{23/2}}+\frac{21656012983 I^{10} \lambda ^8}{534362651099136 k^{13}}-\frac{13733723738215 I^{11} \lambda
   ^9}{692533995824480256 k^{29/2}}+\frac{40804547578795 I^{12} \lambda ^{10}}{4155203974946881536 k^{16}},\\
   \beta_2^{(\pm2)}=& -\frac{I^2}{24 k}+\frac{7 I^3 \lambda }{768 k^{5/2}}-\frac{655 I^4 \lambda ^2}{221184 k^4}+\frac{11855 I^5 \lambda ^3}{10616832 k^{11/2}}-\frac{310421 I^6 \lambda ^4}{679477248 k^7}+\frac{9634829 I^7 \lambda ^5}{48922361856
   k^{17/2}}-\frac{1654242575 I^8 \lambda ^6}{18786186952704 k^{10}}\nonumber\\
   &+\frac{72953410571 I^9 \lambda ^7}{1803473947459584 k^{23/2}}-\frac{273767449847 I^{10} \lambda ^8}{14427791579676672 k^{13}}+\frac{25069842323593 I^{11} \lambda
   ^9}{2770135983297921024 k^{29/2}}\nonumber\\
   &-\frac{872657478596255 I^{12} \lambda ^{10}}{199449790797450313728 k^{16}}, \\
    \beta_2^{(\pm4)}=& \frac{I^2}{96 k}-\frac{I^3 \lambda }{768 k^{5/2}}+\frac{13 I^4 \lambda ^2}{73728 k^4}+\frac{113 I^5 \lambda ^3}{10616832 k^{11/2}}-\frac{257515 I^6 \lambda ^4}{8153726976 k^7}+\frac{4700251 I^7 \lambda ^5}{195689447424
   k^{17/2}}-\frac{10397377 I^8 \lambda ^6}{695784701952 k^{10}}\nonumber\\
   &+\frac{7786939537 I^9 \lambda ^7}{901736973729792 k^{23/2}}-\frac{3338632139569 I^{10} \lambda ^8}{692533995824480256 k^{13}}+\frac{43927833582367 I^{11} \lambda
   ^9}{16620815899787526144 k^{29/2}}\nonumber\\
   &-\frac{2288575869972055 I^{12} \lambda ^{10}}{1595598326379602509824 k^{16}}, \\
    \beta_2^{(\pm6)}=& -\frac{I^3 \lambda }{2304 k^{5/2}}+\frac{17 I^4 \lambda ^2}{73728 k^4}-\frac{779 I^5 \lambda ^3}{7077888 k^{11/2}}+\frac{52805 I^6 \lambda ^4}{1019215872 k^7}-\frac{532487 I^7 \lambda ^5}{21743271936 k^{17/2}}+\frac{18274109 I^8
   \lambda ^6}{1565515579392 k^{10}}\nonumber\\
   &-\frac{1124861431 I^9 \lambda ^7}{200385994162176 k^{23/2}}+\frac{52366851143 I^{10} \lambda ^8}{19237055439568896 k^{13}}-\frac{2456402477509 I^{11} \lambda ^9}{1846757322198614016
   k^{29/2}}\nonumber\\
   &+\frac{1044030329402455 I^{12} \lambda ^{10}}{1595598326379602509824 k^{16}}, \\
    \beta_2^{(\pm8)}=& \frac{5 I^4 \lambda ^2}{442368 k^4}-\frac{I^5 \lambda ^3}{98304 k^{11/2}}+\frac{1741 I^6 \lambda ^4}{254803968 k^7}-\frac{100921 I^7 \lambda ^5}{24461180928 k^{17/2}}+\frac{1854211 I^8 \lambda ^6}{782757789696 k^{10}}-\frac{18650465
   I^9 \lambda ^7}{14089640214528 k^{23/2}}\nonumber\\
   &+\frac{291825893 I^{10} \lambda ^8}{400771988324352 k^{13}}-\frac{412152359593 I^{11} \lambda ^9}{1038800993736720384 k^{29/2}}+\frac{76200897469177 I^{12} \lambda
   ^{10}}{354577405862133891072 k^{16}}, \\
    \beta_2^{(\pm10)}=& -\frac{5 I^5 \lambda ^3}{21233664 k^{11/2}}+\frac{605 I^6 \lambda ^4}{2038431744 k^7}-\frac{49655 I^7 \lambda ^5}{195689447424 k^{17/2}}+\frac{1730915 I^8 \lambda ^6}{9393093476352 k^{10}}-\frac{221147005 I^9 \lambda
   ^7}{1803473947459584 k^{23/2}}\nonumber\\
   &+\frac{1116576365 I^{10} \lambda ^8}{14427791579676672 k^{13}}-\frac{130702718095 I^{11} \lambda ^9}{2770135983297921024 k^{29/2}}+\frac{207488895935 I^{12} \lambda ^{10}}{7387029288794456064 k^{16}}, \\
    \beta_2^{(\pm12)}=& \frac{35 I^6 \lambda ^4}{8153726976 k^7}-\frac{151 I^7 \lambda ^5}{21743271936 k^{17/2}}+\frac{45013 I^8 \lambda ^6}{6262062317568 k^{10}}-\frac{5484323 I^9 \lambda ^7}{901736973729792 k^{23/2}}+\frac{264896183 I^{10} \lambda
   ^8}{57711166318706688 k^{13}}\nonumber\\
   &-\frac{2230866583 I^{11} \lambda ^9}{692533995824480256 k^{29/2}}+\frac{429090687803 I^{12} \lambda ^{10}}{199449790797450313728 k^{16}}, \\
    \beta_2^{(\pm14)}=& -\frac{7 I^7 \lambda ^5}{97844723712 k^{17/2}}+\frac{2653 I^8 \lambda ^6}{18786186952704 k^{10}}-\frac{309169 I^9 \lambda ^7}{1803473947459584 k^{23/2}}+\frac{1591303 I^{10} \lambda ^8}{9618527719784448 k^{13}}\nonumber\\
   &-\frac{2319042901 I^{11}
   \lambda ^9}{16620815899787526144 k^{29/2}}+\frac{3581080741 I^{12} \lambda ^{10}}{33241631799575052288 k^{16}}.
 \end{align}
\end{subequations}

\subsection{Case $k<0$}
Functions $W_{\mu}$ for $\mu=1,\dots,10$:
\begin{subequations}\label{Wnfunctkneg}
\begin{align}
W_1=& \frac{I^{3/2} }{12 \sqrt[4]{2} \sqrt{3}
   (-k)^{3/4}} \left(-9 \cos \left(\phi _0\right)+\cos \left(3 \phi _0\right)\right),\\
W_2=& -\frac{I^2}{384
   \sqrt{2} (-k)^{3/2}} \left(37 \sin \left(2 \phi _0\right)-8 \sin \left(4 \phi _0\right)+\sin \left(6 \phi _0\right)\right),\\
W_3=& -\frac{I^{5/2}}{18432\ 2^{3/4} \sqrt{3} (-k)^{9/4}} \left(2406 \cos \left(\phi _0\right)-888 \cos \left(3 \phi _0\right)+336 \cos \left(5 \phi_0\right)-69 \cos \left(7 \phi _0\right)+7 \cos \left(9 \phi _0\right)\right),\\
W_4=& -\frac{I^3}{442368 k^3} \left(-6504 \sin \left(2 \phi _0\right)+2841 \sin \left(4 \phi _0\right)-1188 \sin \left(6 \phi_0\right)+348 \sin \left(8 \phi _0\right)-60 \sin \left(10 \phi _0\right)+5 \sin \left(12 \phi_0\right)\right),\\
W_5=& \frac{I^{7/2}}{188743680 \sqrt[4]{2}
   \sqrt{3} (-k)^{15/4}} \left(-4906185 \cos \left(\phi _0\right)+2061605 \cos \left(3 \phi _0\right)-1134741 \cos \left(5 \phi
   _0\right)+505545 \cos \left(7 \phi _0\right)\right.\nonumber\\
   &\left.-175085 \cos \left(9 \phi _0\right)+41865 \cos \left(11 \phi
   _0\right)-6105 \cos \left(13 \phi _0\right)+429 \cos \left(15 \phi _0\right)\right),\\
W_6=& -\frac{I^4}{84934656 \sqrt{2} (-k)^{9/2}} \left(601389 \sin \left(2 \phi _0\right)-305226 \sin \left(4 \phi _0\right)+170369 \sin \left(6 \phi
   _0\right)-81906 \sin \left(8 \phi _0\right)\right.\nonumber\\
   &\left.+31413 \sin \left(10 \phi _0\right)-9106 \sin \left(12 \phi
   _0\right)+1839 \sin \left(14 \phi _0\right)-231 \sin \left(16 \phi _0\right)+14 \sin \left(18 \phi
   _0\right)\right), \\
W_7=& -\frac{I^{9/2}}{1217623228416\ 2^{3/4} \sqrt{3} (-k)^{21/4}} \left(17403059940 \cos \left(\phi _0\right)-7561686566 \cos \left(3 \phi _0\right)+4693975146 \cos
   \left(5 \phi _0\right)\right.\nonumber\\
   &\left.-2657276376 \cos \left(7 \phi _0\right)+1330840168 \cos \left(9 \phi _0\right)-552365079
   \cos \left(11 \phi _0\right)+181745025 \cos \left(13 \phi _0\right)\right.\nonumber\\
   &\left.-45127082 \cos \left(15 \phi _0\right)+7885878
   \cos \left(17 \phi _0\right)-867867 \cos \left(19 \phi _0\right)+46189 \cos \left(21 \phi
   _0\right)\right),\\
W_8=& \frac{I^5}{21743271936 k^6} \left(-46426448 \sin \left(2 \phi _0\right)+24902680 \sin \left(4 \phi _0\right)-15580848 \sin \left(6 \phi
   _0\right)+9103525 \sin \left(8 \phi _0\right)\right.\nonumber\\
   &\left.-4689512 \sin \left(10 \phi _0\right)+2060452 \sin \left(12 \phi
   _0\right)-744152 \sin \left(14 \phi _0\right)+213052 \sin \left(16 \phi _0\right)-46216 \sin \left(18 \phi
   _0\right)\right.\nonumber\\
   &\left.+7116 \sin \left(20 \phi _0\right)-696 \sin \left(22 \phi _0\right)+33 \sin \left(24 \phi
   _0\right)\right),\\
W_9=& \frac{I^{11/2}}{6412351813189632 \sqrt[4]{2} \sqrt{3}
   (-k)^{27/4}} \left(-29825420608692 \cos \left(\phi _0\right)+13119866088564 \cos \left(3 \phi
   _0\right)\right.\nonumber\\
   &\left.-8590881168369 \cos \left(5 \phi _0\right)+5419669290273 \cos \left(7 \phi _0\right)-3209105952051 \cos
   \left(9 \phi _0\right)+1696298165091 \cos \left(11 \phi _0\right)\right.\nonumber\\
   &\left.-777910346658 \cos \left(13 \phi
   _0\right)+301429404354 \cos \left(15 \phi _0\right)-95900963886 \cos \left(17 \phi _0\right)+24247831374 \cos
   \left(19 \phi _0\right)\right.\nonumber\\
   &\left.-4668334827 \cos \left(21 \phi _0\right)+642333627 \cos \left(23 \phi _0\right)-56497545
   \cos \left(25 \phi _0\right)+2414425 \cos \left(27 \phi _0\right)\right),\\
W_{10}=& -\frac{I^6}{31310311587840 \sqrt{2} (-k)^{15/2}} \left(45945339825 \sin \left(2 \phi _0\right)-25307205900 \sin \left(4 \phi _0\right)+16767524845 \sin
   \left(6 \phi _0\right)\right.\nonumber\\
   &\left.-10811740260 \sin \left(8 \phi _0\right)+6456248907 \sin \left(10 \phi _0\right)-3480781220
   \sin \left(12 \phi _0\right)+1652572455 \sin \left(14 \phi _0\right)\right.\nonumber\\
   &\left.-676366320 \sin \left(16 \phi
   _0\right)+233384255 \sin \left(18 \phi _0\right)-66206628 \sin \left(20 \phi _0\right)+14975235 \sin \left(22 \phi
   _0\right)-2591380 \sin \left(24 \phi _0\right)\right.\nonumber\\
   &\left.+322245 \sin \left(26 \phi _0\right)-25740 \sin \left(28 \phi
   _0\right)+1001 \sin \left(30 \phi _0\right)\right).
\end{align}
\end{subequations}

Classical functions ${\cal O}_{1}(\phi_{0},I;x)$ and ${\cal O}_{2}(\phi_{0},I;x)$ up to the first order in $\lambda'=\sqrt{\lambda}$ (the complete list can be obtained under request):
\begin{align}\label{O1kneg}
     {\cal O}_{1}=&-\frac{3k}{\left(\lambda'\right)^2}-\frac{2^{3/4} \sqrt{3} \sqrt[4]{-k} \sqrt{I}}{\lambda'}\sin (\phi_0)+\frac{I}{4 \sqrt{2}
   \sqrt{-k}}\left(\sin ^2(\phi_0)+\sin (\phi_0) \sin (3 \phi_0)\right)+\frac{I^{3/2}\lambda'}{64\ 2^{3/4} \sqrt{3} (-k)^{5/4} } \bigg(37 \cos (2 \phi_0) \sin (\phi_0)\nonumber\\
   &-16 \cos (4 \phi_0) \sin
   (\phi_0)+3\cos (6 \phi_0) \sin (\phi_0)+57 \sin ^3(\phi_0)-22 \sin
   ^2(\phi_0) \sin (3 \phi_0)+\sin (\phi_0) \sin ^2(3 \phi_0)\bigg), 
\end{align}

\begin{align}\label{O2kneg}
     {\cal O}_{2}=&\frac{3k^2}{2 \left(\lambda '\right)^4}-\frac{2^{3/4} \sqrt{3} (-k)^{5/4}\sqrt{I}}{\left(\lambda '\right)^3} \sin (\phi_0)+\frac{I \sqrt{-k}}{4 \sqrt{2} \left(\lambda'\right)^2}\left(9 \sin ^2(\phi_0)+\sin (\phi_0) \sin (3\phi_0)\right)+\frac{I^{3/2}}{64\ 2^{3/4} \sqrt{3} \sqrt[4]{-k} \lambda'} \bigg(37  \cos (2 \phi_0) \sin (\phi_0)\nonumber\\
   &-16  \cos (4 \phi_0) \sin (\phi_0)+3  \cos (6 \phi_0)
   \sin (\phi_0)+25  \sin ^3(\phi_0)-54  \sin ^2(\phi_0) \sin (3 \phi_0)+ \sin (\phi_0) \sin ^2(3 \phi_0)\bigg)\nonumber\\
   &+\frac{I^2}{12288(-k)}(-802  \sin ^2(\phi_0)-3996  \cos (2 \phi_0) \sin ^2(\phi_0)+1728  \cos (4 \phi_0) \sin ^2(\phi_0)-324  \cos (6 \phi_0) \sin ^2(\phi_0)\nonumber\\
   &-3692  \sin ^4(\phi_0)+888  \sin (\phi_0) \sin (3 \phi_0)+148  \cos (2 \phi_0) \sin (\phi_0) \sin (3 \phi_0)-64  \cos (4 \phi_0) \sin (\phi_0) \sin (3 \phi_0)\nonumber\\
   &+12  \cos (6 \phi_0) \sin (\phi_0) \sin (3 \phi_0)+1644  \sin^3(\phi_0) \sin (3 \phi_0)-36  \sin ^2(\phi_0) \sin ^2(3 \phi_0)+4  \sin
   (\phi_0) \sin ^3(3 \phi_0)\nonumber\\
   &-560  \sin (\phi_0) \sin (5 \phi_0)+161  \sin(\phi_0) \sin (7 \phi_0)-21  \sin (\phi_0) \sin (9 \phi_0)) +\frac{ I^{5/2} \lambda '}{196608 \sqrt{3} (-k)^{7/4}} \bigg(8672\  \cos (2 \phi_0) \sin (\phi_0)\nonumber\\
   &+1369\  \cos^2(2 \phi_0) \sin (\phi_0)-7576\  \cos (4 \phi_0) \sin (\phi_0)-1184\  \cos (2 \phi_0) \cos (4 \phi_0) \sin (\phi_0)+256\  \cos ^2(4 \phi_0) \sin (\phi_0)\nonumber\\
   &+4752\  \cos (6 \phi_0) \sin(\phi_0)+222\  \cos (2 \phi_0) \cos (6 \phi_0) \sin (\phi_0)-96\  \cos (4 \phi_0) \cos (6 \phi_0) \sin (\phi_0)+9\ \cos ^2(6 \phi_0) \sin (\phi_0)\nonumber\\
   &-1856\  \cos (8 \phi_0) \sin(\phi_0)+400\  \cos (10 \phi_0) \sin (\phi_0)-40\  \cos
   (12 \phi_0) \sin (\phi_0)+21654\  \sin ^3(\phi_0)+30414\ \cos (2 \phi_0) \sin ^3(\phi_0)\nonumber\\
   &-13152\  \cos (4 \phi_0) \sin
   ^3(\phi_0)+2466\  \cos (6 \phi_0) \sin ^3(\phi_0)+5781\  \sin ^5(\phi_0)-24778\  \sin ^2(\phi_0) \sin (3 \phi_0)\nonumber\\
   &-1332\  \cos (2 \phi_0) \sin ^2(\phi_0) \sin (3 \phi_0)+576\  \cos (4 \phi_0) \sin ^2(\phi_0) \sin (3 \phi_0)-108\  \cos (6 \phi_0)
   \sin ^2(\phi_0) \sin (3 \phi_0)\nonumber\\
   &-28\  \sin ^4(\phi_0) \sin (3 \phi_0)+888\  \sin (\phi_0) \sin ^2(3 \phi_0)+222\  \cos (2
   \phi_0) \sin (\phi_0) \sin ^2(3 \phi_0)-96\  \cos (4 \phi_0) \sin (\phi_0) \sin ^2(3 \phi_0)\nonumber\\
   &+18\  \cos (6 \phi_0) \sin (\phi_0) \sin ^2(3 \phi_0)-498\  \sin ^3(\phi_0) \sin ^2(3 \phi_0)-60\
    \sin ^2(\phi_0) \sin ^3(3 \phi_0)+5\  \sin (\phi_0) \sin^4(3 \phi_0)\nonumber\\
   &+15120\  \sin ^2(\phi_0) \sin (5 \phi_0)-560\ \sin (\phi_0) \sin (3 \phi_0) \sin (5 \phi_0)-4347\  \sin^2(\phi_0) \sin (7 \phi_0)+161\  \sin (\phi_0) \sin (3 \phi_0)\sin (7 \phi_0)\nonumber\\
   &+567\  \sin ^2(\phi_0) \sin (9 \phi_0)-21 \sin (\phi_0) \sin (3 \phi_0) \sin (9 \phi_0)\bigg). 
\end{align}

Coefficients $\beta^{(n')}_{1}(I;x)$ and $\beta^{(n')}_{2}(I;x)$ for $n'=0,1,\dots,10$:
\begin{subequations}\label{beta1kneg}
    \begin{align}
    \beta_1^{(0)}=& -\frac{0.707107 I}{\sqrt{-k}}-\frac{3 k}{\left(\lambda '\right)^2}-\frac{0.125 I^2 \left(\lambda '\right)^2}{k^2}-\frac{0.0521737 I^3 \left(\lambda '\right)^4}{(-k)^{7/2}}+\frac{0.0271267 I^4 \left(\lambda '\right)^6}{k^5}-\frac{0.0156621 I^5
   \left(\lambda '\right)^8}{(-k)^{13/2}}\nonumber\\
   &+\frac{0.00212907 I^6 \left(\lambda '\right)^{10}}{k^8}, \\
    \beta_1^{(1)}=& \frac{1.45648 {\rm i} \sqrt[4]{-k} \sqrt{I}}{\lambda '}-\frac{0.128735 {\rm i} I^{3/2} \lambda '}{(-k)^{5/4}}-\frac{0.0393514 {\rm i} I^{5/2} \left(\lambda '\right)^3}{(-k)^{11/4}}-\frac{0.0185784 {\rm i} I^{7/2} \left(\lambda
   '\right)^5}{(-k)^{17/4}}-\frac{0.0103102 {\rm i} I^{9/2} \left(\lambda '\right)^7}{(-k)^{23/4}}\nonumber\\
   &-\frac{0.00622176 {\rm i} I^{11/2} \left(\lambda '\right)^9}{(-k)^{29/4}},\\
    \beta_1^{(2)}=& -\frac{0.353553 I}{\sqrt{-k}}-\frac{0.0520833 I^2 \left(\lambda '\right)^2}{k^2}-\frac{0.0191048 I^3 \left(\lambda '\right)^4}{(-k)^{7/2}}+\frac{0.00896313 I^4 \left(\lambda '\right)^6}{k^5}-\frac{0.00474387 I^5 \left(\lambda '\right)^8}{(-1.
   k)^{13/2}}\nonumber\\
   &-\frac{0.000633553 I^6 \left(\lambda '\right)^{10}}{k^8}\\
    \beta_1^{(3)}=& -\frac{0.0643677 {\rm i} I^{3/2} \lambda '}{(-k)^{5/4}}-\frac{0.0256021 {\rm i} I^{5/2} \left(\lambda '\right)^3}{(-k)^{11/4}}-\frac{0.0130956 {\rm i} I^{7/2} \left(\lambda '\right)^5}{(-k)^{17/4}}-\frac{0.00747685 {\rm i} I^{9/2}
   \left(\lambda '\right)^7}{(-k)^{23/4}}\nonumber\\
   &-\frac{0.00454997 {\rm i} I^{11/2} \left(\lambda '\right)^9}{(-k)^{29/4}},\\
    \beta_1^{(4)}=& \frac{0.0104167 I^2 \left(\lambda '\right)^2}{k^2}+\frac{0.00675189 I^3 \left(\lambda '\right)^4}{(-k)^{7/2}}-\frac{0.00433124 I^4 \left(\lambda '\right)^6}{k^5}+\frac{0.00284286 I^5 \left(\lambda '\right)^8}{(-k)^{13/2}}+\frac{0.00230256 I^6
   \left(\lambda '\right)^{10}}{k^8}, \\
    \beta_1^{(5)}=& \frac{0.00158038 {\rm i} I^{5/2} \left(\lambda '\right)^3}{(-k)^{11/4}}+\frac{0.00142015 {\rm i} I^{7/2} \left(\lambda '\right)^5}{(-k)^{17/4}}+\frac{0.00109183 {\rm i} I^{9/2} \left(\lambda '\right)^7}{(-k)^{23/4}}+\frac{0.000807709 {\rm i} I^{11/2} \left(\lambda '\right)^9}{(-k)^{29/4}},\\
    \beta_1^{(6)}=& -\frac{0.000230178 I^3 \left(\lambda '\right)^4}{(-k)^{7/2}}+\frac{0.000264486 I^4 \left(\lambda '\right)^6}{k^5}-\frac{0.000236772 I^5 \left(\lambda '\right)^8}{(-k)^{13/2}}-\frac{0.000868488 I^6 \left(\lambda '\right)^{10}}{k^8} \\
    \beta_1^{(7)}=& -\frac{0.0000325936 {\rm i} I^{7/2} \left(\lambda '\right)^5}{(-k)^{17/4}}-\frac{0.0000456142 {\rm i} I^{9/2} \left(\lambda '\right)^7}{(-k)^{23/4}}-\frac{0.000046581 {\rm i} I^{11/2} \left(\lambda '\right)^9}{(-k)^{29/4}},\\
    \beta_1^{(8)}=& -\frac{4.521122685185185\times 10^{-6} I^4 \left(\lambda '\right)^6}{k^5}+\frac{7.459471854965145\times 10^{-6} I^5 \left(\lambda '\right)^8}{(-k)^{13/2}}+\frac{0.000701564 I^6 \left(\lambda
   '\right)^{10}}{k^8}, \\
    \beta_1^{(9)}=& \frac{6.173347113384545\times 10^{-7} {\rm i} I^{9/2} \left(\lambda '\right)^7}{(-k)^{23/4}}+\frac{1.1731516942448892\times 10^{-6} {\rm i} I^{11/2} \left(\lambda '\right)^9}{(-k)^{29/4}},\\
    \beta_{1}^{(10)}=& -\frac{8.325303409559314\times 10^{-8} I^5 \left(\lambda '\right)^8}{(-k)^{13/2}}-\frac{0.000568826 I^6 \left(\lambda '\right)^{10}}{k^8} , 
\end{align}
\end{subequations}

\begin{subequations}\label{beta2kneg}
    \begin{align}
    \beta_2^{(0)}=&\frac{0.125 I^2}{k}+\frac{1.5 k^2}{\left(\lambda '\right)^4}-\frac{0.0417389 I^3 \left(\lambda '\right)^2}{(-k)^{5/2}}-\frac{0.0203451 I^4 \left(\lambda '\right)^4}{k^4}-\frac{0.0113906 I^5 \left(\lambda '\right)^6}{(-k)^{11/2}}-\frac{0.00487782
   I^6 \left(\lambda '\right)^8}{k^7}\nonumber\\
   &-\frac{0.000355413 I^7 \left(\lambda '\right)^{10}}{(-k)^{17/2}}, \\
    \beta_2^{(1)}=&\frac{1.45648 {\rm i} (-k)^{5/4} \sqrt{I}}{\left(\lambda '\right)^3}-\frac{0.300383 {\rm i} I^{3/2}}{\sqrt[4]{-k} \lambda '}-\frac{0.0519944 {\rm i} I^{5/2} \lambda '}{(-k)^{7/4}}-\frac{0.0225082 {\rm i} I^{7/2} \left(\lambda
   '\right)^3}{(-k)^{13/4}}-\frac{0.0119265 {\rm i} I^{9/2} \left(\lambda '\right)^5}{(-k)^{19/4}}\nonumber\\
   &-\frac{0.00697415 {\rm i} I^{11/2} \left(\lambda '\right)^7}{(-k)^{25/4}}+\frac{0.000365442 {\rm i} I^{13/2} \left(\lambda
   '\right)^9}{(-k)^{31/4}}, \\
    \beta_2^{(2)}=&-\frac{0.0625 I^2}{k}-\frac{0.707107 \sqrt{-k} I}{\left(\lambda '\right)^2}+\frac{0.0133503 I^3 \left(\lambda '\right)^2}{(-k)^{5/2}}+\frac{0.00576443 I^4 \left(\lambda '\right)^4}{k^4}+\frac{0.0031454 I^5 \left(\lambda '\right)^6}{(-1.
   k)^{11/2}}\nonumber\\
   &-\frac{0.00397671 I^6 \left(\lambda '\right)^8}{k^7}+\frac{0.00217812 I^7 \left(\lambda '\right)^{10}}{(-k)^{17/2}}, \\
    \beta_2^{(3)}=&-\frac{0.236015 {\rm i} I^{3/2}}{\sqrt[4]{-k} \lambda '}-\frac{0.0256021 {\rm i} I^{5/2} \lambda '}{(-k)^{7/4}}-\frac{0.00979904 {\rm i} I^{7/2} \left(\lambda '\right)^3}{(-k)^{13/4}}-\frac{0.00473918 {\rm i} I^{9/2} \left(\lambda
   '\right)^5}{(-k)^{19/4}}\nonumber\\
   &-\frac{0.0025574 {\rm i} I^{11/2} \left(\lambda '\right)^7}{(-k)^{25/4}}+\frac{0.000746187 {\rm i} I^{13/2} \left(\lambda '\right)^9}{(-k)^{31/4}}, \\
    \beta_2^{(4)}=&-\frac{0.0625 I^2}{k}+\frac{0.0208695 I^3 \left(\lambda '\right)^2}{(-k)^{5/2}}+\frac{0.0100731 I^4 \left(\lambda '\right)^4}{k^4}+\frac{0.0055653 I^5 \left(\lambda '\right)^6}{(-k)^{11/2}}-\frac{0.00370461 I^6 \left(\lambda
   '\right)^8}{k^7}\nonumber\\
   &+\frac{0.000739686 I^7 \left(\lambda '\right)^{10}}{(-k)^{17/2}}, \\
    \beta_2^{(5)}=&\frac{0.0142234 {\rm i} I^{5/2} \lambda '}{(-k)^{7/4}}+\frac{0.00812512 {\rm i} I^{7/2} \left(\lambda '\right)^3}{(-k)^{13/4}}+\frac{0.00488775 {\rm i} I^{9/2} \left(\lambda '\right)^5}{(-k)^{19/4}}+\frac{0.00307544 {\rm i} I^{11/2}
   \left(\lambda '\right)^7}{(-k)^{25/4}}\nonumber\\
   &+\frac{0.00190357 {\rm i} I^{13/2} \left(\lambda '\right)^9}{(-k)^{31/4}}, \\
    \beta_2^{(6)}=&-\frac{0.00291559 I^3 \left(\lambda '\right)^2}{(-k)^{5/2}}-\frac{0.00237359 I^4 \left(\lambda '\right)^4}{k^4}-\frac{0.00171874 I^5 \left(\lambda '\right)^6}{(-k)^{11/2}}+\frac{0.00189226 I^6 \left(\lambda '\right)^8}{k^7}-\frac{0.000236153 I^7
   \left(\lambda '\right)^{10}}{(-k)^{17/2}}, \\
    \beta_2^{(7)}=&-\frac{0.000554091 {\rm i} I^{7/2} \left(\lambda '\right)^3}{(-k)^{13/4}}-\frac{0.000587222 {\rm i} I^{9/2} \left(\lambda '\right)^5}{(-k)^{19/4}}-\frac{0.000497969 {\rm i} I^{11/2} \left(\lambda '\right)^7}{(-k)^{25/4}}-\frac{0.000701971 {\rm i} I^{13/2} \left(\lambda '\right)^9}{(-k)^{31/4}}, \\
    \beta_2^{(8)}=&\frac{0.0000994647 I^4 \left(\lambda '\right)^4}{k^4}+\frac{0.000130008 I^5 \left(\lambda '\right)^6}{(-k)^{11/2}}-\frac{0.000819443 I^6 \left(\lambda '\right)^8}{k^7}-\frac{0.000590389 I^7 \left(\lambda '\right)^{10}}{(-k)^{17/2}}, \\
    \beta_2^{(9)}=&\frac{0.0000170796 {\rm i} I^{9/2} \left(\lambda '\right)^5}{(-k)^{19/4}}+\frac{0.0000265642 {\rm i} I^{11/2} \left(\lambda '\right)^7}{(-k)^{25/4}}+\frac{0.000695185 {\rm i} I^{13/2} \left(\lambda '\right)^9}{(-k)^{31/4}}, \\
    \beta_2^{(10)}=&-\frac{2.8306031592501667\times 10^{-6} I^5 \left(\lambda '\right)^6}{(-k)^{11/2}}+\frac{0.000573754 I^6 \left(\lambda '\right)^8}{k^7}+\frac{0.000484516 I^7 \left(\lambda '\right)^{10}}{(-k)^{17/2}}.
\end{align}
\end{subequations}

\end{widetext}

\bibliography{References}

\begin{thebibliography}{39}%
\makeatletter
\providecommand \@ifxundefined [1]{%
 \@ifx{#1\undefined}
}%
\providecommand \@ifnum [1]{%
 \ifnum #1\expandafter \@firstoftwo
 \else \expandafter \@secondoftwo
 \fi
}%
\providecommand \@ifx [1]{%
 \ifx #1\expandafter \@firstoftwo
 \else \expandafter \@secondoftwo
 \fi
}%
\providecommand \natexlab [1]{#1}%
\providecommand \enquote  [1]{``#1''}%
\providecommand \bibnamefont  [1]{#1}%
\providecommand \bibfnamefont [1]{#1}%
\providecommand \citenamefont [1]{#1}%
\providecommand \href@noop [0]{\@secondoftwo}%
\providecommand \href [0]{\begingroup \@sanitize@url \@href}%
\providecommand \@href[1]{\@@startlink{#1}\@@href}%
\providecommand \@@href[1]{\endgroup#1\@@endlink}%
\providecommand \@sanitize@url [0]{\catcode `\\12\catcode `\$12\catcode
  `\&12\catcode `\#12\catcode `\^12\catcode `\_12\catcode `\%12\relax}%
\providecommand \@@startlink[1]{}%
\providecommand \@@endlink[0]{}%
\providecommand \url  [0]{\begingroup\@sanitize@url \@url }%
\providecommand \@url [1]{\endgroup\@href {#1}{\urlprefix }}%
\providecommand \urlprefix  [0]{URL }%
\providecommand \Eprint [0]{\href }%
\providecommand \doibase [0]{https://doi.org/}%
\providecommand \selectlanguage [0]{\@gobble}%
\providecommand \bibinfo  [0]{\@secondoftwo}%
\providecommand \bibfield  [0]{\@secondoftwo}%
\providecommand \translation [1]{[#1]}%
\providecommand \BibitemOpen [0]{}%
\providecommand \bibitemStop [0]{}%
\providecommand \bibitemNoStop [0]{.\EOS\space}%
\providecommand \EOS [0]{\spacefactor3000\relax}%
\providecommand \BibitemShut  [1]{\csname bibitem#1\endcsname}%
\let\auto@bib@innerbib\@empty
\bibitem [{\citenamefont {Provost}\ and\ \citenamefont
  {Vallee}(1980)}]{Provost}%
  \BibitemOpen
  \bibfield  {author} {\bibinfo {author} {\bibfnamefont {J.~P.}\ \bibnamefont
  {Provost}}\ and\ \bibinfo {author} {\bibfnamefont {G.}~\bibnamefont
  {Vallee}},\ }\href {https://doi.org/10.1007/BF02193559} {\bibfield  {journal}
  {\bibinfo  {journal} {Commun. Math. Phys.}\ }\textbf {\bibinfo {volume}
  {76}},\ \bibinfo {pages} {289} (\bibinfo {year} {1980})}\BibitemShut
  {NoStop}%
\bibitem [{\citenamefont {Zanardi}\ \emph {et~al.}(2007)\citenamefont
  {Zanardi}, \citenamefont {Giorda},\ and\ \citenamefont
  {Cozzini}}]{Zanardi2007Information}%
  \BibitemOpen
  \bibfield  {author} {\bibinfo {author} {\bibfnamefont {P.}~\bibnamefont
  {Zanardi}}, \bibinfo {author} {\bibfnamefont {P.}~\bibnamefont {Giorda}},\
  and\ \bibinfo {author} {\bibfnamefont {M.}~\bibnamefont {Cozzini}},\ }\href
  {https://doi.org/10.1103/PhysRevLett.99.100603} {\bibfield  {journal}
  {\bibinfo  {journal} {Phys. Rev. Lett.}\ }\textbf {\bibinfo {volume} {99}},\
  \bibinfo {pages} {100603} (\bibinfo {year} {2007})}\BibitemShut {NoStop}%
\bibitem [{\citenamefont {Kumar}\ \emph {et~al.}(2012)\citenamefont {Kumar},
  \citenamefont {Mahapatra}, \citenamefont {Phukon},\ and\ \citenamefont
  {Sarkar}}]{SarkarGeodesics2012}%
  \BibitemOpen
  \bibfield  {author} {\bibinfo {author} {\bibfnamefont {P.}~\bibnamefont
  {Kumar}}, \bibinfo {author} {\bibfnamefont {S.}~\bibnamefont {Mahapatra}},
  \bibinfo {author} {\bibfnamefont {P.}~\bibnamefont {Phukon}},\ and\ \bibinfo
  {author} {\bibfnamefont {T.}~\bibnamefont {Sarkar}},\ }\href
  {https://doi.org/10.1103/PhysRevE.86.051117} {\bibfield  {journal} {\bibinfo
  {journal} {Phys. Rev. E}\ }\textbf {\bibinfo {volume} {86}},\ \bibinfo
  {pages} {051117} (\bibinfo {year} {2012})}\BibitemShut {NoStop}%
\bibitem [{\citenamefont {Gu}(2010)}]{GuReview}%
  \BibitemOpen
  \bibfield  {author} {\bibinfo {author} {\bibfnamefont {S.-J.}\ \bibnamefont
  {Gu}},\ }\href {https://doi.org/10.1142/S0217979210056335} {\bibfield
  {journal} {\bibinfo  {journal} {Int. J. of Mod. Phys. B}\ }\textbf {\bibinfo
  {volume} {24}},\ \bibinfo {pages} {4371} (\bibinfo {year}
  {2010})}\BibitemShut {NoStop}%
\bibitem [{\citenamefont {Carollo}\ \emph {et~al.}(2020)\citenamefont
  {Carollo}, \citenamefont {Valenti},\ and\ \citenamefont
  {Spagnolo}}]{Carollo2020}%
  \BibitemOpen
  \bibfield  {author} {\bibinfo {author} {\bibfnamefont {A.}~\bibnamefont
  {Carollo}}, \bibinfo {author} {\bibfnamefont {D.}~\bibnamefont {Valenti}},\
  and\ \bibinfo {author} {\bibfnamefont {B.}~\bibnamefont {Spagnolo}},\ }\href
  {http://www.sciencedirect.com/science/article/pii/S0370157319303655}
  {\bibfield  {journal} {\bibinfo  {journal} {Phys. Rep.}\ }\textbf {\bibinfo
  {volume} {838}},\ \bibinfo {pages} {1 } (\bibinfo {year} {2020})}\BibitemShut
  {NoStop}%
\bibitem [{\citenamefont {Zanardi}\ and\ \citenamefont
  {Rasetti}(1999)}]{Zanardi1999}%
  \BibitemOpen
  \bibfield  {author} {\bibinfo {author} {\bibfnamefont {P.}~\bibnamefont
  {Zanardi}}\ and\ \bibinfo {author} {\bibfnamefont {M.}~\bibnamefont
  {Rasetti}},\ }\href
  {https://doi.org/https://doi.org/10.1016/S0375-9601(99)00803-8} {\bibfield
  {journal} {\bibinfo  {journal} {Physics Letters A}\ }\textbf {\bibinfo
  {volume} {264}},\ \bibinfo {pages} {94} (\bibinfo {year} {1999})}\BibitemShut
  {NoStop}%
\bibitem [{\citenamefont {Rezakhani}\ \emph {et~al.}(2010)\citenamefont
  {Rezakhani}, \citenamefont {Abasto}, \citenamefont {Lidar},\ and\
  \citenamefont {Zanardi}}]{Rezakhani2010}%
  \BibitemOpen
  \bibfield  {author} {\bibinfo {author} {\bibfnamefont {A.~T.}\ \bibnamefont
  {Rezakhani}}, \bibinfo {author} {\bibfnamefont {D.~F.}\ \bibnamefont
  {Abasto}}, \bibinfo {author} {\bibfnamefont {D.~A.}\ \bibnamefont {Lidar}},\
  and\ \bibinfo {author} {\bibfnamefont {P.}~\bibnamefont {Zanardi}},\ }\href
  {https://doi.org/10.1103/PhysRevA.82.012321} {\bibfield  {journal} {\bibinfo
  {journal} {Phys. Rev. A}\ }\textbf {\bibinfo {volume} {82}},\ \bibinfo
  {pages} {012321} (\bibinfo {year} {2010})}\BibitemShut {NoStop}%
\bibitem [{\citenamefont {Ozawa}\ and\ \citenamefont {Goldman}(2018)}]{Ozawa1}%
  \BibitemOpen
  \bibfield  {author} {\bibinfo {author} {\bibfnamefont {T.}~\bibnamefont
  {Ozawa}}\ and\ \bibinfo {author} {\bibfnamefont {N.}~\bibnamefont
  {Goldman}},\ }\href {https://doi.org/10.1103/PhysRevB.97.201117} {\bibfield
  {journal} {\bibinfo  {journal} {Phys. Rev. B}\ }\textbf {\bibinfo {volume}
  {97}},\ \bibinfo {pages} {201117} (\bibinfo {year} {2018})}\BibitemShut
  {NoStop}%
\bibitem [{\citenamefont {Ozawa}\ and\ \citenamefont {Goldman}(2019)}]{Ozawa2}%
  \BibitemOpen
  \bibfield  {author} {\bibinfo {author} {\bibfnamefont {T.}~\bibnamefont
  {Ozawa}}\ and\ \bibinfo {author} {\bibfnamefont {N.}~\bibnamefont
  {Goldman}},\ }\href {https://doi.org/10.1103/PhysRevResearch.1.032019}
  {\bibfield  {journal} {\bibinfo  {journal} {Phys. Rev. Res.}\ }\textbf
  {\bibinfo {volume} {1}},\ \bibinfo {pages} {032019} (\bibinfo {year}
  {2019})}\BibitemShut {NoStop}%
\bibitem [{\citenamefont {Tan}\ \emph {et~al.}(2019)\citenamefont {Tan},
  \citenamefont {Zhang}, \citenamefont {Yang}, \citenamefont {Chu},
  \citenamefont {Zhu}, \citenamefont {Li}, \citenamefont {Yang}, \citenamefont
  {Song}, \citenamefont {Han}, \citenamefont {Li}, \citenamefont {Dong},
  \citenamefont {Yu}, \citenamefont {Yan}, \citenamefont {Zhu},\ and\
  \citenamefont {Yu}}]{Tan2019}%
  \BibitemOpen
  \bibfield  {author} {\bibinfo {author} {\bibfnamefont {X.}~\bibnamefont
  {Tan}}, \bibinfo {author} {\bibfnamefont {D.-W.}\ \bibnamefont {Zhang}},
  \bibinfo {author} {\bibfnamefont {Z.}~\bibnamefont {Yang}}, \bibinfo {author}
  {\bibfnamefont {J.}~\bibnamefont {Chu}}, \bibinfo {author} {\bibfnamefont
  {Y.-Q.}\ \bibnamefont {Zhu}}, \bibinfo {author} {\bibfnamefont
  {D.}~\bibnamefont {Li}}, \bibinfo {author} {\bibfnamefont {X.}~\bibnamefont
  {Yang}}, \bibinfo {author} {\bibfnamefont {S.}~\bibnamefont {Song}}, \bibinfo
  {author} {\bibfnamefont {Z.}~\bibnamefont {Han}}, \bibinfo {author}
  {\bibfnamefont {Z.}~\bibnamefont {Li}}, \bibinfo {author} {\bibfnamefont
  {Y.}~\bibnamefont {Dong}}, \bibinfo {author} {\bibfnamefont {H.-F.}\
  \bibnamefont {Yu}}, \bibinfo {author} {\bibfnamefont {H.}~\bibnamefont
  {Yan}}, \bibinfo {author} {\bibfnamefont {S.-L.}\ \bibnamefont {Zhu}},\ and\
  \bibinfo {author} {\bibfnamefont {Y.}~\bibnamefont {Yu}},\ }\href
  {https://doi.org/10.1103/PhysRevLett.122.210401} {\bibfield  {journal}
  {\bibinfo  {journal} {Phys. Rev. Lett.}\ }\textbf {\bibinfo {volume} {122}},\
  \bibinfo {pages} {210401} (\bibinfo {year} {2019})}\BibitemShut {NoStop}%
\bibitem [{\citenamefont {Yu}\ \emph {et~al.}(2019)\citenamefont {Yu},
  \citenamefont {Yang}, \citenamefont {Gong}, \citenamefont {Cao},
  \citenamefont {Lu}, \citenamefont {Liu}, \citenamefont {Zhang}, \citenamefont
  {Plenio}, \citenamefont {Jelezko}, \citenamefont {Ozawa}, \citenamefont
  {Goldman},\ and\ \citenamefont {Cai}}]{Yu2020}%
  \BibitemOpen
  \bibfield  {author} {\bibinfo {author} {\bibfnamefont {M.}~\bibnamefont
  {Yu}}, \bibinfo {author} {\bibfnamefont {P.}~\bibnamefont {Yang}}, \bibinfo
  {author} {\bibfnamefont {M.}~\bibnamefont {Gong}}, \bibinfo {author}
  {\bibfnamefont {Q.}~\bibnamefont {Cao}}, \bibinfo {author} {\bibfnamefont
  {Q.}~\bibnamefont {Lu}}, \bibinfo {author} {\bibfnamefont {H.}~\bibnamefont
  {Liu}}, \bibinfo {author} {\bibfnamefont {S.}~\bibnamefont {Zhang}}, \bibinfo
  {author} {\bibfnamefont {M.~B.}\ \bibnamefont {Plenio}}, \bibinfo {author}
  {\bibfnamefont {F.}~\bibnamefont {Jelezko}}, \bibinfo {author} {\bibfnamefont
  {T.}~\bibnamefont {Ozawa}}, \bibinfo {author} {\bibfnamefont
  {N.}~\bibnamefont {Goldman}},\ and\ \bibinfo {author} {\bibfnamefont
  {J.}~\bibnamefont {Cai}},\ }\href {https://doi.org/10.1093/nsr/nwz193}
  {\bibfield  {journal} {\bibinfo  {journal} {National Science Review}\
  }\textbf {\bibinfo {volume} {7}},\ \bibinfo {pages} {254} (\bibinfo {year}
  {2019})},\ \Eprint
  {https://arxiv.org/abs/https://academic.oup.com/nsr/article-pdf/7/2/254/38881668/nwz193.pdf}
  {https://academic.oup.com/nsr/article-pdf/7/2/254/38881668/nwz193.pdf}
  \BibitemShut {NoStop}%
\bibitem [{\citenamefont {Li}\ \emph {et~al.}(2022)\citenamefont {Li},
  \citenamefont {Chen},\ and\ \citenamefont {Cappellaro}}]{Cappellaro2022}%
  \BibitemOpen
  \bibfield  {author} {\bibinfo {author} {\bibfnamefont {C.}~\bibnamefont
  {Li}}, \bibinfo {author} {\bibfnamefont {M.}~\bibnamefont {Chen}},\ and\
  \bibinfo {author} {\bibfnamefont {P.}~\bibnamefont {Cappellaro}},\
  }\href@noop {} {\bibinfo {title} {A geometric perspective: experimental
  evaluation of the quantum cramer-rao bound}} (\bibinfo {year} {2022}),\
  \Eprint {https://arxiv.org/abs/2204.13777} {arXiv:2204.13777 [quant-ph]}
  \BibitemShut {NoStop}%
\bibitem [{\citenamefont {Bender}\ and\ \citenamefont {Wu}(1969)}]{Bender1}%
  \BibitemOpen
  \bibfield  {author} {\bibinfo {author} {\bibfnamefont {C.~M.}\ \bibnamefont
  {Bender}}\ and\ \bibinfo {author} {\bibfnamefont {T.~T.}\ \bibnamefont
  {Wu}},\ }\href {https://doi.org/10.1103/PhysRev.184.1231} {\bibfield
  {journal} {\bibinfo  {journal} {Phys. Rev.}\ }\textbf {\bibinfo {volume}
  {184}},\ \bibinfo {pages} {1231} (\bibinfo {year} {1969})}\BibitemShut
  {NoStop}%
\bibitem [{\citenamefont {Bender}\ and\ \citenamefont {Wu}(1973)}]{Bender2}%
  \BibitemOpen
  \bibfield  {author} {\bibinfo {author} {\bibfnamefont {C.~M.}\ \bibnamefont
  {Bender}}\ and\ \bibinfo {author} {\bibfnamefont {T.~T.}\ \bibnamefont
  {Wu}},\ }\href {https://doi.org/10.1103/PhysRevD.7.1620} {\bibfield
  {journal} {\bibinfo  {journal} {Phys. Rev. D}\ }\textbf {\bibinfo {volume}
  {7}},\ \bibinfo {pages} {1620} (\bibinfo {year} {1973})}\BibitemShut
  {NoStop}%
\bibitem [{\citenamefont {Lipatov}(1977)}]{Lipatov}%
  \BibitemOpen
  \bibfield  {author} {\bibinfo {author} {\bibfnamefont {L.~N.}\ \bibnamefont
  {Lipatov}},\ }\href@noop {} {\bibfield  {journal} {\bibinfo  {journal} {Sov.
  Phys. JETP}\ }\textbf {\bibinfo {volume} {45}},\ \bibinfo {pages} {216}
  (\bibinfo {year} {1977})}\BibitemShut {NoStop}%
\bibitem [{\citenamefont {Br\'ezin}\ \emph {et~al.}(1977)\citenamefont
  {Br\'ezin}, \citenamefont {Le~Guillou},\ and\ \citenamefont
  {Zinn-Justin}}]{Brezin}%
  \BibitemOpen
  \bibfield  {author} {\bibinfo {author} {\bibfnamefont {E.}~\bibnamefont
  {Br\'ezin}}, \bibinfo {author} {\bibfnamefont {J.~C.}\ \bibnamefont
  {Le~Guillou}},\ and\ \bibinfo {author} {\bibfnamefont {J.}~\bibnamefont
  {Zinn-Justin}},\ }\href {https://doi.org/10.1103/PhysRevD.15.1544} {\bibfield
   {journal} {\bibinfo  {journal} {Phys. Rev. D}\ }\textbf {\bibinfo {volume}
  {15}},\ \bibinfo {pages} {1544} (\bibinfo {year} {1977})}\BibitemShut
  {NoStop}%
\bibitem [{\citenamefont {Seznec}\ and\ \citenamefont
  {Zinn‐Justin}(2008)}]{Zinn1}%
  \BibitemOpen
  \bibfield  {author} {\bibinfo {author} {\bibfnamefont {R.}~\bibnamefont
  {Seznec}}\ and\ \bibinfo {author} {\bibfnamefont {J.}~\bibnamefont
  {Zinn‐Justin}},\ }\href {https://doi.org/10.1063/1.524247} {\bibfield
  {journal} {\bibinfo  {journal} {Journal of Mathematical Physics}\ }\textbf
  {\bibinfo {volume} {20}},\ \bibinfo {pages} {1398} (\bibinfo {year}
  {2008})},\ \Eprint
  {https://arxiv.org/abs/https://pubs.aip.org/aip/jmp/article-pdf/20/7/1398/11050118/1398\_1\_online.pdf}
  {https://pubs.aip.org/aip/jmp/article-pdf/20/7/1398/11050118/1398\_1\_online.pdf}
  \BibitemShut {NoStop}%
\bibitem [{\citenamefont {Zinn‐Justin}(1981)}]{Zinn2}%
  \BibitemOpen
  \bibfield  {author} {\bibinfo {author} {\bibfnamefont {J.}~\bibnamefont
  {Zinn‐Justin}},\ }\href {https://doi.org/10.1063/1.524919} {\bibfield
  {journal} {\bibinfo  {journal} {Journal of Mathematical Physics}\ }\textbf
  {\bibinfo {volume} {22}},\ \bibinfo {pages} {511} (\bibinfo {year} {1981})},\
  \Eprint
  {https://arxiv.org/abs/https://pubs.aip.org/aip/jmp/article-pdf/22/3/511/7428066/511\_1\_online.pdf}
  {https://pubs.aip.org/aip/jmp/article-pdf/22/3/511/7428066/511\_1\_online.pdf}
  \BibitemShut {NoStop}%
\bibitem [{\citenamefont {Zinn-Justin}(1981{\natexlab{a}})}]{Zinn3}%
  \BibitemOpen
  \bibfield  {author} {\bibinfo {author} {\bibfnamefont {J.}~\bibnamefont
  {Zinn-Justin}},\ }\href
  {https://doi.org/https://doi.org/10.1016/0370-1573(81)90016-8} {\bibfield
  {journal} {\bibinfo  {journal} {Physics Reports}\ }\textbf {\bibinfo {volume}
  {70}},\ \bibinfo {pages} {109} (\bibinfo {year}
  {1981}{\natexlab{a}})}\BibitemShut {NoStop}%
\bibitem [{\citenamefont {Zinn-Justin}(2010)}]{Zinn4}%
  \BibitemOpen
  \bibfield  {author} {\bibinfo {author} {\bibfnamefont {J.}~\bibnamefont
  {Zinn-Justin}},\ }\href {https://books.google.com.mx/books?id=bc2xQwAACAAJ}
  {\emph {\bibinfo {title} {Path Integrals in Quantum Mechanics}}},\ Oxford
  Graduate Texts\ (\bibinfo  {publisher} {OUP Oxford},\ \bibinfo {year}
  {2010})\BibitemShut {NoStop}%
\bibitem [{\citenamefont {Zinn-Justin}(1981{\natexlab{b}})}]{Zinn5}%
  \BibitemOpen
  \bibfield  {author} {\bibinfo {author} {\bibfnamefont {J.}~\bibnamefont
  {Zinn-Justin}},\ }\href
  {https://doi.org/https://doi.org/10.1016/0550-3213(81)90197-8} {\bibfield
  {journal} {\bibinfo  {journal} {Nuclear Physics B}\ }\textbf {\bibinfo
  {volume} {192}},\ \bibinfo {pages} {125} (\bibinfo {year}
  {1981}{\natexlab{b}})}\BibitemShut {NoStop}%
\bibitem [{\citenamefont {Zinn-Justin}(1983)}]{Zinn6}%
  \BibitemOpen
  \bibfield  {author} {\bibinfo {author} {\bibfnamefont {J.}~\bibnamefont
  {Zinn-Justin}},\ }\href
  {https://doi.org/https://doi.org/10.1016/0550-3213(83)90369-3} {\bibfield
  {journal} {\bibinfo  {journal} {Nuclear Physics B}\ }\textbf {\bibinfo
  {volume} {218}},\ \bibinfo {pages} {333} (\bibinfo {year}
  {1983})}\BibitemShut {NoStop}%
\bibitem [{\citenamefont {Caswell}(1979)}]{Caswell}%
  \BibitemOpen
  \bibfield  {author} {\bibinfo {author} {\bibfnamefont {W.~E.}\ \bibnamefont
  {Caswell}},\ }\href
  {https://doi.org/https://doi.org/10.1016/0003-4916(79)90269-0} {\bibfield
  {journal} {\bibinfo  {journal} {Annals of Physics}\ }\textbf {\bibinfo
  {volume} {123}},\ \bibinfo {pages} {153} (\bibinfo {year}
  {1979})}\BibitemShut {NoStop}%
\bibitem [{\citenamefont {Turbiner}(2005)}]{Turbiner1}%
  \BibitemOpen
  \bibfield  {author} {\bibinfo {author} {\bibfnamefont {A.}~\bibnamefont
  {Turbiner}},\ }\href {https://doi.org/10.1007/s11005-005-0012-z} {\bibfield
  {journal} {\bibinfo  {journal} {Letters in Mathematical Physics}\ }\textbf
  {\bibinfo {volume} {74}},\ \bibinfo {pages} {169} (\bibinfo {year}
  {2005})}\BibitemShut {NoStop}%
\bibitem [{\citenamefont {Jentschura}\ and\ \citenamefont
  {Zinn-Justin}(2010)}]{Zinn7}%
  \BibitemOpen
  \bibfield  {author} {\bibinfo {author} {\bibfnamefont {U.~D.}\ \bibnamefont
  {Jentschura}}\ and\ \bibinfo {author} {\bibfnamefont {J.}~\bibnamefont
  {Zinn-Justin}},\ }\href
  {https://doi.org/https://doi.org/10.1016/j.apnum.2010.03.015} {\bibfield
  {journal} {\bibinfo  {journal} {Applied Numerical Mathematics}\ }\textbf
  {\bibinfo {volume} {60}},\ \bibinfo {pages} {1332} (\bibinfo {year}
  {2010})},\ \bibinfo {note} {approximation and extrapolation of convergent and
  divergent sequences and series (CIRM, Luminy - France, 2009)}\BibitemShut
  {NoStop}%
\bibitem [{\citenamefont {Turbiner}\ and\ \citenamefont {del
  Valle}(2021)}]{Turbiner2}%
  \BibitemOpen
  \bibfield  {author} {\bibinfo {author} {\bibfnamefont {A.~V.}\ \bibnamefont
  {Turbiner}}\ and\ \bibinfo {author} {\bibfnamefont {J.~C.}\ \bibnamefont {del
  Valle}},\ }\href {https://doi.org/10.1088/1751-8121/ac0733} {\bibfield
  {journal} {\bibinfo  {journal} {Journal of Physics A: Mathematical and
  Theoretical}\ }\textbf {\bibinfo {volume} {54}},\ \bibinfo {pages} {295204}
  (\bibinfo {year} {2021})}\BibitemShut {NoStop}%
\bibitem [{\citenamefont {Turbiner}\ and\ \citenamefont {del
  Valle}(2022)}]{Turbiner3}%
  \BibitemOpen
  \bibfield  {author} {\bibinfo {author} {\bibfnamefont {A.~V.}\ \bibnamefont
  {Turbiner}}\ and\ \bibinfo {author} {\bibfnamefont {J.~C.}\ \bibnamefont {del
  Valle}},\ }\href {https://doi.org/10.14311/AP.2022.62.0208} {\bibfield
  {journal} {\bibinfo  {journal} {Acta Polytechnica}\ }\textbf {\bibinfo
  {volume} {62}},\ \bibinfo {pages} {208–210} (\bibinfo {year}
  {2022})}\BibitemShut {NoStop}%
\bibitem [{\citenamefont {Nader}\ \emph {et~al.}(2023)\citenamefont {Nader},
  \citenamefont {Hernandez-Gonzalez}, \citenamefont {Vazquez-Sanchez},\ and\
  \citenamefont {Lerma-Hernandez}}]{Nader2023}%
  \BibitemOpen
  \bibfield  {author} {\bibinfo {author} {\bibfnamefont {D.~.~J.}\ \bibnamefont
  {Nader}}, \bibinfo {author} {\bibfnamefont {J.~R.}\ \bibnamefont
  {Hernandez-Gonzalez}}, \bibinfo {author} {\bibfnamefont {H.}~\bibnamefont
  {Vazquez-Sanchez}},\ and\ \bibinfo {author} {\bibfnamefont {S.}~\bibnamefont
  {Lerma-Hernandez}},\ }\href@noop {} {\bibinfo {title} {Manifestation of
  instability in the quasiclassical limit of the spectrum of the quartic double
  well}} (\bibinfo {year} {2023}),\ \Eprint {https://arxiv.org/abs/2302.14211}
  {arXiv:2302.14211 [quant-ph]} \BibitemShut {NoStop}%
\bibitem [{\citenamefont {Okun}\ and\ \citenamefont {Burke}(2021)}]{Okun}%
  \BibitemOpen
  \bibfield  {author} {\bibinfo {author} {\bibfnamefont {P.}~\bibnamefont
  {Okun}}\ and\ \bibinfo {author} {\bibfnamefont {K.}~\bibnamefont {Burke}},\
  }\href {https://doi.org/https://doi.org/10.1002/qua.26554} {\bibfield
  {journal} {\bibinfo  {journal} {International Journal of Quantum Chemistry}\
  }\textbf {\bibinfo {volume} {121}},\ \bibinfo {pages} {e26554} (\bibinfo
  {year} {2021})},\ \Eprint
  {https://arxiv.org/abs/https://onlinelibrary.wiley.com/doi/pdf/10.1002/qua.26554}
  {https://onlinelibrary.wiley.com/doi/pdf/10.1002/qua.26554} \BibitemShut
  {NoStop}%
\bibitem [{\citenamefont {Turbiner}(1984)}]{Turbiner1984}%
  \BibitemOpen
  \bibfield  {author} {\bibinfo {author} {\bibfnamefont {A.~V.}\ \bibnamefont
  {Turbiner}},\ }\href
  {http://refhub.elsevier.com/S0370-1573(16)30132-6/sbref51} {\bibfield
  {journal} {\bibinfo  {journal} {Sov. Phys. Usp.}\ }\textbf {\bibinfo {volume}
  {27}},\ \bibinfo {pages} {668} (\bibinfo {year} {1984})}\BibitemShut
  {NoStop}%
\bibitem [{\citenamefont {Gonzalez}\ \emph {et~al.}(2023)\citenamefont
  {Gonzalez}, \citenamefont {Guti\'errez-Ruiz},\ and\ \citenamefont
  {Vergara}}]{Diego5}%
  \BibitemOpen
  \bibfield  {author} {\bibinfo {author} {\bibfnamefont {D.}~\bibnamefont
  {Gonzalez}}, \bibinfo {author} {\bibfnamefont {D.}~\bibnamefont
  {Guti\'errez-Ruiz}},\ and\ \bibinfo {author} {\bibfnamefont {J.~D.}\
  \bibnamefont {Vergara}},\ }\bibfield  {journal} {\bibinfo  {journal} {Work in
  preparation}\ }\href@noop {} {} (\bibinfo {year} {2023})\BibitemShut
  {NoStop}%
\bibitem [{\citenamefont {Alvarez-Jimenez}\ \emph {et~al.}(2017)\citenamefont
  {Alvarez-Jimenez}, \citenamefont {Dector},\ and\ \citenamefont
  {Vergara}}]{Alvarez2017}%
  \BibitemOpen
  \bibfield  {author} {\bibinfo {author} {\bibfnamefont {J.}~\bibnamefont
  {Alvarez-Jimenez}}, \bibinfo {author} {\bibfnamefont {A.}~\bibnamefont
  {Dector}},\ and\ \bibinfo {author} {\bibfnamefont {J.~D.}\ \bibnamefont
  {Vergara}},\ }\href {https://doi.org/10.1007/JHEP03(2017)044} {\bibfield
  {journal} {\bibinfo  {journal} {J. High Energy Phys.}\ }\textbf {\bibinfo
  {volume} {2017}}\bibinfo  {number} { (03)},\ \bibinfo {pages}
  {044}}\BibitemShut {NoStop}%
\bibitem [{\citenamefont {Campos~Venuti}\ and\ \citenamefont
  {Zanardi}(2007)}]{Zanardi2007Scaling}%
  \BibitemOpen
\bibfield  {number} {  }\bibfield  {author} {\bibinfo {author} {\bibfnamefont
  {L.}~\bibnamefont {Campos~Venuti}}\ and\ \bibinfo {author} {\bibfnamefont
  {P.}~\bibnamefont {Zanardi}},\ }\href
  {https://doi.org/10.1103/PhysRevLett.99.095701} {\bibfield  {journal}
  {\bibinfo  {journal} {Phys. Rev. Lett.}\ }\textbf {\bibinfo {volume} {99}},\
  \bibinfo {pages} {095701} (\bibinfo {year} {2007})}\BibitemShut {NoStop}%
\bibitem [{\citenamefont {Sokolnikoff}(1951)}]{Sokol}%
  \BibitemOpen
  \bibfield  {author} {\bibinfo {author} {\bibfnamefont {I.}~\bibnamefont
  {Sokolnikoff}},\ }\href@noop {} {\emph {\bibinfo {title} {Tensor Analysis:
  Theory and Applications}}},\ Applied mathematics series\ (\bibinfo
  {publisher} {Wiley},\ \bibinfo {year} {1951})\BibitemShut {NoStop}%
\bibitem [{\citenamefont {Gonzalez}\ \emph {et~al.}(2019)\citenamefont
  {Gonzalez}, \citenamefont {Guti\'errez-Ruiz},\ and\ \citenamefont
  {Vergara}}]{GonzalezPRE}%
  \BibitemOpen
  \bibfield  {author} {\bibinfo {author} {\bibfnamefont {D.}~\bibnamefont
  {Gonzalez}}, \bibinfo {author} {\bibfnamefont {D.}~\bibnamefont
  {Guti\'errez-Ruiz}},\ and\ \bibinfo {author} {\bibfnamefont {J.~D.}\
  \bibnamefont {Vergara}},\ }\href {https://doi.org/10.1103/PhysRevE.99.032144}
  {\bibfield  {journal} {\bibinfo  {journal} {Phys. Rev. E}\ }\textbf {\bibinfo
  {volume} {99}},\ \bibinfo {pages} {032144} (\bibinfo {year}
  {2019})}\BibitemShut {NoStop}%
\bibitem [{\citenamefont {Alvarez-Jimenez}\ \emph {et~al.}(2020)\citenamefont
  {Alvarez-Jimenez}, \citenamefont {Gonzalez}, \citenamefont
  {Guti\'errez-Ruiz},\ and\ \citenamefont {Vergara}}]{GonzalezAnnalen}%
  \BibitemOpen
  \bibfield  {author} {\bibinfo {author} {\bibfnamefont {J.}~\bibnamefont
  {Alvarez-Jimenez}}, \bibinfo {author} {\bibfnamefont {D.}~\bibnamefont
  {Gonzalez}}, \bibinfo {author} {\bibfnamefont {D.}~\bibnamefont
  {Guti\'errez-Ruiz}},\ and\ \bibinfo {author} {\bibfnamefont {J.~D.}\
  \bibnamefont {Vergara}},\ }\href {https://doi.org/10.1002/andp.201900215}
  {\bibfield  {journal} {\bibinfo  {journal} {Ann. Phys. (Berlin)}\ }\textbf
  {\bibinfo {volume} {532}},\ \bibinfo {pages} {1900215} (\bibinfo {year}
  {2020})}\BibitemShut {NoStop}%
\bibitem [{\citenamefont {Inc.}()}]{Mathematica}%
  \BibitemOpen
  \bibfield  {author} {\bibinfo {author} {\bibfnamefont {W.~R.}\ \bibnamefont
  {Inc.}},\ }\href {https://www.wolfram.com/mathematica} {\bibinfo {title}
  {Mathematica, {V}ersion 13.2}},\ \bibinfo {note} {{C}hampaign, IL,
  2022}\BibitemShut {NoStop}%
\bibitem [{\citenamefont {Dittrich}\ and\ \citenamefont
  {Reuter}(2020)}]{dittrich2020}%
  \BibitemOpen
  \bibfield  {author} {\bibinfo {author} {\bibfnamefont {W.}~\bibnamefont
  {Dittrich}}\ and\ \bibinfo {author} {\bibfnamefont {M.}~\bibnamefont
  {Reuter}},\ }\href@noop {} {\emph {\bibinfo {title} {Classical and Quantum
  Dynamics}}}\ (\bibinfo  {publisher} {Springer Nature},\ \bibinfo {year}
  {2020})\BibitemShut {NoStop}%
\bibitem [{\citenamefont {Goldstein}\ \emph {et~al.}(2000)\citenamefont
  {Goldstein}, \citenamefont {Poole},\ and\ \citenamefont
  {Safko}}]{goldstein2000}%
  \BibitemOpen
  \bibfield  {author} {\bibinfo {author} {\bibfnamefont {H.}~\bibnamefont
  {Goldstein}}, \bibinfo {author} {\bibfnamefont {C.}~\bibnamefont {Poole}},\
  and\ \bibinfo {author} {\bibfnamefont {J.}~\bibnamefont {Safko}},\
  }\href@noop {} {\emph {\bibinfo {title} {Classical Mechanics}}},\ \bibinfo
  {edition} {3rd}\ ed.\ (\bibinfo  {publisher} {Addison Wesley, Boston, MA},\
  \bibinfo {year} {2000})\BibitemShut {NoStop}%
\end{thebibliography}%

\end{document}